\documentclass[12pt]{article}
\usepackage{eqsection,subeqnarray,indent,amsfonts,amssymb}
\usepackage{pstricks,pst-node,pst-text,pst-3d,pst-plot}
\usepackage{amsmath}     
\usepackage{bm}          
\usepackage{graphicx}
\usepackage{cite}

\def\today{Mar 9, 2007} 
\begin{document}

\bibliographystyle{plain}

\date{\today}

\title{Phase diagram of the chromatic polynomial on a torus} 

\author{
  {\small Jesper Lykke Jacobsen${}^{1,2}$ and
          Jes\'us Salas${}^3$}                           \\[1mm]
  {\small\it ${}^1$Laboratoire de Physique Th\'eorique
  et Mod\`eles Statistiques}                             \\[-0.2cm]
  {\small\it Universit\'e Paris-Sud}                     \\[-0.2cm]
  {\small\it B\^atiment 100}                             \\[-0.2cm]
  {\small\it 91405 Orsay, FRANCE}                        \\[-0.2cm]
  {\small\tt JESPER.JACOBSEN@U-PSUD.FR}                  \\[1mm]
  {\small\it ${}^2$Service de Physique Th\'eorique}      \\[-0.2cm]
  {\small\it CEA Saclay}                                 \\[-0.2cm]
  {\small\it Orme des Merisiers}                         \\[-0.2cm]
  {\small\it 91191 Gif sur Yvette, FRANCE}               \\[1mm]
  {\small\it ${}^3$Grupo de Modelizaci\'on, Simulaci\'on Num\'erica 
                   y Matem\'atica Industrial}  \\[-0.2cm]
  {\small\it Universidad Carlos III de Madrid} \\[-0.2cm]
  {\small\it Avda.\  de la Universidad, 30}    \\[-0.2cm]
  {\small\it 28911 Legan\'es, SPAIN}           \\[-0.2cm]
  {\small\tt JSALAS@MATH.UC3M.ES}              \\[-0.2cm]
  {\protect\makebox[5in]{\quad}}  
  \\
}

\maketitle
\thispagestyle{empty}   

\begin{abstract} 
We study the zero-temperature partition function of the
Potts antiferromagnet (i.e., the chromatic polynomial) 
on a torus using a transfer-matrix approach.
We consider square- and triangular-lattice strips with fixed width 
$L$, arbitrary length $N$, and fully periodic boundary conditions.
On the mathematical side, we obtain exact expressions for the 
chromatic polynomial of widths $L=5,6,7$ for the square and triangular 
lattices. 
On the physical side, we obtain the exact {\em phase diagrams} for these
strips of width $L$ and infinite length, and from these results we 
extract useful information about the infinite-volume phase diagram
of this model: in particular, the number and position of the
different phases. 
\end{abstract} 


\clearpage

\newcommand{\be}{\begin{equation}}
\newcommand{\ee}{\end{equation}}
\newcommand{\<}{\langle}
\renewcommand{\>}{\rangle}
\newcommand{\widebar}{\overline}
\def\reff#1{(\protect\ref{#1})}
\def\spose#1{\hbox to 0pt{#1\hss}}
\def\ltapprox{\mathrel{\spose{\lower 3pt\hbox{$\mathchar"218$}}
 \raise 2.0pt\hbox{$\mathchar"13C$}}}
\def\gtapprox{\mathrel{\spose{\lower 3pt\hbox{$\mathchar"218$}}
 \raise 2.0pt\hbox{$\mathchar"13E$}}}
\def\textprime{${}^\prime$}
\def\proof{\par\medskip\noindent{\sc Proof.\ }}
\def\qed{\hbox{\hskip 6pt\vrule width6pt height7pt depth1pt \hskip1pt}\bigskip}
\def\proofof#1{\bigskip\noindent{\sc Proof of #1.\ }}
\def\half{ {1 \over 2} }
\def\third{ {1 \over 3} }
\def\twothird{ {2 \over 3} }
\def\smfrac#1#2{\textstyle{#1\over #2}}
\def\smhalf{ \smfrac{1}{2} }
\newcommand{\real}{\mathop{\rm Re}\nolimits}
\renewcommand{\Re}{\mathop{\rm Re}\nolimits}
\newcommand{\imag}{\mathop{\rm Im}\nolimits}
\renewcommand{\Im}{\mathop{\rm Im}\nolimits}
\newcommand{\sgn}{\mathop{\rm sgn}\nolimits}
\newcommand{\tr}{\mathop{\rm tr}\nolimits}
\newcommand{\diag}{\mathop{\rm diag}\nolimits}
\newcommand{\Gal}{\mathop{\rm Gal}\nolimits}
\newcommand{\mycup}{\mathop{\cup}}
\newcommand{\Arg}{\mathop{\rm Arg}\nolimits}
\def\hboxscript#1{ {\hbox{\scriptsize\em #1}} }
\def\zhat{ {\widehat{Z}} }
\def\phat{ {\widehat{P}} }
\def\qtilde{ {\widetilde{q}} }
\renewcommand{\mod}{\mathop{\rm mod}\nolimits}
\renewcommand{\emptyset}{\varnothing}

\def\scra{\mathcal{A}}
\def\scrb{\mathcal{B}}
\def\scrc{\mathcal{C}}
\def\scrd{\mathcal{D}}
\def\scrf{\mathcal{F}}
\def\scrg{\mathcal{G}}
\def\scrl{\mathcal{L}}
\def\scro{\mathcal{O}}
\def\scrp{\mathcal{P}}
\def\scrq{\mathcal{Q}}
\def\scrr{\mathcal{R}}
\def\scrs{\mathcal{S}}
\def\scrt{\mathcal{T}}
\def\scrv{\mathcal{V}}
\def\scrz{\mathcal{Z}}

\def\q{{\sf q}}

\def\Z{{\mathbb Z}}
\def\R{{\mathbb R}}
\def\C{{\mathbb C}}
\def\Q{{\mathbb Q}}
\def\N{{\mathbb N}}

\def\T{{\mathsf T}}
\def\H{{\mathsf H}}
\def\V{{\mathsf V}}
\def\D{{\mathsf D}}
\def\J{{\mathsf J}}
\def\P{{\mathsf P}}
\def\QQ{{\mathsf Q}}
\def\RR{{\mathsf R}}

\def\bsigma{{\boldsymbol{\sigma}}}
\def\bone{\bm{1}}
\def\vv{\bm{v}}
\def\uu{\bm{u}}
\def\ww{\bm{w}}

\newtheorem{theorem}{Theorem}[section]
\newtheorem{definition}[theorem]{Definition}
\newtheorem{proposition}[theorem]{Proposition}
\newtheorem{lemma}[theorem]{Lemma}
\newtheorem{corollary}[theorem]{Corollary}
\newtheorem{conjecture}[theorem]{Conjecture}


\newenvironment{sarray}{
          \textfont0=\scriptfont0
          \scriptfont0=\scriptscriptfont0
          \textfont1=\scriptfont1
          \scriptfont1=\scriptscriptfont1
          \textfont2=\scriptfont2
          \scriptfont2=\scriptscriptfont2
          \textfont3=\scriptfont3
          \scriptfont3=\scriptscriptfont3
        \renewcommand{\arraystretch}{0.7}
        \begin{array}{l}}{\end{array}}

\newenvironment{scarray}{
          \textfont0=\scriptfont0
          \scriptfont0=\scriptscriptfont0
          \textfont1=\scriptfont1
          \scriptfont1=\scriptscriptfont1
          \textfont2=\scriptfont2
          \scriptfont2=\scriptscriptfont2
          \textfont3=\scriptfont3
          \scriptfont3=\scriptscriptfont3
        \renewcommand{\arraystretch}{0.7}
        \begin{array}{c}}{\end{array}}

\clearpage

%
%
\section{Introduction} \label{sec.intro} 

The two--dimensional (2D) $q$--state Potts model \cite{Potts} is one
of the most studied models in Statistical Mechanics. Despite many
efforts over more than 50 years, its {\em exact} free energy and phase
diagram are still unknown.  The ferromagnetic regime of the Potts
model is the best understood case: exact (albeit not always rigorous)
results have been obtained for the ferromagnetic-paramagnetic phase
transition temperature $T_{\rm c}(q)$ (at least for several regular
lattices), the order of the transition (continuous for $0 \le q \le
4$), the phase diagram, and the characterization in terms of conformal
field theory (CFT) of the universality classes. (See e.g.,
Ref.~\cite{Baxter_82}.)

The antiferromagnetic regime is less understood, partly because universality 
is not expected to hold in general (in contrast with the ferromagnetic 
regime). We know the exact free energy along some curves of the
phase diagram $(q,T)$ (where $T$ is the temperature) for some regular 
lattices \cite{Baxter_82}; but in this regime, this (partial) solubility 
of the model does not imply criticality (as it is been observed for the 
ferromagnetic regime with $0 \le q \le 4$). 

The standard $q$-state Potts model at temperature $T$ can be defined on any
undirected finite graph $G=(V,E)$ as follows: on every vertex of the graph
$x\in V$, we place a spin $\sigma_x$ taking an integer value in the set
$\{1,2,\ldots,q\}$, where $q\in\N$.
The spins interact via the Hamiltonian
\be
\mathcal{H}_\text{Potts}( \{\sigma\}) \;=\; 
-J \sum\limits_{\<x,y\> \in E} \delta_{\sigma_x,\sigma_y}
\label{def_H_Potts}
\ee
where the sum is over all edges $e=\<x,y\>\in E$ joining nearest-neighbor
sites, $J$ is the coupling constant, and $\delta$ is the Kronecker delta. 
The ferromagnetic (resp.\ antiferromagnetic) regime corresponds to $J>0$ 
(resp.\  $J<0$). The partition function of the Potts model 
on a graph $G$ at inverse temperature $\beta = 1/T\geq 0$ can then be 
written as
\be
Z_G(q;J\beta) \;=\; 
\sum\limits_{\{\sigma\}} e^{-\beta \mathcal{H}_\text{Potts}}
\label{def_Z_Potts} 
\ee

A very useful representation of the Potts model is due to Fortuin and
Kasteleyn \cite{Kasteleyn}. They showed that the partition function
\reff{def_Z_Potts} can be rewritten as
\be
Z_G(q;v) \;=\; \sum\limits_{E'\subseteq E} q^{k(E')}\, v^{|E'|} 
\label{def_Z_FK} 
\ee
where $v=e^{J\beta}-1$, the sum is over the $2^{|E|}$ spanning 
subgraphs $(V,E')$ of $G$, and $k(E')$ is the number of connected components 
(including isolated vertices) of the spanning subgraph $(V,E')$. As a result, 
the partition function \reff{def_Z_FK} is a polynomial in both 
variables $q$ and $v$. Thus, we can promote \reff{def_Z_FK} to the definition  
of the model, which allows us to consider $q$ as an arbitrary {\em complex}
number.

We shall here focus on the zero-temperature antiferromagnetic case
$J = -\infty$ (i.e., $v = -1$), whose restriction to $q\in\N$ can be 
interpreted as a coloring problem. In this case, $P_G(q) = Z_G(q,-1)$ 
is known as the {\em chromatic polynomial} of $G$. 
Its evaluation for a general graph is a hard problem. More precisely,
Oxley and Welsh have proven that the computation of the coefficients 
$a_k$ of the chromatic polynomial $P_G(q)=\sum_{k=1}^{|V|} a_k \, q^k$
of a general graph $G$ (including bipartite graphs) is  
\#-P hard \cite{Oxley_02}. 

In this paper, we will consider only strip graphs of certain regular 
lattices (i.e., square and triangular).
For a family of strip graphs $\{G_n\}$ (or, more generally, recursive 
families of graphs \cite{Noy_04}), the chromatic polynomial $P_{G_n}(q)$ 
of any member of the family $G_n$ can be computed from a transfer matrix 
$\T$ and certain boundary condition vectors $\uu$ and $\vv$:
\be
P_{G_n}  \;=\; \uu^{\text{T}} \cdot \T^n \cdot \vv \,.
\label{def_PGn}
\ee
Thus, the computation time grows as a polynomial in $n$ for {\em fixed}
strip width $L$. However, it still grows exponentially in the strip width 
(due to the $L$ dependence of $\dim \T$). It is therefore of interest to
devise new and efficient algorithms to be able to handle as large widths
$L$ as possible. 

\bigskip

\noindent
{\bf Remark.}
It is important to stress that the Potts spin model has a probabilistic 
interpretation (i.e., has non-negative Boltzmann weights) only when $q$ is 
a positive integer and $v \geq -1$ (i.e., positive temperature). 
The Fortuin-Kasteleyn random-cluster model \reff{def_Z_FK}, which extends
the Potts model to non-integer $q$, has non-negative weights only when 
$q,v\geq  0$. 
In all other cases, the model belongs to the ``unphysical'' regime (i.e., some
weights are negative or complex), and some familiar properties of 
statistical mechanics need not hold. For example, even for integer 
$q \geq 2$ and real $v$ in the antiferromagnetic range $-1 \leq  v < 0$, 
where the spin representation exists and has non-negative weights,
the dominant eigenvalue $\lambda_\star$ of the transfer matrix in the 
cluster representation need not be simple (i.e., the eigenvector may not 
be unique); and even if simple, it may not play any role in determining 
$Z_{G_n}(q, v)$ because the corresponding amplitude
may vanish. Both these behaviors are of course impossible for any transfer
matrix with positive weights, by virtue of the Perron-Frobenius theorem. It is
important to note that the dominant eigenvalues in the cluster and spin 
representations need not be equal, because the former may have a 
vanishing amplitude (see below). In any case, the Potts model can thus 
make probabilistic sense in the cluster representation at parameter values 
where it fails to make probabilistic sense in the spin representation, and 
vice versa. It is worth mentioning that for
$q = 4\cos^2(\pi/n)$ with integer $n \geq 3$ and planar graphs $G$, 
the Potts-model partition function $Z_G(q, v)$ admits a third representation, 
in terms of a restricted solid-on-solid (RSOS) model
\cite{Pasquier_alone,ABF} (See also Ref.~\cite{RSOS} for an recent study
with a more comprehensive list of references). 

Even though the zero-temperature limit of the Potts antiferromagnet may
not have a probabilistic interpretation in the Fortuin--Kasteleyn 
representation \cite{Kasteleyn} for non-integer $q$, it can nevertheless 
being studied with the standard tools of Statistical Mechanics and CFT 
with appropriate modifications. In particular, we expect that critical 
``unphysical regions'' can be described with non-unitary field theories.
(See \cite{forests,transfer4} for a more detailed discussion on these points.) 

\bigskip

It is well established that antiferromagnetic models in general, and the 
chromatic polynomial in particular, are very sensitive to the choice of 
boundary conditions. Indeed, different choices may lead to quite different 
thermodynamic limits. In our earlier publications 
\cite{transfer1,transfer2,transfer3} we studied free and cylindrical
boundary conditions: free boundary conditions in the longitudinal direction
and free or periodic boundary conditions in the transverse direction, 
respectively. More recently \cite{transfer4}, we considered cyclic boundary
conditions: namely, free boundary conditions in the transverse direction,
and periodic boundary conditions in the longitudinal direction. In particular,
we observed that in the latter case, the phase diagram of the 
triangular-lattice chromatic polynomial was more involved than for free or 
cylindrical boundary conditions. In this paper we shall deal with fully
periodic boundary conditions (i.e., periodic in both directions). 
Periodic boundary conditions in the longitudinal direction (i.e., toroidal and
cyclic) can be easily implemented in the spin representation as 
$P_{G_n}=\tr \T^n$. In the Fortuin-Kasteleyn representation, this 
implementation is less obvious as shown in Ref.~\cite{transfer4}. For the
triangular-lattice model with periodic boundary conditions on the transverse
direction we should also use some additional tricks 
\cite{Baxter_86_87,transfer1,transfer3}. 

The results existing in the literature about exact chromatic 
polynomials for strips 
of the square and triangular lattices with toroidal boundary conditions 
are limited to $L\leq 4$. The earlier results were obtained by
recursive use of the contraction-deletion theorem. These works include
the pioneering paper by Biggs, Damerell, and Sands \cite{Biggs_72}
(where the solution for a square-lattice strip of width $L=2$ was presented), 
and those of Chang and Shrock \cite{Shrock_01a,Shrock_01b,Shrock_01c}. 
In Ref.~\cite{Shrock_01b} the latter authors gave the exact solution 
for the full 
Potts-model partition function (i.e., with arbitrary $v$) for 
square-lattice strips of widths $L=2,3$. 
In Ref.~\cite{Shrock_01c} the solution for the triangular-lattice chromatic 
polynomial for width $L=3$ was presented, while in  
Ref.~\cite{Shrock_01a} they gave the solution for both lattice strips of width  
$L=4$. Finally, in  2006 Chang and Shrock exhibited a
transfer-matrix formalism to compute the full Potts-model partition function 
for regular lattice strips with toroidal boundary conditions 
\cite{Shrock_06}.\footnote{
  The transfer-matrix methods of Ref.~\protect\cite{Shrock_06} are 
  essentially the same as
  ours; but our implementation allows us to achieve larger values of 
  the strip width $L$.
} 
In particular, they gave the exact partition function for square-lattice 
strips of widths $L\leq 4$, and triangular-lattice strips of widths 
$L=2,3$.\footnote{
 We believe the solution of the chromatic polynomial for triangular-lattice
 strips of width $L=2$ was obtained by Chang and Shrock earlier 
 than 2006. In fact, they computed the more involved cases $L=3,4$ 
 in 2001!
} 
In Ref.~\cite{Shrock_06}, they also obtained structural properties of the 
partition function and the chromatic polynomial. Additional structural 
properties were discussed by one of us \cite{Jesper_06a}. 

We have several motivations for this work. As explained above, the
exact solution of the 2D $q \neq 2$ state Potts model at general $v$
is still unknown, even for the square lattice (and {\em a fortiori}
for more complicated regular lattices), and even in the thermodynamic
limit (and {\em a fortiori} on general finite lattices).  Therefore,
exact solutions of this model on strips of finite fixed width and
arbitrary length are valuable in their own right (even in the
particular case $v=-1$).  In addition, the chromatic polynomial
$P_G(q)$ is an object of great interest to mathematicians (see e.g.,
Ref.~\cite{Sokal_05} for a recent survey).  As the computation of the
chromatic polynomial is \#--P hard \cite{Oxley_02}, new and efficient
methods for computing this object on large graphs are also of great
interest to this community.

The second motivation of this work is to deepen our physical understanding 
of the critical points of the $q$-state Potts model in the antiferromagnetic 
$-1\leq v<0$ and unphysical regimes $v< -1$. 
Indeed, we can extract this information from our finite-width strips 
using CFT and finite-size scaling (FSS) \cite{Bloete_86,Cardy_88,Privman}.  
In particular, we will consider the following way to attain the thermodynamic
limit. From the chromatic polynomial of a strip of size $L\times N$ we can 
compute its chromatic zeros in the complex $q$-plane. From these zeros
it is very hard to obtain infinite-volume quantities, as they have strong 
FSS corrections. Thus, we consider the infinite-length limit
$N\to\infty$. Then, the chromatic zeros accumulate along certain limiting 
curves $\mathcal{B}_L$ and around isolated limiting points. This phenomenon
is a consequence of the Beraha--Kahane--Weiss theorem \cite{BKW}. Our 
methods allow us to obtain the {\em exact} values of the relevant physical 
quantities in this limit. Their FSS corrections are found to be
smaller than when both $L$ and $N$ are finite. The extrapolation to the 
true infinite-volume limit $L\to\infty$, using standard FSS 
techniques, therefore becomes more precise by employing this order of limits.

\bigskip

\noindent
{\bf Remark.} 
Toroidal boundary conditions are rather special compared to the other
boundary conditions studied in previous papers: namely, free, cylindrical, 
and cyclic \cite{transfer1,transfer2,transfer3,transfer4}.
Strip graphs with the latter boundary conditions are {\em planar}; thus,
the four-color theorem \cite{Appel_76} applies. However, strip graphs with
the former boundary conditions can be regarded as graphs embedded on a 
torus. In this case, we have only a seven-color theorem. This follows from the
upper bound obtained by Heawood \cite{Heawood} in 1890 on the chromatic number 
of a graph $G$ embedded on an orientable surface of genus $g\geq 1$
\be
\chi(G) \;\leq\; H(g) \;=\; 
\left\lfloor \frac{1}{2} \left( 7 + \sqrt{1 + 48 g} \right) \right\rfloor
\label{Heawood}
\ee
combined with the fact that $K_7$ (with $\chi(K_7)=7$) can be 
embedded on a torus, 
as shown in Figure~\ref{figure_k7}. Ringel and Youngs \cite{Ringel_68}
finally proved that the maximum chromatic number of any graph $G$ on an 
orientable surface of genus $g\geq 1$ is given by $H(g)$. This 
graph-theoretic difference between graphs embedded on a surface or on a 
torus may have physical implications that are worth studying: in particular,  
is there any additional structure (e.g., phase-transition lines) in 
the interval $q\in(4,7)$ for the chromatic polynomial of a strip 
graph embedded on a torus? 

\bigskip

The main part of our {\em physical} (as opposed to mathematical) understanding
of the mechanism for generating chromatic zeros stems from Saleur's
analysis of the so-called 
Berker--Kadanoff (BK) phase \cite{Saleur_90_91}. This is a massless phase with
algebraic decay of correlations. 
For the {\em square lattice}, this phase is bounded by the curves
\be
v_\pm(q) \;=\; -2 \pm \sqrt{4-q}
\label{def_vpm_sq}
\ee
and all this region is attracted (at fixed $q$) by the curve 
\be
v(q) \;=\; -\sqrt{q}\,, \qquad q\in [0,4]
\label{def_v_BK_sq}
\ee
It is important to stress four points here: 
First, the BK phase corresponds to generic values of $q\in(0,4)$. 
More precisely, for the Beraha numbers $B_p$ 
\be
B_p \;=\; 4\cos^2 \left(\frac{\pi}{p}\right) \,, \qquad p\in \N\,, \quad 
p\ge 2 
\label{def_Bp}
\ee
the BK phase does not exists due to the vanishing
of the amplitude of some of the leading eigenvalues and/or cancellations among
them \cite{Saleur_90_91,AFpaper}.
Second, the chromatic-polynomial subspace $v=-1$ intersects the BK phase for 
$q\in[0,3]$. Thus, because of the attractive nature of the BK critical curve
\reff{def_v_BK_sq}, we can study its conformal properties by considering 
the chromatic-polynomial case in the window $q\in[0,3]$. 
Third, the upper boundary of the BK phase $v_+(q)$ \reff{def_vpm_sq} can be 
identified as the antiferromagnetic critical curve \cite{Baxter_82b}.  
Fourth, the exact free energy of the chromatic-polynomial line is not known.
However, from our previous studies \cite{transfer1,transfer2,transfer4},
we can draw a qualitative picture of the chromatic-polynomial phase diagram for
real $q$: the system is disordered in the region 
$q\in(-\infty,0)\cup(3,\infty)$, while it is critical in the interval
$q\in(0,3)$. The nature of the system in this critical interval depends on the
boundary conditions: for free and cylindrical boundary conditions there is a
single phase \cite{transfer1,transfer2}; but for cyclic boundary conditions,
we find two different phases: the intervals $q\in(0,2)$, and $q\in(2,3)$ 
corresponding to two distinct values of a topological order parameter
(See Section~\ref{sec.structure}) \cite{transfer4}.   
If we define the value $q_c$ as the largest value of $q$ such that
the system is disordered for all $q>q_c$ and the system is critical at $q=q_c$
\cite{Sokal_97}, then for the square lattice we have that $q_c(\text{sq})=3$  
\cite{Lenard}.

The situation for the {\em triangular lattice} is slightly different. 
This model is solvable on the curve \cite{Baxter_78}
\be
v^3 + 3v^2 \;=\; q
\label{def_solvable_tri}
\ee
The lower branch of the curve \reff{def_solvable_tri} can be identified
with the lower bound of the BK phase $v_{-}(q)$, while the middle branch of 
\reff{def_solvable_tri} is expected to play the same role as
the curve \reff{def_v_BK_sq} [i.e., it is the renormalization-group (RG) 
attractor governing the BK phase]. Thus, for the triangular lattice at 
least one curve is missing (if we assume a RG flow similar to that of the 
square lattice): 
the upper bound of the BK phase [i.e., the analogue of the curve $v_{+}(q)$
\reff{def_v_BK_sq}], and the antiferromagnetic critical curve. 
In the limit $q\to 0$, we have found \cite{forests} that there is a single
phase-transition line separating the (critical) BK phase 
and the (non-critical) high-temperature phase. This line is interpreted as
the antiferromagnetic critical curve for this model, and its slope is given
approximately by   
\be
\left.\frac{{\rm d}v}{{\rm d}q}\right|_{q=0} \;=\; -0.1753\pm 0.0002
\ee
The position of these two curves is an open problem (Some preliminary numerical results
about the antiferromagnetic critical curve were published in 
\cite[Figure 2]{forests}. The full curve will be published elsewhere
\cite{AFtri}). 
Theoretical \cite{AFpaper} and numerical \cite{forests} studies show that the BK
phase does exist in the triangular-lattice model, and the results suggest
that universality holds for this phase (i.e., the critical properties of the
square- and triangular-lattice BK phase coincide).  

Even though some important curves of the phase diagram of the 
triangular-lattice chromatic polynomial are not known, 
Baxter \cite{Baxter_86_87} found its exact free energy, albeit
with peculiar boundary conditions that amount to {\em not} giving
a weight $q$ to clusters with non-trivial homotopy. We comment further
on that solution in Sections~\ref{sec.discussion}--\ref{sec.conclusions}
(see also Ref.~\cite{transfer3}).
The resulting phase diagram consists of three different phases in the 
complex $q$-plane: the disordered phase contained the interval
$q\in(-\infty,0)\cup(4,\infty)$, and there are two critical phases.
One contains the interval $q\in(0,q_2(\text{tri}))$, where $q_2(\text{tri})$ 
is given by  
\be 
q_2(\text{tri}) \;=\; q_\text{B} \;\simeq\; 3.8196717312
\label{def_qB}
\ee
and the third one contains the interval $q\in(q_2(\text{tri}),4)$. From this
phase diagram, we obtain that $q_c(\text{tri})=4$.
 
Our previous studies \cite{transfer3,transfer4},
gave a similar qualitative picture of that phase diagram. In these papers,  
we have defined $q_0(L)$ as the largest value of $q$ at which the
limiting curve $\mathcal{B}_L$ crosses the real $q$-axis. We expect that
the limit $q_0 = \lim_{L\to\infty} q_0(L)$ exists, and it is equal to 
$q_2(\text{tri})$.\footnote{
  For the square lattice, on the contrary, we found that 
  $q_0(\text{sq})=q_c(\text{sq})=3$ \protect\cite{transfer2,transfer4}. 
} 
In Ref.~\cite{transfer3} we reanalyzed Baxter's solution and found that
{\em if} his three analytical expressions for the free energy are correct
and exhaustive, the phase transition should not occur at 
$q_2(\text{tri})=q_\text{B}$, but rather at
\be
q_2(\text{tri}) \;=\; q_\text{G} \;=\; 2 + \sqrt{3} \;\approx\; 3.7320508076
\label{def_qG}
\ee
Unfortunately, our numerical data were not conclusive enough to 
judge which is the correct value of the transition point. 
For free and cylindrical boundary conditions \cite{transfer3},
the corresponding phase diagram contains only the above three phases. 
For cyclic boundary conditions \cite{transfer4}, we found a richer 
phase structure within the interval $q\in(0,4)$. In particular, in the
interval $q\in (0,q_2(\text{tri}))$, there should be phase transitions
at the Beraha numbers $q=B_4,B_6,\ldots\le q_2({\rm tri})$ 
[cf.~Eq.~\reff{def_Bp}]. If 
$q_2(\text{tri})=q_\text{B}$, then the last such transition occurs at
$q=B_{14}\approx 3.8019377358$; but if $q_2(\text{tri})=q_\text{G}$, 
then the last transition is at $q=B_{12}=q_\text{G}$.   
  
Finally, the chromatic-polynomial line $v=-1$ has a non-void intersection
with the BK phase for the triangular lattice. In particular, we
expect that the intersection should include the whole interval
$q\in(0,q_2(\text{tri}))$. Thus, the critical properties extracted
from this interval are related directly to those of the BK phase. 
In the interval $q\in(q_2(\text{tri}),4)$, the universality class should
be different from that of the BK phase. For other boundary conditions,
we have been unable to extract any meaningful physical information for
this interval; so this question remains open. In the ferromagnetic regime,
it is well-known that results obtained from toroidal boundary conditions
are closer to the true thermodynamic limit that for the other boundary 
conditions. Although the argument leading to this conclusion may not apply
for antiferromagnetic or ``unphysical'' models, it opens an opportunity to
try to answer the above-posed questions.  

This paper is organized as follows: in Section~\ref{sec.chromatic} we
show in detail how to obtain the chromatic polynomial of a strip with
toroidal boundary conditions, and some structural properties: 
dimensionality of the transfer matrix and properties of the amplitudes.
Sections~\ref{sec.sq.results} and~\ref{sec.tri.results} contain the 
detailed description (when possible) of the chromatic polynomial for 
square- and triangular-lattice strips, respectively. In these two
sections we also describe the limiting curve for each regular lattice. The 
less-mathematically inclined readers can skip most of these sections.
In Section~\ref{sec.discussion} we discuss the finite-width 
results and extract the physically interesting quantities: phase 
diagram of the chromatic polynomial for the square and triangular lattices,
and critical properties of the different phases.
Finally, in Section~\ref{sec.conclusions}, we draw our conclusions.
In Appendix~\ref{sec.combin} we prove two useful combinatorial identities 
needed in Section~\ref{sec.structure}. 

%
%
\section{Chromatic polynomials with toroidal boundary conditions}
\label{sec.chromatic}

We want to build the transfer matrix for square- and a triangular-lattice 
strips of width $m$ and arbitrary length $n$ with toroidal boundary conditions. 
The goal is to write the transfer matrix $\T(m)$ as a product of two
matrices
\be
\T(m) \;=\; \H(m) \cdot \V(m) 
\label{def_T}
\ee
where $\H$ (resp.\ $\V$) represents the horizontal (resp.\ vertical)
bonds of the lattice, and to write its chromatic polynomial 
in the form 
\be
P_{m_\text{P}\times n_\text{P}}(q) \;=\; 
\uu^{\text{T}}\cdot \H(m)\cdot \T(m)^n \cdot \vv_{\text{id}}
\label{def_Pmn}
\ee
for some left and right vectors $\uu$ and $\vv_\text{id}$. The subindex P
denotes the periodic boundary condition in the corresponding direction. 

%
%
\subsection{The transfer-matrix formalism for toroidal boundary conditions}
\label{sec.tm}

The main problem in writing a transfer matrix in the Fortuin-Kasteleyn 
representation is the non-local factor $q^{k(E')}$ in \reff{def_Z_FK}
\cite{transfer1}. For free boundary conditions in the longitudinal direction
(i.e., free and cylindrical), this can be efficiently handled by 
keeping track of the connectivity state $\vv_{\mathcal{P}}$ of the top
layer. This connectivity is related to a certain partition $\mathcal{P}$ of
the single-layer vertex set $\{1,2,\ldots,m\}$ \cite{transfer1,transfer3}.  

For periodic boundary conditions in the longitudinal direction (i.e., cyclic 
and toroidal) we must keep track of the connectivity state of both the top
and bottom rows: each connectivity state $\vv_{\mathcal{P}}$ is now
given by a certain partition $\mathcal{P}$ of the vertex set of the 
top row $\{1,2,\ldots,m\}$ and the vertex set of the bottom row 
$\{1',2',\ldots,m'\}$. In Figure~\ref{figure_sq_cyclic_bc} we show an
example of a square-lattice strip of size $4\times 2$. As for cyclic 
boundary conditions \cite{transfer4}, initially the top and bottom rows
are identical. Then, we enlarge the strip by adding new layers via the 
action of the transfer-matrix operator $\T(m)$ exclusively on the top-row 
vertices (while the bottom-row sites remain unchanged). At the end, when we
have built a strip of $n+1$ rows, we identify the top and bottom rows. 

Let us now explain in more detail how to obtain the chromatic polynomial 
for a square-lattice strip of width $m$, length $n$, and toroidal boundary 
conditions. 

We characterize the state of the top and bottom rows by a connectivity 
state $\vv_{\mathcal{P}}$, which is associated to a partition
$\mathcal{P} = \{\mathcal{P}_1,\mathcal{P}_2,\ldots,\mathcal{P}_k\}$ of the
set $\{1,2,\ldots,m,1',2',\ldots,m'\}$. We then introduce the operators
\begin{subeqnarray}
\P_x      &=& v I + \D_x \slabel{def_Dx} \\
\QQ_{x,y} &=& I + v\J_{x,y} \slabel{def_Qxy}
\end{subeqnarray}
acting on the connectivity space $\{\vv_{\mathcal{P}}\}$. Here $I$ is the
identity, $\D_x$ is the detach operator that detaches site $x$ from the
block it currently belongs to, and $\J_{x,y}$ is the join operator that
amalgamates the blocks containing sites $x$ and $y$, if they were not already
in the same block. (Further details can be found in Ref.~\cite{transfer1}.)

The matrices $\V$ and $\H$ are defined in terms of the above operators as
follows:
\begin{subeqnarray}
 \H_{\text{sq}}(m) &=& \left(\prod\limits_{i=1}^{m-1} \QQ_{i,i+1} \right)
    \cdot \QQ_{m,1} \slabel{def_Hsq} \\
 \V_{\text{sq}}(m) &=& \prod\limits_{i=1}^{m} \P_i 
    \slabel{def_Vsq} 
\end{subeqnarray} 
Notice that periodic boundary conditions have been implemented by adding the
operator $\QQ_{m,1}$ in \reff{def_Hsq}. We can similarly define a 
matrix $\H'_{\text{sq}}$ acting only on the bottom row.  

Then, the chromatic polynomial for a square-lattice strip of size $m\times n$
with toroidal boundary conditions can be written as in 
\reff{def_T}/\reff{def_Pmn}, where the two vectors are those of cyclic 
boundary conditions \cite{transfer4}. The vector $\vv_\text{id}$ denotes 
the partition
\be
\vv_{\text{id}} \;=\; \{\{1,1'\},\{2,2'\},\ldots,\{m,m'\}\} 
\label{def_vid}
\ee
(i.e., we start with the top and bottom rows identified). The left vector
$\uu^\text{T}$ acts on a connectivity state $\vv_\mathcal{P}$ as
\be
\uu^\text{T}\cdot \vv_\mathcal{P} \;=\; q^{|\mathcal{P}'|}
\label{def_u}
\ee
where $|\mathcal{P}'|$ denotes the number of blocks in the connectivity state
$\vv_\mathcal{P}'$ obtained from $\vv_\mathcal{P}$ by 
\be
\vv_\mathcal{P}' \;=\; \left( \prod\limits_{i=1}^m \J_{i,i'} \right) 
 \cdot \vv_\mathcal{P}
\label{def_Pprime}
\ee
Thus, $\uu^\text{T}$ acts on $\vv_\mathcal{P}$ by identifying the top and
bottom rows, and then by assigning a factor of $q$ to each block in the 
resulting partition. 

The definition of the transfer matrix for a triangular-lattice
strip of width $m$, length $n$, and toroidal boundary conditions is more
involved because of the periodic boundary conditions in the transverse 
direction. As explained in Refs.~\cite{transfer1,transfer3}, in this case one
has to consider a triangular lattice of width $m+1$ with free boundary 
conditions in the transverse direction, and then identify the columns $1$ and
$m+1$. (See Figure~\ref{figure_tri_cyclic_bc} for an example of size
$4\times 2$.) Then, the matrices $\H$ and $\V$ take the form
\begin{subeqnarray}
 \H_{\text{tri}}(m) &=& \left( \prod\limits_{i=1}^{m} \QQ_{i,i+1} \right)
    \cdot \J_{m+1,1} \slabel{def_Htri} \\
 \V_{\text{tri}}(m) &=& \left( \prod\limits_{i=1}^{m} \P_i \cdot 
    \QQ_{i,i+1} \right)  \cdot \P_{m+1} \slabel{def_Vtri} 
\end{subeqnarray} 
where the operators $\QQ_{i,i+1}$ in \reff{def_Vtri} represent the oblique
bonds $e=\<(i,j+1),(i+1,j)\>$. With this definitions, the chromatic polynomial
can be written as in \reff{def_T}/\reff{def_Pmn} with the same right and 
left vectors as for the square-lattice case \reff{def_vid}/\reff{def_u}. 

A further simplification can be obtained if we take into account the fact
that $v=-1$. Then, $\H$ is a projector (i.e., $\H^2 = \H$), and we can 
use instead of the transfer matrix $\T$ and the basis vectors 
$\vv_\mathcal{P}$, the modified transfer matrix $\T'$
\be
\T'(m) \;=\; \H(m) \cdot \V(m) \cdot \H(m)  
\label{def_Tprime}
\ee
and the modified basis vectors
\be
\ww_\mathcal{P} \;=\; \H(m)\cdot \vv_\mathcal{P} 
\label{def_wP}
\ee
Note that $\ww_\mathcal{P}=0$ if $\mathcal{P}$ has any pair of nearest-neighbor
sites of the top row in the same block. Thus, the space $\{\ww_\mathcal{P}\}$
has lower dimension than the original connectivity space 
$\{\vv_\mathcal{P}\}$.  

Notice that we start with the state $\ww_\text{id}=\H\cdot\vv_\text{id}$,
and that at the end we identify the top and bottom rows. Then, it is 
very useful to work directly with the modified connectivity basis
\be
\widehat{\ww}_\mathcal{P} \;=\; \H'(m)\cdot \H(m)\cdot \vv_\mathcal{P}
\label{def_whatP} 
\ee
This choice implies that $\widehat{\ww}_\mathcal{P}=0$ also if $\mathcal{P}$
contains any pair of nearest-neighbor sites of the bottom row
in the same block. For simplicity, we will drop hereafter the 
prime in the modified transfer matrix \reff{def_Tprime} and the hat in the
modified basis vectors \reff{def_whatP}. Then, the chromatic polynomial
can be written as
\be
P_{m_\text{P}\times n_\text{P}}(q) \;=\; 
\uu^{\text{T}}\cdot \T(m)^n \cdot \ww_{\text{id}}
\label{def_Pmn_final} 
\ee
Finally, to lighten the notation, we will write the connectivity states 
$\vv_\mathcal{P}$ using Kronecker delta-functions. For instance,
$\vv_{\{\{1,2\},\{1',3'\},\{3\},\{2'\}\}}$ will be written as
$\delta_{1,2}\delta_{1',3'}$. The identity connectivity state 
\reff{def_vid} will be denoted simply as $1$. 

%
%
\subsection{Structural properties of the chromatic polynomial}
\label{sec.structure}

Our first goal is to compute the number $C_{m,m}$ of non-crossing 
connectivities in the interior of an annulus with $m$ points on the inner 
rim and $m$ points on the outer rim (See Figure~\ref{figure_sq_cyclic_bc_bis}).
This problem can be reduced to a {\em simpler} one by introducing the
quantities $d(m,\ell)$. 
Given $m$ points on a circle and a non-crossing connectivity ${\cal P}$
exterior to that circle, we can compute the
number of ways $d(m,\ell,{\cal P})$ we can connect that particular connectivity 
${\cal P}$ to infinity by means of $\ell$ mutually non-connected paths 
(bridges). 
Then, the $d(m,\ell)$ are the sum over all possible 
connectivities ${\cal P}$ 
\be
d(m,\ell) = \sum\limits_{{\cal P}} d(m,\ell,{\cal P})
\ee
The values of $d(m,\ell)$ are given by the following formula
\cite{Shrock_06}\footnote{
 The numbers $d(m,\ell)$ \protect\reff{def_dml} are denoted as 
 $n_{Z,tor}(m,\ell)$ by Chang and Shrock 
 \protect\cite[Eqs.~(2.19)--(2.22)]{Shrock_06}, and as $n_\text{tor}(m,\ell)$ 
 in Ref.~\protect\cite[Eq.~(2.1)]{Jesper_06a}.
}
\be
d(m,\ell) \;=\;  \begin{cases} 
\qquad               C_m               &\qquad \text{if $\ell=0$} \\[5mm]
\qquad \displaystyle \binom{2m-1}{m-1} &\qquad \text{if $\ell=1$} \\[5mm]
\qquad \displaystyle \binom{2m}{m-\ell}&\qquad \text{if $2\leq \ell \leq m$} 
                                                                  \\[5mm]
\qquad               0                 &\qquad \text{if $\ell > m$}  
  \end{cases} 
\label{def_dml}
\ee
where $C_m$ is the Catalan number
\be
C_m = \frac{1}{m+1}\, \binom{2m}{m}
\label{def_Cm}
\ee
which gives the number of non-crossing connectivities with $m$ points on 
a circle (or on a line). 

The number of connectivities $C_{m,m}$ can be obtained easily from the
$d(m,\ell)$ \reff{def_dml} by using the $\ell$ paths as {\em bridges} and by
matching two such {\em bridge} connectivities with the same value of 
$0\leq \ell\leq m$
\be
C_{m,m} \;=\; C_m^2 + \sum\limits_{\ell=1}^m \ell \,  d(m,\ell)^2
\label{def_Cmm}
\ee
If we insert \reff{def_dml} for the $d(m,\ell)$, then we obtain a 
closed expression for $C_{m,m}$: 
\be
C_{m,m} \;=\; \frac{m^2-2m+5}{m+1} \, \binom{2m-1}{m}^2
\label{def_CmmBis}
\ee
where we have used the identity \reff{lemma1} proven in 
Appendix~\ref{sec.combin}.
The asymptotic growth of $C_{m,m}$ as $m\to \infty$ is exponentially 
fast with a large constant:
\be
C_{m,m} \;\sim\; \frac{16^m}{4\pi} \left[ 1 + O\left(\frac{1}{m}\right)\right]
\,, \quad \text{as $m\to\infty$} 
\label{CmmBis_asym}
\ee

In practice, we build the $C_{m,m}$ non-crossing connectivities in a
recursive way.
We start with all the $C_{m-1,m-1}$ non-crossing connectivities on an annulus
with $m-1$ points on its inner and outer rims, and then we add a new site on
the inner rim. The number of non-crossing connectivities for this annulus
is obtained again by matching the $\ell$-bridge connectivities
\be
C_{m,m-1} \;=\; C_m \, C_{m-1} + 
  \sum\limits_{\ell=1}^{m-1} \ell\, d(m,\ell)\, d(m-1,\ell)
\label{def_Cmm-1}
\ee
Then, we add a new site on the outer rim, and we obtain the $C_{m,m}$
connectivities \reff{def_CmmBis}. A closed formula for $C_{m,m-1}$ 
\reff{def_Cmm-1} can be
obtained if we use the combinatorial identity \reff{lemma2} proven in
Appendix~\ref{sec.combin}. After some algebra, we obtain
\be
C_{m,m-1} \;=\; \frac{2m^4-6m^3+11m^2+3m-4}{2m^2(m+1)} \,
                \binom{2m-2}{m-1}^2  
\label{def_Cmm-1Bis}
\ee
valid for all $m\geq 2$. This formula also grows exponentially fast in $m$
in the large-$m$ limit:
\be
C_{m,m-1} \;\sim\; \frac{16^m}{16\pi} 
\left[ 1 + O\left(\frac{1}{m}\right)\right]
\,, \quad \text{as $m\to\infty$}
\label{Cmm-1Bis_asym}
\ee
In Table~\ref{table_dim_T_torus} we show the values of $C_{m,m}$ and
$C_{m,m-1}$ for $1\leq m \leq 7$. 

The condition $v=-1$ implies that connectivities $\mathcal{P}$ containing
nearest-neighbor sites in any sub-block do not contribute. 
The number of non-crossing non-nearest-neighbor connectivities
for an annulus with $m$ points on its inner and outer rims will be denoted
by $D_{m,m}$. This number gives therefore the dimension of the modified 
transfer matrix \reff{def_Tprime} for a square- or triangular-lattice strip 
of width $m$ and toroidal boundary conditions. These numbers are
displayed for $1\leq m\leq 7$ in Table~\ref{table_dim_T_torus}. 

As in previous works \cite{transfer1,transfer2,transfer3,transfer4}, we can
reduce this dimension by considering the equivalence classes modulo
symmetries of non-crossing non-nearest-neighbor connectivity states.
For the triangular (resp.\ square) lattice, we should consider equivalence 
classes modulo rotations (resp.\ rotations and reflections with respect to 
any diameter of the strip).  
The number of generated equivalence classes modulo rotations 
(resp.\ reflections and rotations) of non-crossing non-nearest-neighbor 
connectivity states for a strip of width $m$ will be denoted by TriTorus$'(m)$ 
(resp.\  SqTorus$'(m)$). The adjective ``generated'' means that we only take
into account those classes of connectivity states that can be produced by
applying the various operators $\P_i$ and $\QQ_{i,j}$ on the right vector
$\ww_\text{id}$. This distinction is relevant for the square lattice, 
as there are perfectly legal non-crossing non-nearest-neighbor partitions
that cannot be generated when we start from the right vector 
$\ww_\text{id}$. (See e.g., the discussion on this issue in 
Ref.~\cite[end of Section~2.1]{Jesper_06a}.) 
The simplest example corresponds to $m=2$ and the connectivity state 
$\delta_{2,1'}\delta_{1,2'}$. In general, out of the $m$ possible 
connectivity states with $\ell=m$ bridges, only the state $\vv_\text{id}$
\reff{def_vid} 
can be realized on a square-lattice strip. The other $m-1$ non-crossing
connectivity states cannot be realized because of the lattice structure. 
However, on the triangular lattice, which has a larger set of ``vertical'' 
edges, one can produce those $m-1$ connectivity states without violating the
non-crossing condition. 
 
Finally, as for cyclic boundary conditions, for a given width $m$ we
find many repeated eigenvalues. Thus, we define TriTorus$(m)$  
(resp.\ SqTorus$(m)$) as the number of {\em distinct} eigenvalues 
(with non-zero amplitude) for a triangular-lattice (resp.\ square-lattice)
strip of width $m$ with fully periodic boundary
conditions. In Table~\ref{table_dim_T_torus} we have displayed 
the quantities TriTorus$'$, SqTorus$'$, TriTorus and SqTorus for $m\leq 7$ 
(when available). 

The symbolic computation of the transfer matrix and the $\uu^\text{T}$ and 
$\ww_\text{id}$ vector has been carried out using a {\tt mathematica}
script for the smallest widths $2\leq m \leq 4$, and a C program for
the larger widths $m\leq 7$. Indeed, for $m\leq 4$, both methods perfectly
agree. We refer to Refs.~\cite{transfer2,transfer3,transfer4} for details
about the symbolic computation. For larger widths $m\leq 15$, we have 
computed {\em numerically} the transfer matrix using another C program.

\bigskip

\noindent
{\bf Remarks.} 1) With our transfer-matrix approach we have checked all
previous results by Biggs, Damerell and Sands \cite{Biggs_72}, and  
Chang and Shrock \cite{Shrock_01a,Shrock_01b,Shrock_01c,Shrock_06}.

2) Even though the final dimensions of the transfer matrices are 
TriTorus$'(m)$ or SqTorus$'(m)$, the actual computations are carried out
in the larger space of non-crossing connectivities (of dimension
$C_{m,m}$). This fact limits the maximum width we are able to handle. 

3) We have cross-checked our results using the trivial identity
\be
P_{m_\text{P}\times n_\text{P}}(q) \;=\; P_{n_\text{P}\times m_\text{P}}(q)
\label{check_Pmn}
\ee
for all $2\leq m,n\leq 7$.

\medskip

%
%
\subsection{Amplitudes for toroidal boundary conditions}

The transfer matrix for a strip with toroidal boundary conditions has
a block structure. This property also holds for cyclic boundary conditions
\cite{transfer4}, and in both cases the origin is the same. Given a 
certain connectivity $\mathcal{P}$ of the top and bottom rows (see e.g., 
Figure~\ref{figure_sq_cyclic_bc_bis}), we can regard it
as a bottom-row connectivity $\mathcal{P}_\text{bot}$, a top-row
connectivity $\mathcal{P}_\text{top}$ joined by $\ell$ bridges. Notice that
each top-row block can be connected to a bottom-row block by {\em at most}
one bridge, and vice versa. Thus, the number of bridges is an integer
$0\leq \ell \leq m$. 

As explained in Ref.~\cite{transfer4}, the full transfer matrix has a 
lower-triangular block form (if we order the connectivity states in
decreasing order of bridges).
\be
\T(m) \;=\; \left( \begin{array}{cccc} 
   T_{m,m}   & 0           & \ldots & 0 \\
   T_{m-1,m} & T_{m-1,m-1} & \ldots & 0 \\
   \vdots    & \vdots      &        & \vdots\\
   T_{0,m}   & T_{0,m-1}   & \ldots & T_{0,0}
                           \end{array}
                    \right)
\label{T_blocks1}
\ee
The reason is the application of the $\H$ and $\V$ operators cannot increase 
the number of bridges of a given connectivity state. Furthermore, as the
transfer matrix does not act on the bottom-row connectivity and they act on
the space of non-crossing connectivities, each diagonal block $T_{\ell,\ell}$
has also a diagonal-block form
\be
T_{\ell,\ell} \; = \; \left( \begin{array}{cccc}
   T_{\ell,\ell,1}  & 0               & \ldots & 0 \\
   0                & T_{\ell,\ell,2} & \ldots & 0 \\
   \vdots           & \vdots          &        & \vdots\\
   0                & 0               & \ldots & T_{\ell,\ell,N_\ell}
                             \end{array}
                       \right)
\label{T_blocks2}
\ee
where
each sub-block $T_{\ell,\ell,j}$ is characterized by a certain bottom-row
connectivity $\mathcal{P}_\text{bot}$ and a position of the $\ell$
bridges. Its dimension is given by the number of top-row connectivities that
are compatible with $\mathcal{P}_\text{bot}$, and the number $\ell$ and 
relative positions of the bridges. The number of blocks $N_\ell$
($0\leq \ell\leq m$) for square- and triangular-lattice strips of width $m$ 
are displayed in Tables~\ref{table_blocks_sq} and~\ref{table_blocks_tri}, 
respectively. 

In particular, as for cyclic boundary conditions \cite{transfer4}, this 
structure means that the characteristic polynomial of the
full transfer matrix $\T$ can be factorized as follows
\be
  \chi(\T) \; = \; \prod\limits_{\ell=0}^m \left[
          \prod\limits_{j=1}^{N_\ell}
          \chi \left( T_{\ell,\ell,j} \right) \right]
  \label{chi_block_T_sq}
\ee

In practice, for each strip width $2\leq m\leq 7$, we have performed
the following procedure:

\begin{enumerate}

\item Using a C program, we compute the full transfer matrix
      $\T(m)$ and the left $\uu$ and right vectors
      $\ww_\text{id}$. These objects are obtained in the basis of
      non-crossing non-nearest-neighbor classes of connectivities that are
      invariant  under the right symmetries (depending on the lattice 
      structure). The dimension of this transfer matrix is SqTorus$'(m)$ or
      TriTorus$'(m)$. With these objects, we can compute the chromatic 
      polynomial of any 
      strip of finite length $n$ using Eq.~\reff{def_Pmn_final}. 

\item Given the full transfer matrix $\T(m)$, we compute the
      the SqTorus$(m)$ or TriTorus$(m)$ distinct eigenvalues 
      $\lambda_i(q)$. To do this, we
      split the transfer matrix into blocks $T_{\ell,\ell,j}$, each block
      characterized by a bottom-row connectivity and the number $\ell$ and
      position of the bridges (modulo the corresponding symmetries). 
      By diagonalizing these
      blocks we obtain the whole set of distinct eigenvalues $\lambda_i(q)$
      and their multiplicities $k_i$.

\item The final goal is to express the chromatic polynomial in the form
\be
  P_{m_\text{P}\times n_\text{P}}(q) \; = \; 
        \sum\limits_i \alpha_i(q) \, \lambda_i(q)^n
  \label{def_Pmn_eigen}
\ee
     where the $\{\lambda_i\}$ are the {\em distinct} eigenvalues 
     of the transfer
     matrix $\T(m)$ and the $\{\alpha_i\}$ are some amplitudes we
     have to determine. The calculation of the amplitudes can be achieved by
     solving SqTorus$(m)$ or TriTorus$(m)$ linear equations of the type 
     \reff{def_Pmn_eigen},
     where the l.h.s.\ (i.e., the true chromatic polynomials) have been
     obtained via Eq.~\reff{def_Pmn_final} 
     (i.e., by using the full transfer matrix and vectors).\footnote{
  In practice, to obtain the amplitudes $\alpha_i(q)$ we assume that they
  are polynomials in $q$. The order of the polynomial $\alpha_i(q)$ 
  corresponds to the largest $\ell$ sector the eigenvalue $\lambda_i$ 
  belongs to. 
  Unlike for cyclic boundary conditions, a given eigenvalue for toroidal 
  boundary conditions can appear in several sectors (each of them 
  characterized by a different value of $\ell$. See text.) 
  Once the Ansatz for the amplitudes is fixed, we
  solve the linear equations for the numerical coefficients in
  $\alpha_i(q)$ using as many chromatic polynomials  
  $P_{m_\text{P}\times n_\text{P}}(q)$ as needed.  
  After finding the solution, we check that it can reproduce chromatic
  polynomials not used in the determination of the amplitudes. 
}
\end{enumerate}

We have used this procedure
to compute the transfer matrix up to width $m=7$ for both the square and
the triangular lattices. The next cases were unmanageable with
our current computer resources.\footnote{
 The largest transfer matrix we have computed corresponds to the $m=7$ 
 triangular lattice of dimension $40996$. We needed $\sim 4.3$Gb of 
 RAM memory and $\sim 140$ hours of CPU to perform this computation. 
 In addition, the raw {\tt mathematica} output file was very large 
 $\sim 355$Mb. 
} 
The number of distinct eigenvalues found 
in each block $T_{\ell,\ell}$ can be found in Tables~\ref{table_eigen_sq} 
and~\ref{table_eigen_tri} for the square and triangular lattices, respectively. 

In Ref.~\cite{transfer4}, we found several properties of the eigenvalues
$\lambda_i$ and their corresponding amplitudes $\alpha_i$ for {\em cyclic}
boundary conditions.  
In particular, that all eigenvalues within a block $T_{\ell,\ell}$
have the same amplitude $\alpha^{(\ell)}$ (see below), and that any two
distinct blocks $T_{\ell,\ell}$ and $T_{\ell',\ell'}$ (with $\ell\ne\ell'$) 
have no 
common eigenvalues. These properties were essentially due to the presence 
of a quantum-group symmetry in the Potts model with cyclic boundary conditions.
These properties were also used to simplify the algorithm to compute 
symbolically the transfer-matrix eigenvalues $\lambda_i$.
Unfortunately, this quantum-group symmetry is broken when toroidal boundary
conditions are considered, and none of these properties hold in our case. 

Chang and Shrock \cite{Shrock_06} have found that the chromatic polynomial
for toroidal boundary conditions \reff{def_Pmn_eigen} 
has a structure similar (but not identical) to that for cyclic boundary
conditions \cite{Saleur_90_91,Shrock_05,transfer4,Jesper_06b}. 
For a square- or triangular-lattice 
strip of width $m$, length $n$ and toroidal boundary conditions, the
chromatic polynomial can be written as 
\be
P_{m_\text{P}\times n_\text{P}}(q) \;=\; \sum\limits_{\ell=0}^m 
                       \sum\limits_{j=1}^{L_\ell}
                        b_j^{(\ell)} \, \lambda_{\ell,j}(m)^n
\label{def_Pmn_eigen_final}
\ee
where the amplitudes $b_j^{(\ell)}$ are polynomials in $q$ of order $\ell$,
and in contrast with cyclic boundary conditions, are not the same within
a given $\ell$ sector. The eigenvalues $\lambda_{\ell,j}(m)$ are algebraic
functions of $q$ and, in contrast with cyclic boundary conditions, a given
eigenvalue can appear in more than one $\ell$ sector.

Chang and Shrock \cite{Shrock_06} found a family of basic amplitudes 
$b^{(\ell)}$ given by
\begin{subeqnarray}
b^{(0)} &=& \alpha^{(0)} \;=\; 1   \slabel{def_b0} \\
b^{(1)} &=& \alpha^{(1)} \;=\; q-1 \slabel{def_b1} 
\label{def_b0-b1}
\end{subeqnarray}
and for $\ell\geq 2$, by 
\begin{subeqnarray}
b^{(\ell)} &=& \alpha^{(\ell)} - \alpha^{(\ell-1)} + (-1)^\ell \alpha^{(1)}\\ 
           &=& q^\ell + \sum\limits_{j=1}^\ell (-1)^j \, \frac{2\ell}{2\ell-j}
               \, \binom{2\ell-j}{j} \, q^{\ell-j} \;+\; (-1)^\ell \, (q-1)
\end{subeqnarray}
where the coefficients $\alpha^{(\ell)}$ are the amplitudes obtained for
cyclic boundary conditions \cite{Saleur_90_91}  
\begin{subeqnarray}
\alpha^{(\ell)} &=& U_{2\ell}\left( \frac{\sqrt{q}}{2} \right) \\
                &=& \sum\limits_{j=0}^\ell (-1)^j \, \binom{2\ell-j}{j} \,
                     q^{\ell-j}
\end{subeqnarray}
where $U_n(x)$ is the Chebyshev polynomial of second kind.
The first non-trivial $b^{(\ell)}$ are \cite[Eqs.~(2.1)--(2.4)]{Shrock_06}:
\begin{subeqnarray}
b^{(2)} &=& q^2-3q+1 \slabel{def_b2}\\
b^{(3)} &=& q^3-6q^2+8q-1 \slabel{def_b3}\\
b^{(4)} &=& q^4-8q^3+20q^2-15q+1 \slabel{def_b4}\\
b^{(5)} &=& q^5-10q^4+35q^3-50q^2+24q-1 \slabel{def_b5}\\
b^{(6)} &=& q^6 -12q^5 +54q^4-112 q^3 +105 q^2 -35q+1 \slabel{def_b6}\\
b^{(7)} &=& q^7 -14q^6 +77 q^5 -210 q^4 + 294 q^3 - 196q^2 +48 q -1
            \slabel{def_b7} 
\label{def_b2-b7}
\end{subeqnarray} 
Finally, Chang and Shrock argue that the coefficients $b^{(\ell)}_j$ 
cannot be equal to the above $b^{(\ell)}$. For $m=2,3$ they find that
the coefficients $b^{(2)}$ and $b^{(3)}$ split into two coefficients 
(up to a positive integer factor):\footnote{
  Note that the expression for $b_2^{(3)}$ \protect\reff{def_b32} is twice the
  expression given by Chang and Shrock \protect\cite[Eq.~(2.36)]{Shrock_06}.
} 
\begin{subeqnarray}
b_1^{(2)} &=& \frac{1}{2} q(q-3) \slabel{def_b21}\\
b_2^{(2)} &=& \frac{1}{2} (q-1)(q-2) \slabel{def_b22}\\[3mm]
b_1^{(3)} &=& \frac{1}{3} (q-1)(q^2-5q+3)\slabel{def_b31} \\
b_2^{(3)} &=& \frac{1}{3} q (q-2)(q-4)\slabel{def_b32} 
\label{def_b2-b3}
\end{subeqnarray}

In our numerical study, we have found that this also true for $m=4,5$.
The basic amplitudes are given by 
\begin{subeqnarray}
b_1^{(4)} &=& \frac{1}{4} q(q-2)(q-3)^2 \slabel{def_b41}\\
b_2^{(4)} &=& \frac{1}{4} q(q-1)(q-3)(q-4) \slabel{def_b42}\\
b_3^{(4)} &=& \frac{1}{4} (q-1)(q^3-7q^2+14q-4)\slabel{def_b43} \\[3mm]
b_1^{(5)} &=& \frac{1}{5} (q^5-10q^4+35q^3-50q^2+24q-5) \slabel{def_b51}\\
b_2^{(5)} &=& \frac{1}{5} q (q-1)(q-2)(q-3)(q-4)\slabel{def_b52} 
\label{def_b4-b5}
\end{subeqnarray}

In general, an eigenvalue $\lambda_i$ which only appear in a single 
$T_{\ell,\ell}$ block has an amplitude given by a positive integer multiple
of either \reff{def_b0-b1}/\reff{def_b2-b7} or 
\reff{def_b2-b3}/\reff{def_b4-b5}. All the amplitudes  
$b_j^{(\ell)}$ and $b^{(\ell)}$ \reff{def_b2-b7} are simply polynomials
in $q$ of order $\ell$.  
When an eigenvalue $\lambda_i$ appears 
in more than one block $T_{\ell,\ell}$, then its amplitude is just a 
certain linear combination (with integer coefficients) of the 
corresponding amplitudes given above. Thus, the order of this  
polynomial amplitude is just the largest value of $\ell$ involved. 

\bigskip

\noindent
{\bf Remarks}. 1. After we empirically obtained the amplitudes 
\reff{def_b4-b5}, one of us obtained the general formula for the
different amplitudes $b_j^{(\ell)}$ \cite{Jesper_06a}. Indeed, our
results fully agree with the general formula. The number of distinct
amplitudes $b_j^{(\ell)}$ for a given $\ell$ was shown to be the
number of integer divisors of $\ell$ \cite{Jesper_06a}.

2. As we shall discuss in the next Section, the sub-block $T_{L,L}$ of 
the square-lattice strip of width $L$ contains the eigenvalue 
$\lambda=(-1)^L$ with amplitude $b^{(L)}$ \reff{def_b2-b7}. 
Indeed, each $b^{(\ell)}$ can be written as a linear combination of 
the amplitudes $b_j^{(\ell)}$ \reff{def_b2-b3}/\reff{def_b4-b5}:
\begin{subeqnarray}
b^{(2)} &=& b_1^{(2)} + b_2^{(2)} \\
b^{(3)} &=& b_1^{(3)} + 2 b_2^{(3)}\\
b^{(4)} &=& b_1^{(4)} + 2 b_2^{(4)} + b_3^{(4)} \\
b^{(5)} &=& b_1^{(5)} + 4 b_2^{(5)} 
\label{def_b_linear_combin}
\end{subeqnarray} 
This fact was first noted by Chang and Shrock \cite{Shrock_06}; and has been
proven in Ref.~\cite[c.f., Eq.~(4.7)]{Jesper_06a} by showing how many times
each $b_j^{(\ell)}$ appears in a given $b^{(\ell)}$.    

3. In our previous study of the chromatic polynomial for cyclic boundary
conditions \cite{transfer4}, we could only obtain the exact chromatic 
polynomial up to widths $m=6$. The reason was that the computation of the 
characteristic polynomial of a {\em symbolic} matrix of dimension 
$\gtapprox 100$ was unfeasible with {\tt mathematica}'s 
standard built-in functions (based essentially on Gaussian elimination). 
In this paper, we have been able to overcome this problem by using a different 
{\em division--free} algorithm, which computes the characteristic 
polynomial of a matrix with elements belonging to a ring (not to a field). 
We have used a {\tt mathematica} implementation \cite{Nakos_00} of the
Samuelson--Berkowitz--Abdeljaoued algorithm \cite{charpoly}.\footnote{
  We thank Alan Sokal for bringing to our attention several important 
  references on this topic.
}
With this algorithm we have been able to deal with symbolic matrices 
of dimensions up to $308$. Unfortunately, for $L=7$ we need to compute
the symbolic characteristic polynomial of much larger matrices (of 
dimensions $483$ and $532$). These computations are beyond our current 
computer facilities, even with the help of the above-described improved 
algorithm.\footnote{ 
  The computation of the characteristic polynomial of the sub-block of 
  dimension $308$ took approximately two months of CPU time.
}   
Further details will be published elsewhere.

%
%
\section{Square-lattice numerical results}\label{sec.sq.results}

In this section we will analyze the results for the square-lattice strips of
widths $2\leq L \leq 7$. We have checked our results using (in addition
to the already known cases $L\leq 4$ \cite{Shrock_01a,Shrock_01b,Shrock_06})
the identity \reff{check_Pmn}. 

For each $L$, we give a detailed (when possible) description of the 
eigenvalues $\lambda_i$ and amplitudes $\alpha_i$. We have included
the already known cases $L\leq 4$ for completeness and, more importantly, 
to show in detail how our method works. In particular, for $L=2,3$ we give 
the explicit expression of the transfer matrix and the left and right vectors.
The exact expressions for the new cases $L=5,6$ are very lengthy, 
so we refrain from writing down all the needed formulae. Instead, 
we have included a {\tt mathematica} file named {\tt transfer6\_sqT.m} 
available as part of the on-line version of this paper in the {\tt cond-mat} 
archive at {\tt arXiv.org}. For $L=7$, even though we have obtained the 
symbolic form of the transfer matrix and the left and right vectors, 
we have been unable to compute some of the eigenvalues (because of the large 
dimensionality of some sub-blocks), and thus, the expression of the 
amplitudes. 

Finally, for each strip, we compute the 
corresponding limiting curve and isolated limiting points (where the 
chromatic zeros accumulate in the infinite-length limit), 
and analyze their main properties.  

%
%
\subsection{$\bm{L=2}$}

The chromatic polynomial for this strip is exactly the same as the chromatic
polynomial for a square-lattice strip of with $L=2$ with cyclic boundary 
conditions \cite{transfer4}: in the basis    
$\{ \delta_{1,1'}\delta_{2,2'}, \delta_{1,1'}+\delta_{2,2'},
    \delta_{1,2'}+\delta_{2,1'}, 1 \}$, it takes the form
\be
\T(2_\text{T}) \; = \; \left( \begin{array}{c|cc|c}
 1   & 0      & 0      & 0 \\
\hline
-1   & 2-q    & 1      & 0 \\
 0   & 1      & 2-q    & 0 \\
\hline
 1   & 2(q-2) & 2(q-2) & q^2 -3q + 3\\
            \end{array}\right)
\label{Tsq2T}
\ee
where we show by vertical and horizontal lines the lower-triangular 
block structure of this matrix. The right $\bm{w}_\text{id}$ and left 
$\bm{u}$ vectors are given by
\begin{subeqnarray}
 \bm{u}^T          &=& (A,2A,0,A)\,, \qquad \text{with $A=q (q-1)$}\\
 \bm{w}_\text{id}^T &=& (1,0,0,0)
 \label{UVsq2T}
\end{subeqnarray}

As in \cite{transfer4}, we find that the three diagonal blocks for
$\ell=0,1,2$ bridges give the following eigenvalues and amplitudes:
\begin{itemize}
\item $\bm{\ell=0}$: The eigenvalue is $\lambda_0=q^2-3q+3$ with 
      amplitude $b^{(0)}=1$.
      
\item $\bm{\ell=1}$: There are two eigenvalues $\lambda_{1,1}=1-q$ and 
      $\lambda_{1,2}=3-q$, with the common amplitude $b^{(1)}=q-1$. 
                
\item $\bm{\ell=2}$: The eigenvalue is $\lambda_2=1$ with amplitude
      $b^{(2)}=q^2-3q+1$.
\end{itemize} 
Each diagonal block $T_{\ell,\ell}$ is characterized by the  
bottom-row connectivity $\mathcal{P}_\text{bot}=1$. 

In Figure~\ref{figure_sq_1}(a) we have shown the chromatic zeros for
the strips of lengths $n=10$ and $n=20$. We also show the limiting curve
$\mathcal{B}_2$ where the chromatic zeros accumulate in the infinite-length
limit. 

The limiting curve $\mathcal{B}_2$ splits the complex $q$-plane into four 
regions. The regions inside the outer parts of $\mathcal{B}_2$ are 
characterized by a dominant eigenvalue belonging to the $\ell=1$ sector, 
while in the outer region the eigenvalue $\lambda_0$ dominates. 

The curve $\mathcal{B}_2$ crosses the real $q$ axis at two points:
$q=0$ and $q=2$. This latter point is in fact a multiple point. Finally,
there is a single isolated limiting point at $q=1$.

%
%
\subsection{$\bm{L=3}$}

The chromatic polynomial for this strip was computed by Chang and 
Shrock \cite{Shrock_01b} (in fact, they computed the full Potts-model
partition function \cite{Shrock_06}). 

The full transfer matrix $\T(3_\text{T})$ has dimension eight. 
In the basis
$\{ \delta_{1,1'}\delta_{2,2'}\delta_{3,3'}, 
    \delta_{1,2'}\delta_{2,1'}+\ldots,
    \delta_{1,2'}\delta_{3,3'}+\ldots, 
    \delta_{1,1'}\delta_{2,2'}+\ldots,
    \delta_{1,2'}\delta_{1,2'}+\ldots,
    \delta_{1,2'}+\ldots,
    \delta_{1,1'}+\ldots,1 \}$ (where the dots ``$\ldots$'' mean all possible 
    states equivalent under reflections and/or rotations), it takes the form
\be
\T(3_\text{T}) \; = \; \left( \begin{array}{c|cccc|cc|c}
-1   & 0    & 0    & 0    & 0    & 0     & 0   & 0    \\
\hline                                               
 0   & s_3  & 0    & 0    & -2   & 0     & 0   & 0    \\
 0   & 0    & s_3  & -1   & -1   & 0     & 0   & 0    \\
 1   & 0    & -2   & s_3  & 0    & 0     & 0   & 0    \\
 0   &-1    & -1   & 0    & s_3  & 0     & 0   & 0    \\
\hline                                              
 0   &-s_4  &-s_4  & 1    & 7-2q & s_3-S & s_3 & 0    \\
-1   & 0    &-2s_4 &-2s_3 & 2    & 2s_3  & -S  & 0    \\
\hline                                               
 1   & 3s_3 & 6s_3 &3s_3  &6s_3  &6S     &3S   & \lambda_0 \\
            \end{array}\right)
\label{Tsq3T}
\ee
where we by vertical and horizontal lines show the lower-triangular
block structure of this matrix, and 
\begin{subeqnarray}
s_k &=& q- k                        \slabel{def_sk}\\
S   &=& q^2 -5q + 7                 \slabel{def_S}\\
\lambda_0 &=& q^3 -6q^2 +14q-13     \slabel{def_l0_Tsq3T}
\end{subeqnarray}
The vectors $\bm{w}_\text{id}$ and $\bm{u}$ take the form 
\begin{subeqnarray}
 \bm{u}^T          &=& (A,0,0,3A,0,0,3A,A)\,, \qquad 
                       \text{with $A=q (q-1)(q-2)$}\\
 \bm{w}_\text{id}^T &=& (1,0,0,0,0,0,0,0)
 \label{UVsq3T}
\end{subeqnarray}

The eigenvalues can be obtained from the diagonal blocks $T_{\ell,\ell}$. 
Each block $T_{\ell,\ell}$ is characterized by the trivial 
bottom-row connectivity state $\mathcal{P}_\text{bot}=1$. 
We find the following eigenvalue structure for $\T(3_\text{T})$ 
\reff{Tsq3T}: 
\begin{itemize}
\item $\bm{\ell=0}$: There is a single eigenvalue 
$\lambda_0$ \reff{def_l0_Tsq3T} with amplitude $b^{(0)}=1$. 

\item $\bm{\ell=1}$: 
We find two eigenvalues: $\lambda_{1,1}=-(q^2-7q+13)$ with amplitude 
$b^{(1)}=q-1$ and $\lambda_{1,2}=-(q-2)^2$ with amplitude
$2b^{(1)}=2(q-1)$. 
                
\item $\bm{\ell=2}$: 
The four eigenvalues are $\lambda_{2,1}=q-1$, $\lambda_{2,2}=q-2$,
$\lambda_{2,3}=q-4$, and $\lambda_{2,4}=q-5$. The amplitudes are
given respectively by 
$b^{(2)}_2$, $2b_1^{(2)}$, $2b^{(2)}_2$, and $b_1^{(2)}$ 
[c.f., \reff{def_b2-b3}]. 

\item $\bm{\ell=3}$: The eigenvalue is $\lambda_3 = -1$ with amplitude 
$b^{(3)}$ \reff{def_b3}.
\end{itemize} 

In Figure~\ref{figure_sq_1}(b) we have shown the chromatic zeros for
the strips of lengths $n=15$ and $n=30$, as well as the limiting curve
$\mathcal{B}_3$ where the chromatic zeros accumulate in the infinite-length
limit. This curve splits the complex $q$-plane into three 
regions. The region containing the real segment $(0,2)$ is dominated 
by an eigenvalue from the $\ell=1$ sector. The region containing the
real segment $(2,3)$ is dominated by the $\ell=2$ sector. The rest of the
complex $q$-plane is dominated by the $\ell=0$ eigenvalue $\lambda_0$  
\reff{def_l0_Tsq3T}.

The curve $\mathcal{B}_3$ crosses the real $q$ axis at three points:
$q=0$, $q=2$, and $q=3$. We also find a pair of complex conjugate T points
at $q\approx 2.4692972375 \pm 1.4013927796\, i$. 
Finally, there is a single isolated limiting point at $q=1$.

%
%
\subsection{$\bm{L=4}$}

The chromatic polynomial for this strip was computed by Chang and
Shrock \cite{Shrock_01a}. 
The full transfer matrix $\T(4_\text{T})$ has dimension $68$. 
In this case, we restrict ourselves to report the eigenvalue 
structure:
\begin{itemize}

\item $\bm{\ell=0}$: This block has dimension five with two sub-blocks
   of dimensions three and two. There are three distinct eigenvalues:
   $\lambda_{0,1} = (q-1)(q-3)$ and the solutions of a second-order 
   equation \cite[Eq.~(2.12)]{Shrock_01a}. All of them have the
   amplitude $b^{(0)}=1$.

\item $\bm{\ell=1}$: This block has dimension $22$,  
   and it contains three sub-blocks of dimensions six and eight (two of them). 
   We find eight different eigenvalues. 
   Two of them are the solutions of the second-order equation 
   \cite[Eq.~(2.13)]{Shrock_01a} with amplitudes $2b^{(1)}=2(q-1)$;  
   The other six eigenvalues come from two distinct third-order 
   equations \cite[Eqs.~(2.18)/(2.19)]{Shrock_01a}, with 
   amplitudes $b^{(1)}=q-1$. 

\item $\bm{\ell=2}$: This block has dimension $33$ containing three 
   sub-blocks of dimensions $8$, $12$, and $13$. 
   Among these $33$ eigenvalues, we find $15$ distinct eigenvalues. 
   Three of them are simple: $\lambda_{2,1} =q^2 - 5q + 5$ with 
   amplitude $b_2^{(2)}$, $\lambda_{2,2}=q^2 - 5q + 7$ with amplitude
   $b_1^{(2)}$, and the special one $\lambda_{(2,3)} = 2 - q$ 
   which also appears in the $\ell=3$ sector (see below). 
   The next six eigenvalues are solutions of three second-order equations 
   \cite[Eqs.~(2.14)--(2.16)]{Shrock_01a}, with amplitudes
   $b_1^{(2)}$, $2b_1^{(2)}$, and $b_2^{(2)}$, respectively
   [c.f., \reff{def_b2-b3}].
   The last six eigenvalues come from two third-order equations 
   \cite[Eqs.~(2.20)/(2.21)]{Shrock_01a}, with amplitudes
   $b_1^{(2)}$ and $2b_2^{(2)}$, respectively. 

\item $\bm{\ell=2,3}$: The eigenvalue $\lambda_{(2,3)} = 2 - q$ 
appears in these two blocks with an amplitude 
$(q-2)(4q^2-13q-3)/6=b_2^{(2)}+2b_2^{(3)}$.
 
\item $\bm{\ell=3}$: This block has dimension seven:  
in addition to $\lambda_{(2,3)}$, we find six other eigenvalues: 
$\lambda_{3,1}=5-q$ with amplitude $b_1^{(3)}$ \cite[Eq.~(2.8)]{Shrock_01a};  
$\lambda_{3,2}=4-q$ with amplitude $2b_2^{(3)}$ \cite[Eq.~(2.7)]{Shrock_01a};
$\lambda_{3,3}=3-q$ with amplitude $2b_1^{(3)}$ \cite[Eq.~(2.6)]{Shrock_01a};
$\lambda_{3,4}=1-q$ with amplitude $b_1^{(3)}$ \cite[Eq.~(2.4)]{Shrock_01a};
and
$\lambda_{3,5,\pm}=3-q\pm\sqrt{3}$ with common amplitude $2b_2^{(3)}$ 
\cite[Eq.~(2.17)]{Shrock_01a}.

\item $\bm{\ell=4}$: This one-dimensional block gives 
$\lambda_4= 1$ with amplitude $b^{(4)}$ \reff{def_b4}.  
 
\end{itemize}

Thus, for this strip we find $33$ different eigenvalues. All of them but
$\lambda_{(2,3)}=2-q$, belong to a single $\ell$ sector. 

In Figure~\ref{figure_sq_1}(c) we have shown the chromatic zeros for
the strips of lengths $n=20$ and $n=40$, as well as the limiting curve
$\mathcal{B}_4$. This curve splits the complex $q$-plane into five 
regions. The region containing the real segment $(0,2)$ is dominated 
by an eigenvalue from the $\ell=1$ sector. The region containing the
real segment $(2,q_0(4))$ with $q_0(4)\approx 2.7827657401$ 
is dominated by the $\ell=2$ sector. The rest of the
complex $q$-plane (i.e., the other three regions) is dominated by the 
$\ell=0$ sector. Notice that the special eigenvalue
$\lambda_{(2,3)}=q-2$ is not dominant in any open set of the complex $q$-plane. 

The curve $\mathcal{B}_4$ crosses the real $q$ axis at three points:
$q=0$, $q=2$, and $q=q_0(4)\approx 2.7827657401$. 
We also find three pairs of complex conjugate T points at 
$q\approx 2.4184666383 \pm 1.7281014476\,i$,
$q\approx 2.6228997762 \pm 1.5279208067\,i$,
$q\approx 2.7784542034 \pm 0.9708262424\,i$.
Finally, there is a single isolated limiting point at $q=1$.

%
%
\subsection{$\bm{L=5}$}

The full transfer matrix $\T(5_\text{T})$ has dimension $347$.  
The relevant eigenvalues can be extracted from the diagonal blocks
$T_{\ell\ell}(5_\text{T})$. We find the following eigenvalue 
structure:

\begin{itemize}

\item $\bm{\ell=0}$: This block has dimension six. In the basis
 $\{\delta_{1',3'}+\ldots,\delta_{1',3'}\delta_{1,3}+\ldots, 
    \delta_{1',3'}\delta_{1,4}+\ldots,\delta_{1',3'}\delta_{2,4}+\ldots,
    1,\delta_{1,3}+\ldots\}$ (where the dots ``\ldots'' mean all 
    equivalent states under rotations and reflections), it takes the form:
\be
T_{0,0}(5_\text{T}) \;=\; \left( \begin{array}{cccc|cc}
T_1   & T_2      & 2T_2        & 2T_2       & 0 & 0\\ 
S-s_3 & S s_2    & -2s_2 s_3   & 2s_3       & 0 & 0\\
S-s_3 & -s_2 s_3 & S s_2 + s_3 & -s_3^2     & 0 & 0\\
S-s_3 & s_3      & -s_3^2      & s_2(S-s_3) & 0 & 0\\
\hline
0 & 0 & 0 & 0 & T_3  & S-s_3 \\
0 & 0 & 0 & 0 & 5T_2 & T_1  
\end{array}\right)
\ee
where $s_k$ and $S$ are defined in \reff{def_sk}/\reff{def_S}, and 
\begin{subeqnarray}
T_1 &=& q^5-10q^4+45q^3-115q^2+169q-116 \\
T_2 &=& q^4-9q^3+34q^2-63q + 47 \\
T_3 &=& q^3-9q^2+29q+32  
\end{subeqnarray}
This matrix is block diagonal: the first four-dimensional block corresponds
to the bottom-row connectivity $\mathcal{P}_\text{bot}=\delta_{1',3'}$,
while the second two-dimensional block corresponds to the trivial 
bottom-row connectivity $\mathcal{P}_\text{bot}=1$.

We find four distinct eigenvalues coming from two second-order equations: 
\begin{subeqnarray}
0 &=& 
\xi^2 - \xi (2q^3-13q^2+28q-19) + q^6 -13q^5 +69q^4-191q^3 \nonumber \\
 & & \quad +292q^2-236q+79 \slabel{def_sq_5T_eq1} \\
0&=& \xi^2 - \xi (q^5 -10q^4+46q^3-124q^2+198q -148)  + q^8-19q^7+159 q^6 
\nonumber \\
& & \quad 
-767 q^5 + 2339q^4 -4627 q^3+5800q^2-4212 q + 1362 
\slabel{def_sq_5T_eq2}
\end{subeqnarray} 
Their amplitudes are $2b^{(0)}=2$ and $b^{(0)}=1$, respectively. 

\item $\bm{\ell=1}$: This block has dimension $67$ 
and it contains four sub-blocks of dimension $14$ (three of them) and 
$25$. We find $15$ different eigenvalues:
$\lambda_{(1,2)}=-(q^2-4q+3)$ also appears in the $\ell=2$ block (see below), 
and the rest come from polynomial equations of order ten 
(with amplitude $2b^{(1)}=2(q-1)$) 
and four (with amplitude $b^{(1)}=q-1$). 

\item $\bm{\ell=1,2}$: The eigenvalue $\lambda_{(1,2)} = -(q^2-4q+3)$,
appears in these two blocks with amplitude $(q+1)(q-2)/2 = b^{(1)}+b_1^{(2)}$. 

\item $\bm{\ell=2}$: This block has dimension $190$ and it contains 
five sub-blocks of dimensions $31$, $33$ (three of them), and $60$.
We find $36$ distinct eigenvalues, including $\lambda_{(1,2)}$. 
Two eigenvalues come from the second-order equation:
\be
\xi^2 + \xi (2q^2-11q +16) + q^4-11q^3 +44q^2 -77q+51 \;=\; 0
\ee
with amplitude $b^{(2)}_2$ \reff{def_b22}.
The other eigenvalues are the solutions of polynomial equations of order
four, five, and twelve (two of them). Their amplitudes are 
are $b^{(2)}_2$, $b^{(2)}_1$, $2b^{(2)}_1$, and $2b^{(2)}_2$, respectively  
[c.f., \reff{def_b2-b3}].

\item $\bm{\ell=3}$: This block has dimension $72$, and it contains 
three sub-blocks of dimension $24$. The $24$ distinct eigenvalues
come from polynomial equations of order three (two of them), six, and twelve. 
The corresponding amplitudes are
$2b^{(3)}_2$, $b^{(3)}_1$, $2b^{(3)}_1$, and $2b^{(3)}_2$, respectively 
[c.f., \reff{def_b2-b3}].

\item $\bm{\ell=4}$: This block has dimension $11$. 
Three eigenvalues are rather simple: $\lambda_{4,1}=q-5$ with amplitude
$b_1^{(4)}$; $\lambda_{4,2} = q-3$ with amplitude $2b_2^{(4)}$; and
$\lambda_{4,3} = q-1$ with amplitude $b_3^{(4)}$ [c.f., \reff{def_b4-b5}].. 
The next four are solutions of two second-order equations
\begin{subeqnarray}
0&=& \xi^2 - \xi(2q-5) +q^2 -5q +5 \\ 
0&=& \xi^2 - \xi(2q-7) + q^2 -7q+ 11 
\end{subeqnarray}
with amplitudes $2b_1^{(4)}$ and $2b_3^{(4)}$, respectively.
The last eigenvalues come from a fourth-order equation
with amplitudes $2b_2^{(4)}$. 
 
\item $\bm{\ell=5}$: This one-dimensional block gives 
$\lambda_{5}=-1$ with amplitude $b^{(5)}$ \reff{def_b5}.

\end{itemize}

For this strip we find $90$ different eigenvalues. All of them but
one belong to a single $\ell$ sector. The only exception is the
eigenvalue $\lambda_{(1,2)}=-(q^2-4q+3)$ which appears in the sectors
$\ell=1,2$.

In Figure~\ref{figure_sq_1}(d) we have shown the chromatic zeros for
the strips of lengths $n=25$ and $n=50$, and the limiting curve
$\mathcal{B}_5$. 
This curve splits the complex $q$-plane into three 
regions. The region containing the real segment $(0,2)$ is dominated 
by an eigenvalue from the $\ell=1$ sector. The region containing the
real segment $(2,3)$ is dominated by the $\ell=2$ sector. The rest of the
complex $q$-plane is dominated by the $\ell=0$ sector. 
Notice that the special eigenvalue
$\lambda_{(1,2)}$ is not dominant in any open set of the complex $q$-plane. 

The curve $\mathcal{B}_5$ crosses the real $q$ axis at three points:
$q=0$, $q=2$, and $q=3$. 
We also find two pairs of complex conjugate T points at 
$q\approx 2.2834116867 \pm 2.0404211832\,i$,
$q\approx 2.4349888915 \pm 1.9935503628\,i$.
We also find a pair of complex-conjugate endpoints at
$q\approx 2.5034648020 \pm 2.0851731763\,i$.
Finally, there is a single isolated limiting point at $q=1$.

%
%
\subsection{$\bm{L=6}$}

The full transfer matrix $\T(6_\text{T})$ has dimension $2789$. 
The relevant eigenvalues can be extracted from the diagonal blocks
$T_{\ell\ell}(6_\text{T})$. We find the following eigenvalue 
structure:

\begin{itemize}

\item $\bm{\ell=0}$: This block has dimension 39, 
   and it contains five sub-blocks of dimensions five, seven, eight 
   (two of them), and eleven. We find eleven distinct eigenvalues. 
The simplest one is $\lambda_{0,1}=q^4 -9q^3+30q^2 -44q+25$ with amplitude 
$2b^{(0)}=2$. Two eigenvalues are the solutions of the second-order equation
\be
\xi^2 - \xi (q^4-10q^3+39q^2-68q+38) 
       -q^7+15q^6-93q^5+309q^4-598q^3+681q^2-422q+109 \;=\; 0  
\ee
with amplitude $b^{(0)}=1$. The other eigenvalues come from polynomial equations
of order three (with amplitude $2b^{(0)}=2$) and five 
(with amplitude $b^{(0)}=1$).

\item $\bm{\ell=1}$: This block has dimension $494$ 
   and it contains eleven sub-blocks of dimensions ranging from $21$ to $71$.
   We find $48$ different eigenvalues in this block. The simplest ones are
   given by $\lambda_{(1,2)} = -(q^3-8q^2+20q-14)$ and 
   $\lambda_{(1,2,3)} = -(q^3-6q^2+12q-8)$,   
   which also they appear in other blocks (see below).
Two eigenvalues are the solutions of the second-order equation:
\be
\xi^2 + \xi (2q^3-16q^2+42q-38) + 
        q^6-16q^5+104q^4-352q^3+656q^2-642q+262 \;=\; 0
\ee
with amplitude $b^{(1)}=q-1$. The other eigenvalues come from polynomial
equations of order ten, eleven (two of them), and twelve.

\item $\bm{\ell=1,2}$: The eigenvalue 
$\lambda_{(1,2)} = -(q^3-8q^2+20q-14)$ appears in these two blocks with an
amplitude $(q-2)(q+1)/2=b^{(1)}+b_1^{(2)}$. 

\item $\bm{\ell=2}$: This block has dimension $1432$ and it contains 
   $13$ sub-blocks of dimensions between $56$ and $180$.  
   We have found $120$ distinct eigenvalues: 
   $\lambda_{(1,2)}$, $\lambda_{(1,2,3)}$,  
   $\lambda_{(2,3)} = q^2-5q+5$   
   which also appears in the $\ell=3$ sector (see below),
   and $117$ eigenvalues coming from equations of order five (two equations), 
   seven (two equations), nine, eleven, $13$, $14$ (two equations), and 
   $16$ (two equations). 
   
\item $\bm{\ell=1,2,3}$: The eigenvalue 
$\lambda_{(1,2,3)} = -(q^3-6q^2+12q-8)$ 
appears in these three blocks with amplitude 
$(q-3)(2q^2 -3q+4)/6=b^{(1)}+b_1^{(2)}+b_1^{(3)}$.

\item $\bm{\ell=2,3}$: The eigenvalue 
$\lambda_{(2,3)} = q^2 -5q+5$ appears in these two blocks with amplitude
$(2q^3-9q^2+7q-6)/6=b_1^{(2)}+b_1^{(3)}$.

\item $\bm{\ell=3}$: This block has dimension $644$ and it contains 
    seven sub-blocks of dimensions between $31$ and $150$. 
We have found $82$ distinct eigenvalues coming from equations of
order three, six, seven, eight (six equations), and nine (two equations).
In addition, we obtain the already discussed eigenvalues  
$\lambda_{(1,2,3)}$ and $\lambda_{(2,3)}$. 
    
\item $\bm{\ell=4}$: This block has dimension $163$ and it contains four
   sub-blocks of dimensions $26$, $45$, and $46$ (two of them). We have found
   $49$ distinct eigenvalues. The simplest one is
   $\lambda_{4,1} = -(q-2)(q-3)$ with amplitude $b_1^{(4)}$ \reff{def_b41}.  
   Six eigenvalues are solutions of three second-order equations
\begin{subeqnarray}
0&=& \xi^2 - \xi(2q^2-11q+13) +q^4-11q^3+43q^2-71q+40 \\ 
0&=& \xi^2 - \xi(q^2-7q+8) -q^3+8q^2-18q+11 \\ 
0&=& \xi^2 - \xi(q^2-7q+10) -q^3+8q^2-20q+15 
\end{subeqnarray}
The other eigenvalues come from equations of order three (four equations),
four (four equations), six, and eight. 

\item $\bm{\ell=5}$: This block has dimension $16$. 
The simplest eigenvalues are
$\lambda_{5,1} = 5-q$ (with amplitude $b_1^{(5)}$),   
$\lambda_{5,2} = 4-q$ (with amplitude $2b_1^{(5)}$),  
$\lambda_{5,3} = 2-q$ (with amplitude $2b_1^{(5)}$), and  
$\lambda_{5,4} = 1-q$ (with amplitude $b_1^{(5)}$). 
Four eigenvalues are solutions of two second-order equations:
\begin{subeqnarray}
0&=& \xi^2 + \xi(2q-7)+q^2-7q+11\\
0&=& \xi^2 + \xi(2q-5)+q^2-5q+5 
\end{subeqnarray}
The remaining eight eigenvalues come from two fourth-order equations. 
The amplitude of all these twelve eigenvalues is $2b_2^{(5)}$ \reff{def_b52}.

\item $\bm{\ell=6}$: This one-dimensional block gives
$\lambda_{6}=1$ with amplitude $b^{(6)}$ \reff{def_b6}.
\end{itemize}

For this strip we find $325$ different eigenvalues; all of them but
three ($\lambda_{(1,2)}$, $\lambda_{(1,2,3)}$, and $\lambda_{(2,3)}$) 
belong to a single $\ell$ sector. 

In Figure~\ref{figure_sq_2}(a) we have shown the chromatic zeros for
the strips of lengths $n=30$ and $n=60$, and the limiting curve
$\mathcal{B}_6$. This curve splits the complex $q$-plane into four 
regions. The region containing the real segment $(0,2)$ is dominated
by an eigenvalue from the $\ell=1$ sector. The region containing the
real segment $(2,5/2)$ is dominated by the $\ell=2$ sector. The two small 
complex conjugate regions are dominated by the $\ell=1$ sector. The rest of 
the complex $q$-plane is dominated by the $\ell=0$ sector.

The curve $\mathcal{B}_5$ crosses the real $q$ axis at three points:
$q=0$, $q=2$, and $q\approx 2.9078023424$.
We also find four pairs of complex conjugate T points at
$q\approx 2.1180081660 \pm 2.2566257839\,i$,
$q\approx 2.1632612447 \pm 2.2209172289\,i$,
$q\approx 2.8895175995 \pm 1.1648906031\,i$, and 
$q\approx 2.9197649745 \pm 0.7583704656\,i$.
We also find a pair of complex-conjugate endpoints at
$q\approx 2.0571168133 \pm 2.3885607275\,i$.
Finally, there is a single isolated limiting point at $q=1$.

%
%
\subsection{$\bm{L=7}$}

The full transfer matrix $\T(7_\text{T})$ has dimension $20766$.
We have been unable to obtain the full eigenvalue structure, as some
of the blocks for $\ell=2,3$ are very large. Thus, we cannot obtain the
amplitudes.  We find the following (partial) eigenvalue structure: 
\begin{itemize}

\item $\bm{\ell=0}$: This block has dimension $111$, 
   and it contains six sub-blocks of dimensions six and $21$ (five of them). 
   We find $21$ different eigenvalues coming from polynomial equations of
   order six and $15$.  

\item $\bm{\ell=1}$: This block has dimension $2794$,
   and it contains $19$ sub-blocks of dimensions $103$ and $196$. 
   We find $112$ different eigenvalues coming from polynomial equations of 
   order nine, $19$, and $84$. 

\item $\bm{\ell=2}$: This block has dimension $10224$, 
   and it contains $26$ sub-blocks of dimensions $270$, $276$, and $532$. 
   We find $292$ different eigenvalues coming from polynomial equations of 
   order $16$, $22$, $26$, and $114$ (two of them) arising from 
   the characteristic polynomials of the sub-blocks of dimensions $\leq 276$. 
   This description is not complete, as we cannot obtain the characteristic
   polynomial of the sub-blocks of dimension $532$. However, we can numerically
   diagonalize this sub-block and find that it provides $12$ additional
   distinct eigenvalues; hence, we expect to have $304$ distinct eigenvalues 
   in this sector.

\item $\bm{\ell=3}$: This block has dimension $5615$, 
   and it contains $17$ sub-blocks of dimensions $247$ and $483$. 
   We find $247$ different eigenvalues coming from the polynomial equations of 
   order $17$, $23$, $69$, and $138$ that arise from the characteristic 
   polynomials of the smaller sub-blocks. Furthermore, we numerically 
   find that the larger sub-blocks of dimension $483$ contains only six
   additional distinct eigenvalues. Thus, we expect $253$ distinct 
   eigenvalues for this sector. 

\item $\bm{\ell=4}$: This block has dimension $2711$, 
   and it contains eight sub-blocks of dimensions $155$, $158$, and $308$. 
   We find $165$ different eigenvalues coming from polynomial equations of 
   order three, four, seven, eight, eleven, $33$ (two of them), and $66$. 

\item $\bm{\ell=5}$: This block has dimension $288$, 
   and it contains four sub-blocks of dimension $72$. 
   We find $72$ different eigenvalues coming from polynomial equations of 
   order four, eight, twelve, and $48$. 

\item $\bm{\ell=6}$: This block has dimension $22$. We find four simple
  eigenvalues $\lambda_{6,1}=q-5$, $\lambda_{6,2}=q-4$, $\lambda_{6,3}=q-2$,
  and $\lambda_{6,4}=q-1$. The other come from equations of order three
  and six (two of each). 

\item $\bm{\ell=7}$: This one-dimensional block gives $\lambda_7=-1$.  
\end{itemize}

In Figure~\ref{figure_sq_2}(b) we have shown the chromatic zeros for
the strips of lengths $n=35$ and $n=70$, and the limiting curve
$\mathcal{B}_7$. 

The curve $\mathcal{B}_7$ crosses the real $q$ axis at three points:
$q=0$, $q=2$, and $q=3$.
We also find six pairs of complex conjugate T points at
$q\approx 1.4876491726 \pm 2.5577806355\,i$,
$q\approx 1.6498030004 \pm 2.5002880707\,i$, 
$q\approx 1.9765852319 \pm 2.3664338035\,i$,
$q\approx 2.5531414480 \pm 1.8775702667\,i$,
$q\approx 2.8395564367 \pm 1.4513813377\,i$, and 
$q\approx 2.8734078918 \pm 1.3454749268\,i$.
There are two complex conjugate oval-shaped regions between the T points
at $q\approx 1.488 \pm 2.558\,i$ and $q\approx 1.650\pm 2.500\,i$. And
there is another pair of small bulb-like regions between the T points
at $q\approx 2.840 \pm 1.451\,i$ and $q\approx 2.873\pm 1.345\,i$. 

Thus, the limiting curve $\mathcal{B}_7$ splits the complex $q$-plane into 
seven regions. The region containing the real segment $(0,2)$ is dominated
by an eigenvalue from the $\ell=1$ sector. The region containing the
real segment $(2,3)$ is dominated by the $\ell=2$ sector. The two small
complex conjugate oval-shaped regions are dominated by the $\ell=0$ sector;
and the two small complex conjugate bulb-like regions are dominated by the
$\ell=1$ sector. The rest of the complex $q$-plane is dominated by the 
$\ell=0$ sector.

We also find a pair of complex-conjugate endpoints at
$q\approx 1.6947007027 \pm 2.5327609879\,i$.
Finally, there is a single isolated limiting point at $q=1$.

%
%
\section{Triangular-lattice numerical results}\label{sec.tri.results}

In this section we will analyze the results for the triangular-lattice strips 
of widths $2\leq L \leq 7$. We have checked our results using (in addition
to the already known cases $L\leq 4$ \cite{Shrock_01a,Shrock_01c,Shrock_06})
the identity \reff{check_Pmn}.

As for the square lattice, for each $L$ we give a detailed description 
of the eigenvalues and amplitudes. For the already known cases $L=2,3$,
we include for the sake of clarity the full expressions for the transfer
matrices and the left and right vectors. The exact expressions for the 
new cases $L=5,6$ can be found in the {\tt mathematica} file 
{\tt transfer6\_triT.m} available as part of the on-line version of this
paper in the {\tt cond-mat} archive at {\tt arXiv.org}.
As in the square-lattice case, for $L=7$  we can only present a partial
description of the eigenvalue structure, as some of the sub-blocks are
very large. 

%
%
\subsection{$\bm{L=2}$}

The chromatic polynomial for this strip was computed by Chang and 
Shrock \cite{Shrock_06} (in fact, they computed the full Potts-model
partition function). 

The full transfer matrix $\T(2_\text{T})$ has dimension five. In the basis
$\{\delta_{1',2}\delta_{2',1},\delta_{1',1}\delta_{2,2'},
   \delta_{1',2}+\delta_{2',1},\delta_{1',1}+\delta_{2',2},1\}$,  
we find the following transfer matrix with a lower-triangular block 
structure
\be
\T(2_\text{T})\;=\; \left( \begin{array}{cc|cc|c}
   1 & 1 & 0 & 0 & 0 \\
   1 & 1 & 0 & 0 & 0 \\
   \hline 
  -1 &-1 & 3-q & 3-q & 0 \\
  -1 &-1 & 3-q & 3-q & 0 \\
   \hline
   1 & 1 & 2(q-3) & 2(q-3) & q^2 -5q+6 
   \end{array} \right)
\label{Ttri2T}
\ee
The right $\bm{w}_\text{id}$ and left $\bm{u}$ vectors are given by
\begin{subeqnarray}
 \bm{u}^T          &=& (0,A,0,2A,A)\,, \qquad \text{with $A=q (q-1)$}\\
 \bm{w}_\text{id}^T &=& (0,1,0,0,0)
 \label{UVtri2T}
\end{subeqnarray}
The eigenvalues can be obtained from the diagonal blocks $T_{\ell,\ell}$. 
(each of them associated to $\mathcal{P}_\text{bot}=1$). 

After some algebra, we find the following eigenvalues and amplitudes: 
\begin{itemize}
\item $\bm{\ell=0}$: 
The eigenvalue is $\lambda_0=q^2-5q+6$ with amplitude $b^{(0)}=1$.
      
\item $\bm{\ell=1}$: 
There are two eigenvalues: $\lambda_1=-2(q-3)$ with amplitude 
$b^{(1)}=q-1$, and $\lambda_\odot=0$, which is common to the $\ell=2$ block. 

\item $\bm{\ell=1,2}$: The eigenvalue $\lambda_\odot=0$ appears in these two
blocks. Its amplitude is irrelevant for the
computation of both the limiting curve $\mathcal{B}_2$ and the chromatic
polynomials $P_{2_\text{P}\times n_\text{P}}(q)$. 

\item $\bm{\ell=2}$: We find $\lambda_\odot=0$, and the eigenvalue  
$\lambda_2=2$ with amplitude $b_1^{(2)}$ \reff{def_b21}. 
\end{itemize} 

In Figure~\ref{figure_tri_1}(a) we have shown the chromatic zeros for
the strips of lengths $n=10$ and $n=20$, and the limiting curve  
$\mathcal{B}_2$. This curve splits the complex $q$-plane into three 
regions. The outer one is dominated by the $\ell=0$ sector. The one 
containing the real interval $q\in (0,2)$ is dominated by the $\ell=1$
sector; and the one containing the real interval $q\in(2,4)$ is dominated
by the $\ell=2$ sector.  

The curve $\mathcal{B}_2$ crosses the real $q$ axis at three points:
$q=0$, $q=2$, and $q=4$. This latter point is in fact a multiple point. 
Finally, there are two isolated limiting points at $q=1$ and $q=3$.

%
%
\subsection{$\bm{L=3}$}

The chromatic polynomial for this strip was first computed by Chang and 
Shrock \cite{Shrock_01c}. 
The full transfer matrix $\T(3_\text{T})$ has dimension $13$. 
In the basis 
$\{\delta_{1,3'}\delta_{2,1'}\delta_{3,2'},
\delta_{1,2'}\delta_{2,3'}\delta_{3,1'},$
$\delta_{1,1'}\delta_{2,2'}\delta_{3,3'},
\delta_{1,2'}\delta_{2,1'}+\ldots,
\delta_{1,2'}\delta_{3,1'}+\ldots,
\delta_{1,1'}\delta_{2,2'}+\ldots,
\delta_{2,1'}\delta_{3,2'}+\ldots,
\delta_{2,1'}\delta_{3,3'}+\ldots,
\delta_{1,2'}\delta_{3,3'}+\ldots,
\delta_{1,2'}+\ldots,\delta_{1,1'}+\ldots,
\delta_{1,3'}+\ldots,1\}$ (where the dots ``\ldots'' mean all possible 
connectivity states symmetric under rotations), it takes the form
\be
\T(3_\text{T})\;=\; \left( \begin{array}{ccc|cccccc|ccc|c}
  -1 & 0 &-1 & 0 & 0 & 0 & 0 & 0 & 0 & 0 & 0 & 0 & 0\\
  -1 &-1 & 0 & 0 & 0 & 0 & 0 & 0 & 0 & 0 & 0 & 0 & 0\\
   0 &-1 &-1 & 0 & 0 & 0 & 0 & 0 & 0 & 0 & 0 & 0 & 0\\
   \hline 
   1 & 0 & 0 & s_4 & -1  & -1  & s_4 & s_4 & 0   & 0 & 0 & 0 & 0 \\
   1 & 1 & 0 & s_4 & s_4 &  0  & s_4 & -1  & -1  & 0 & 0 & 0 & 0 \\
   0 & 1 & 1 & -1  & s_4 & s_4 &  0  & -1  & s_4 & 0 & 0 & 0 & 0 \\
   1 & 0 & 1 & -1  &  0  & s_4 & s_4 & s_4 & -1  & 0 & 0 & 0 & 0 \\
   0 & 0 & 1 &  0  &  -1 & s_4 &  -1 & s_4 & s_4 & 0 & 0 & 0 & 0 \\
   0 & 1 & 0 & s_4 & s_4 &  -1 &  -1 & 0   & s_4 & 0 & 0 & 0 & 0 \\
   \hline
  -1 &-1 & 0 &-2s_4&-2s_4& 2   &-2s_4& -s_5& -s_5& -R&2s_3&-R & 0 \\ 
   0 &-1 & -1&-s_5 &-2s_4&-2s_4& 2   & -s_5&-2s_4& -R& -R &2s_3&0 \\ 
  -1 & 0 & -1&-s_5 & 2   &-2s_4&-2s_4&-2s_4& -s_5&2s_3&-R&-R & 0 \\ 
   \hline
   1 & 1 & 1 &3s_4 & 3s_4&3_s4 &3s_4 &3s_4 & 3s_4&3R & 3R & 3R &\lambda_0 
   \end{array} \right)
\label{Ttri3T}
\ee
where $s_k$ is defined in \reff{def_sk} and 
\begin{subeqnarray}
R   &=& q^2 -7q +  13               \slabel{def_R}\\
\lambda_0 &=& q^3 -9q^2 +29q-32     \slabel{def_l0_Ttri3T}
\end{subeqnarray}
The right $\bm{w}_\text{id}$ and left $\bm{u}$ vectors are given by
\begin{subeqnarray}
 \bm{u}^T          &=& (0,0,A,0,0,3A,0,0,0,0,3A,0A)\,, 
\qquad \text{with $A=q (q-1)(q-2)$}\\
 \bm{w}_\text{id}^T &=& (0,0,1,0,0,0,0,0,0,0,0,0,0)
 \label{UVtri3T}
\end{subeqnarray}
The eigenvalues can be obtained from the diagonal blocks $T_{\ell,\ell}$. 
(each of them associated to $\mathcal{P}_\text{bot}=1$).  
We find the following eigenvalue structure: 
\begin{itemize}
\item $\bm{\ell=0}$: There is a single eigenvalue 
$\lambda_0=q^3-9q^2+29q-32$ with amplitude $b^{(0)}=1$. 

\item $\bm{\ell=1}$: 
There are three distinct eigenvalues $\lambda_{1,1}=-2(q-4)^2$, 
and $\lambda_{1,2\pm}=(q^2-5q+7)\, e^{ \pm 2\pi i/3}$, 
with amplitude $b^{(1)}=q-1$.    
                
\item $\bm{\ell=2}$: The eigenvalues $\lambda_{(2,3\pm)}=e^{\pm 2\pi i/3}$ 
also appear in the $\ell=3$ (see below). We also find 
$\lambda_{2,1}=q-2$, $\lambda_{2,2}=3q-14$, and 
$\lambda_{2,3\pm}=(7-2q)e^{\pm 2\pi i/3}$. 
The corresponding amplitudes are 
$b^{(2)}_2$, $b_1^{(2)}$, and $b^{(2)}_2$, respectively  
[c.f., \reff{def_b2-b3}]. 

\item $\bm{\ell=2,3}$: 
The complex-conjugate pair $\lambda_{(2,3\pm)}=e^{\pm 2\pi i/3}$ appears 
in these two blocks with an amplitude 
$q(q-1)(2q-7)/6=b_1^{(2)} + b_2^{(3)}=b_2^{(2)} + b_1^{(3)}$. Please note
that in this case the linear combination for the amplitude is {\em not} 
unambiguously determined. 

\item $\bm{\ell=3}$: 
In addition to $\lambda_{(2,3\pm)}$,   
we find $\lambda_3 = -2$ with amplitude $b_1^{(3)}$ \reff{def_b31}. 
\end{itemize} 

In Figure~\ref{figure_tri_1}(b) we have shown the chromatic zeros for
the strips of lengths $n=15$ and $n=30$, and the limiting curve
$\mathcal{B}_3$. This curve splits the complex $q$-plane into three 
regions. The region containing the real segment $(0,2)$ is dominated 
by an eigenvalue from the $\ell=1$ sector. The region containing the
real segment $(2,3]$ is dominated by the $\ell=2$ sector. The rest of the
complex $q$-plane is dominated by the $\ell=0$ eigenvalue $\lambda_0$  
\reff{def_l0_Ttri3T}.

The curve $\mathcal{B}_3$ crosses the real $q$ axis at three points:
$q=0$, $q=2$, and $q\approx 3.7240755514$.
We also find a pair of complex conjugate T points
at $q\approx 3.9777108575 \pm 1.6284204463\,i$. 
Finally, there are two isolated limiting points at $q=1,3$.

%
%
\subsection{$\bm{L=4}$}

The chromatic polynomial for this strip was computed by Chang and
Shrock \cite{Shrock_01a}. 
The full transfer matrix $\T(4_\text{T})$ has dimension $99$. 
In this case, we restrict ourselves to report the eigenvalue 
structure:

\begin{itemize}
\item $\bm{\ell\geq 0}$: The eigenvalue $\lambda_\odot =0$   
  is common to all blocks $0\leq \ell\leq 4$. Its amplitude cannot be computed
  from our analysis; but it is unimportant to compute the chromatic polynomials
  and the limiting curve.  

\item $\bm{\ell=0}$: This block has dimension five with two sub-blocks of 
  dimensions three and two. In addition to $\lambda_\odot =0$, 
  we find two distinct eigenvalues coming from a 
  second-order equation \cite[Eq.~(4.9)]{Shrock_01a} and with amplitude
  $b^{(0)}=1$.     

\item $\bm{\ell=1}$: This block has dimension $26$,  
  and it contains three sub-blocks of dimensions
  six, and ten (two of them). We find eight different eigenvalues, one 
  of them being $\lambda_\odot=0$.
  The other eigenvalues come from a third-order equation 
  \cite[Eq.~(4.12)]{Shrock_01a} and from a fourth-order equation 
  \cite[Eq.~(4.14)]{Shrock_01a}. The amplitudes are in all cases $b^{(1)}=q-1$. 

\item $\bm{\ell=2}$: This block has dimension $52$,  
  and it contains three sub-blocks of dimensions $12$, and $20$ (two of them).
  There are $18$ different eigenvalues. In addition to 
  $\lambda_\odot=0$, we find two simple ones:
  $\lambda_{2,1}=2(3-q)$ and $\lambda_{2,2} = 2(q-3)^2$, both with 
  amplitudes $b_2^{(2)}$.    
  The other eigenvalues come from polynomial equations
  of order two, three, four, and six, respectively
  \cite[Eqs.~(4.11)/(4.13)/(4.15)/(4.17)]{Shrock_01a}. 
  The corresponding amplitudes are $b_1^{(2)}$ for the first three, and
  $b_2^{(2)}$ for the last one [c.f.~\reff{def_b2-b3}]. 

\item $\bm{\ell=3}$: This block has dimension $12$. 
We find three simple eigenvalues, $\lambda_\odot=0$,
$\lambda_{3,1}=3-q$ with amplitude $2b_2^{(3)}$, and $\lambda_{3,2}= 2(9-2q)$ 
with amplitude $b_1^{(3)}$ \reff{def_b31}.
We also find six eigenvalues coming from polynomial equations of order
two \cite[Eq.~(4.10)]{Shrock_01a} and six \cite[Eq.~(4.16)]{Shrock_01a},
with common amplitude $b_2^{(3)}$.
The remaining two eigenvalues $\lambda_{(3,4\pm)}$ are common to the 
$\ell=4$ block (see below). 

\item $\bm{\ell=3,4}$: There are two eigenvalues common to the 
$\ell=3,4$ sectors: $\lambda_{(3,4\pm)} = 1 \pm i$ with amplitude 
$(q-1)(q-2)(3q^2-11q-6)/12=b_1^{(3)} + b_2^{(4)}$.

\item $\bm{\ell=4}$: This is a four-dimensional block. In addition to
$\lambda_\odot=0$ and $\lambda_{(3,4\pm)}$, we find $\lambda_4=2$  
with amplitude $b_1^{(4)}$ \reff{def_b41}. 
\end{itemize}

Thus, for this strip we find $38$ distinct eigenvalues. All of them but
three ($\lambda_\odot$ and $\lambda_{(3,4\pm)}$) 
belong to a single $\ell$ sector.  

In Figure~\ref{figure_tri_1}(c) we have shown the chromatic zeros for
the strips of lengths $n=20$ and $n=40$, and the limiting curve
$\mathcal{B}_4$. This curve splits the complex $q$-plane into three 
regions. The region containing the real segment $(0,2)$ is dominated 
by an eigenvalue from the $\ell=1$ sector. The region containing the
real segment $(2,4)$ is dominated by the $\ell=2$ sector. The rest of the
complex $q$-plane is dominated by the $\ell=0$ sector. 

The curve $\mathcal{B}_4$ crosses the real $q$ axis at three points:
$q=0$, $q=2$, and $q=4$. 
We also find a pair of complex conjugate T points at 
$q\approx 3.3943280448 \pm 2.1041744504\,i$. 
Finally, there are two isolated limiting points at $q=1,3$.

%
%
\subsection{$\bm{L=5}$}

The full transfer matrix $\T(5_\text{T})$ has dimension $653$. 
The relevant eigenvalues can be extracted from the diagonal blocks
$T_{\ell\ell}(5_\text{T})$. We find the following eigenvalue 
structure:

\begin{itemize}

\item $\bm{\ell=0}$: This block has dimension eight, 
and it contains two sub-blocks of dimensions two and six.  
We find six distinct eigenvalues (with amplitude $b^{(0)}=1$)
coming from a fourth-order equation and the following second-order equation:
\begin{eqnarray}
0 &=& \xi^2 - \xi ( q^5 -15 q^4 + 98 q^3 - 355 q^2 +711q -614) 
  + 3q^8 -75q^7 +817 q^6 \nonumber \\ 
  & & -5062q^5 +19492 q^4 - 47696 q^3 + 72273 q^2 - 61822q + 22764 
\end{eqnarray}

\item $\bm{\ell=1}$: This block has dimension $125$ 
and it contains seven sub-blocks of dimension $25$. 
We find $24$ different eigenvalues in this block.
The eigenvalue $\lambda_{(1,2)}=-(q^2-5q+6)$ also appears in the  
$\ell=2$ block (see below).
The other eigenvalues come from polynomial equations of order three and $20$.
They all have the same amplitude $b^{(1)}=q-1$.

\item $\bm{\ell=1,2}$: The eigenvalue 
$\lambda_{(1,2)} = -(q^2-5q+6)$ with amplitude 
$2q(q-2)=2(b^{(1)} + b_1^{(2)} + b_2^{(2)})$ appears in these
two blocks.  

\item $\bm{\ell=2}$: This block has dimension $360$ and it contains six 
sub-blocks of dimension $60$. We find $57$ distinct eigenvalues, including
$\lambda_{(1,2)}$. We also find the simple eigenvalue
$\lambda_{2,1}=-(q-4)(2q-7)$ with amplitude $b_2^{(2)}$ \reff{def_b22}.
The other eigenvalues come from polynomial equations of order three, four,
and $24$ (two of them). 

\item $\bm{\ell=3}$: This block has dimension $135$ and it contains 
three sub-blocks of dimension $45$. The $42$ distinct eigenvalues
come from the solutions of polynomial equations of order: three 
(two of them), $12$, and $24$. 

\item $\bm{\ell=4}$: This block has dimension $20$. 
Three eigenvalues are rather simple: 
$\lambda_{4,1} = q-4$ with amplitude $2b_2^{(4)}$,
$\lambda_{4,2} = q-2$ with amplitude $b_3^{(4)}$, and  
$\lambda_{4,3} = 5q-22$ with amplitude $b_1^{(4)}$.
Four eigenvalues $\lambda_{(4,5,i)}$ are common to the $\ell=5$ block 
(see below). The other eigenvalues come from polynomial equations of order
four and eight, with amplitudes $b_3^{(4)}$ and $b_2^{(4)}$, respectively   
[c.f., \reff{def_b4-b5}]. 
 
\item $\bm{\ell=4,5}$: The four eigenvalues $\lambda_{(4,5,i)}$ are the 
solution of the equation
\be
\xi^4 + 3\xi^3 + 4\xi^2 + 2\xi +1 \;=\; 0
\ee
Their amplitude is 
$q(q-3)(q-2)(4q^2-15q+1)/20=b_1^{(4)} + b_2^{(5)}=b_3^{(4)} + b_1^{(5)}$.
Again, we find an ambiguity in the linear combination of basic amplitudes.

\item $\bm{\ell=5}$: This block is five-dimensional. In addition to 
$\lambda_{(4,5,i)}$, we find  
$\lambda_{5}=-2$ with amplitude $b_1^{(5)}$ \reff{def_b51}.
\end{itemize}

Thus, for this strip we find $148$ different eigenvalues. All of them but
five ($\lambda_{(1,2)}=-(q^2-5q+6)$ and $\lambda_{(4,5,i)}$) 
belong to a single $\ell$ sector. 

In Figure~\ref{figure_tri_1}(d) we have shown the chromatic zeros for
the strips of lengths $n=25$ and $n=50$, and the limiting curve
$\mathcal{B}_5$. This curve splits the complex $q$-plane into six 
regions. The region containing the real segment $(0,2)$ is dominated 
by an eigenvalue from the $\ell=1$ sector. The regions containing the
real segment $(2,4)$ is dominated by the $\ell=2$ sector. 
The rest of the complex $q$-plane (including the two complex-conjugate
triangular-shaped regions) is dominated by the $\ell=0$ sector. 

The curve $\mathcal{B}_5$ crosses the real $q$ axis at four points:
$q=0$, $q=2$, $q\approx 3.8482632901$ and $q=4$. 
We also find four pairs of complex conjugate T points at 
$q\approx 2.7382976052 \pm 2.8142366786\,i$,
$q\approx 2.7643522733 \pm 2.3252257137\,i$,
$q\approx 3.7492171129 \pm 1.9879721106\,i$, and
$q\approx 3.9634199872 \pm 0.3751435740\,i$.
Finally, there are two isolated limiting points at $q=1,3$.

%
%
\subsection{$\bm{L=6}$}

The full transfer matrix $\T(6_\text{T})$ has dimension $5194$. 
The relevant eigenvalues can be extracted from the diagonal blocks
$T_{\ell\ell}(6_\text{T})$. 

\begin{itemize}

\item $\bm{\ell\geq 0}$: There is a single eigenvalue common to all
blocks $\lambda_\odot=0$. As for the cases $L=2,4$, its amplitude cannot
be determined; but it is irrelevant to compute both the chromatic polynomials
and the limiting curve. 

\item $\bm{\ell=0}$: This block has dimension $49$,
and it contains five sub-blocks of dimensions ranging from five to $15$. 
We find $14$ distinct eigenvalues: $\lambda_\odot =0$, the solutions of  
the second-order equation
\begin{eqnarray}
0 &=& \xi^2 - \xi(3q^4 -33q^3+132q^2-225q+135) + 3q^8 -66q^7 +627 q^6 
      \nonumber \\
  & & -3354 q^5 + 11028 q^4 -22770 q^3 + 28755 q^2 -20250 q + 6075 
\end{eqnarray}
and the solutions of equations of order five and six. 
All these eigenvalues have the amplitude $b^{(0)}=1$.

\item $\bm{\ell=1}$: This block has dimension $843$
and it contains $13$ sub-blocks of dimensions ranging from $25$ to $71$. 
We find $62$ different eigenvalues in this block: 
$\lambda_\odot =0$, the solutions of the second-order equation
\be
\xi^2 + \xi (3q^3 -27q^2 + 79q -75) + q^6 -18 q^5 +131 q^4 
-492 q^3 +1000 q^2 -1032 q + 414 =0
\ee
and the solutions of equations of order four, nine, $22$, and $24$. 

\item $\bm{\ell=2}$: This block has dimension $2712$
and it contains $21$ sub-blocks of dimensions ranging from $96$ to $180$. 
We find $165$ different eigenvalues in this block: 
$\lambda_\odot =0$, and the solutions of polynomial equations of order
three (two of them), four (two of them), ten (three of them), 
$28$ (two of them), and $32$ (two of them).
 
\item $\bm{\ell=3}$: This block has dimension $1254$
and it contains nine sub-blocks of dimensions ranging from $54$ to $150$. 
We find $136$ different eigenvalues in this block: 
$\lambda_\odot =0$, $\lambda_{3,1}=(q-2)(q-3)$ with amplitude $5b_1^{(3)}$,
and solutions of equations of order four, six, eight, $16$ (five of them), and
$18$ (two of them).  

\item $\bm{\ell=4}$: This block has dimension $300$
and it contains five sub-blocks of dimensions ranging from $48$ to $84$. 
We find $76$ different eigenvalues in this block: 
$\lambda_\odot =0$, 
$\lambda_{4,1} = 3q^2-18q+26$ with amplitude $b_3^{(4)}$ \reff{def_b43},
$\lambda_{4,2} = 2(3-q)$ with amplitude $3b_3^{(4)}$, 
the solutions of the second-order equation
\be
\xi^2 + 4q^4-44 q^3 +176 q^2 -304 q+192 \;=\; 0
\ee
with amplitude $b_1^{(4)}$, 
and the solutions of equations of order three, four, six (two of them),
eight (three of them), twelve, and $16$. 

\item $\bm{\ell=5}$: This block has dimension $30$, and it contains
$28$ different eigenvalues: $\lambda_\odot =0$, 
$\lambda_{5,1}=26-6q$ with amplitude $b_1^{(5)}$ \reff{def_b51}, the 
solutions of the second-order equation
\be
\xi^2 + \xi (2q-7) + q^2 -7q +11 \;=\; 0 
\ee
with amplitude $2b_2^{(5)}$, four eigenvalues $\lambda_{(5,6,i)}$ common
to the $\ell=6$ sector (see below), and the solutions of equations of order
four and eight (two of them). 

\item $\bm{\ell=5,6}$: There are four common non-zero eigenvalues to these 
two sectors, coming from two second-order equations: 
$\lambda_{(5,6,1\pm)}=\sqrt{3}\, e^{\pm i\pi/6}$ with amplitude
$(q-3)(5q^5-39q^4+93q^3-76q^2+22q+10)$, and 
$\lambda_{(5,6,2\pm)}=e^{\pm i\pi/3}$ with amplitude
$(q-1)(5q^5-49q^4+161q^3-184q^2+6q+30)$. 

\item $\bm{\ell=6}$: This block has dimension $6$. The eigenvalues
are $\lambda_\odot=0$, $\lambda_{(5,6,i)}$, and  
$\lambda_6=2$ with amplitude $q(q-3)(q-1)(q^3-8q+19q-11)/6$. 
\end{itemize}

Thus, in this strip we find $476$ distinct eigenvalues. All of them but five
($\lambda_\odot=0$ and $\lambda_{(5,6,i)}$) belong to a single sector. 

In Figure~\ref{figure_tri_2}(a) we have shown the chromatic zeros for
the strips of lengths $n=30$ and $n=60$, and the limiting curve
$\mathcal{B}_6$. This curve splits the complex $q$-plane into seven  
regions. The region containing the real segment $(0,2)$ is dominated
by an eigenvalue from the $\ell=1$ sector. The region containing the
real segment $(2,3)$ is dominated by the $\ell=2$ sector. 
The two bulb-like complex-conjugate regions appearing on the right hand-side
of the limiting curve belong to the $\ell=1$ sector and they are rather
special (see below).  
The rest of the complex $q$-plane (including the two complex-conjugate
triangular-shaped regions) is dominated by the $\ell=0$ sector.

The curve $\mathcal{B}_6$ crosses the real $q$ axis at three points:
$q=0$, $q=2$, and $q\approx 3.7943886378$. 
We also find five pairs of complex conjugate T points at
$q\approx 2.0352758561 \pm 3.1492288108\,i$,
$q\approx 3.3399899103 \pm 2.4428522934\,i$,
$q\approx 2.5037195029 \pm 2.3940503612\,i$,
$q\approx 4.1131475777 \pm 0.7636584371\,i$, and
$q\approx 4.1295836790 \pm 0.8010545279\,i$.
Finally, there are two isolated limiting points at $q=1,3$.

\bigskip

\noindent
{\bf Important technical remark.} 
As stated above, the two bulb-like complex-conjugate protruding from T points
$q\approx 4.11 \pm 0.76\,i$ and $q \approx 4.13\pm 0.80\,i$,  are rather
special. Their peculiarity arises from the fact that inside these two
bulb-like regions there is a pair of {\em exactly} equimodular dominant 
eigenvalues [with non-zero amplitudes $\sim (q-1)$]. Indeed, we do not see
chromatic zeros becoming dense in the interior of these regions, but only on
their boundaries.  

The explanation of this phenomenon is rather simple: the two dominant 
eigenvalues come from a polynomial equation of order $24$ belonging to the
$\ell=1$ sector. A closer look at this equation reveals that its roots $z_k$ 
can be written as $z_k = z_{0,k} e^{\pm \pi i/3}$ where the 
$z_{0,k}$ are the roots of another polynomial equation of order $12$ with 
integer coefficients (that can be easily computed).  
It is important to stress that the existence of this pair of equimodular 
eigenvalues does not mean that the chromatic zeros accumulate in this open
set in the infinite-length limit. 
According to the Beraha--Kahane--Weiss theorem \cite{BKW}, 
the eigenvalues should satisfy a ``no-degenerate-dominance'' condition: 
there do not exist indexes $k\neq k'$ such that 
$\lambda_k = \omega \lambda_{k'}$ with $|\omega|=1$, and such that the 
region where $\lambda_k$ and $\lambda_{k'}$ are
dominant has non-empty interior. This is precisely the condition violated by
the leading eigenvalues in the interior of the above bulb-like regions. 
Thus, in order to compute the limiting curve in these regions, we should 
only consider the roots $z_{0,k}$ of the $12$-order equation. 
In this way, we ensure the ``no-degenerate-dominance'' condition.  
In conclusion, the chromatic zeros cannot accumulate in the interior of these 
bulk-like regions; but on their boundary (where the dominant $z_{0,k}$ root 
from the $\ell=1$ sector becomes equimodular to the leading eigenvalue from the
$\ell=0$ sector).  

In other strips we have already found the existence of two different 
eigenvalues differing by a phase [i.e., $\lambda$ and $\lambda e^{i\phi}$].
However, this is the first case where one of these ``degenerate'' pairs 
is dominant on an open set of the complex $q$-plane. In order to correctly
compute the limiting curve, one should eliminate one of eigenvalues of each 
``degenerate'' pair to ensure the hypothesis of the 
Beraha--Kahane--Weiss theorem.

%
%
\subsection{$\bm{L=7}$}

The full transfer matrix $\T(7_\text{T})$ has dimension $40996$.\footnote{  
  This matrix is so large that its {\tt mathematica} representation cannot be
  fitted in 12Gb of RAM. In order to deal with it, we have used 
  {\tt mathematica}'s {\tt SparseMatrix} function, which saves a lot of
  memory due to the block-diagonal structure of our matrix.
}
We have been unable to obtain the full eigenvalue structure, as some
of the blocks for $\ell=2,3$ are very large. 
We find the following (partial) eigenvalue structure: 

\begin{itemize}
\item $\bm{\ell=0}$: This block has dimension $186$, and contains five 
   sub-blocks of dimension $36$, and a single block of dimension six. 
   We find $36$ distinct eigenvalues coming from two polynomial equations 
   of order six and $30$, respectively.

\item $\bm{\ell=1}$: This block has dimension $5488$, and contains $28$ 
   sub-blocks of dimension $196$. We find $196$ distinct eigenvalues 
   coming from a single polynomial equation of order $196$.

\item $\bm{\ell=2}$: This block has dimension $20216$, and contains $38$ 
   sub-blocks of dimension $532$. A numerical study of these sub-blocks
   reveals that there are $508$ distinct eigenvalues in this sector. 

\item $\bm{\ell=3}$: The dimension of this block is $11109$ and it contains
   $23$ sub-blocks of dimensions $483$. We numerically find that there
   are $454$ distinct eigenvalues for this sector.
 
\item $\bm{\ell=4}$: This block has dimension $3388$, and contains eleven 
   sub-blocks of dimension $308$. There are $289$ distinct eigenvalues:
   two of them are rather simple: $\lambda_{4,1}=-(q^2-5q+6)$ and 
   $\lambda_{4,1}=-(6q^2-44q+80)$, while the rest come from polynomial equations
   of order three, four, five, eleven, $66$ (two of them), and $132$.

\item $\bm{\ell=5}$: The dimension of this block is $560$ and it contains
   four sub-blocks of dimensions $140$. There are $122$ distinct eigenvalues
   coming from polynomial equations of order four, eight, $24$, and $96$.

\item $\bm{\ell=6}$: This block has dimension $42$. We find four simple
   eigenvalues $\lambda_{6,1}=q-5$, $\lambda_{6,2}=q-3$, $\lambda_{6,3}=q-2$,
   and $\lambda_{6,4}=7q-30$. There are $30$ additional eigenvalues
   coming from equations of order six and twelve (two of them). Finally,
   there are six other eigenvalues common to the $\ell=7$ block.

\item $\bm{\ell=6,7}$: We find six eigenvalues common to these blocks. They
  are the solutions of the equation
\be
\xi^6 + 5 \xi^5 +11\xi^4 +13\xi^3 + 9\xi^2 + 3\xi + 1 \;=\; 0 
\ee

\item $\bm{\ell=7}$: This block has dimension seven, and in addition to the 
  eigenvalues common to the $\ell=6$ block, we find $\lambda_{7}=-2$.

\end{itemize}

In Figure~\ref{figure_tri_2}(b) we have shown the chromatic zeros for
the strips of lengths $n=35$ and $n=70$, and the limiting curve
$\mathcal{B}_7$. This curve splits the complex $q$-plane into eight  
regions. The region containing the real segment $(0,2)$ is dominated
by an eigenvalue from the $\ell=1$ sector; the region containing the
real segment $(2,3)$ is dominated by the $\ell=2$ sector; and the small
region containing the real segment $(3.8,4)$ is dominated also by the 
$\ell=2$ sector. There are two small lens-like complex-conjugate regions 
which belong to the $\ell=1$ sector. The rest of the complex $q$-plane 
(including the two complex-conjugate triangular-shaped regions) is 
dominated by the $\ell=0$ sector.

The curve $\mathcal{B}_7$ crosses the real $q$ axis at four points:
$q=0$, $q=2$, $q\approx 3.8605508375$, and $q=4$.
We also find six pairs of complex conjugate T points at
$q\approx 1.4835869410 \pm 3.2225993236\,i$,
$q\approx 2.4516540074 \pm 2.4976908148\,i$, 
$q\approx 3.0454262812 \pm 2.6881696318\,i$,
$q\approx 3.3660354065 \pm 2.4484944601\,i$,
$q\approx 3.8749646144 \pm 1.8155509876\,i$, 
and
$q\approx 3.9442064557 \pm 0.2720862591\,i$.
Finally, there are two isolated limiting points at $q=1,3$.

%
%
\section{Discussion}
\label{sec.discussion}

%
%
\subsection{Square lattice limiting curves}
\label{sec.sq.discussion}

A summary of qualitative results for the limiting curves $\mathcal{B}$, as
obtained in Section~\ref{sec.sq.results}, is shown in
Table \ref{table_summary_sq}. In particular, we have computed $q_0(L)$, the
maximum real value in $\mathcal{B}$ for strip widths $L \le 7$ from the
symbolic transfer matrices. In Figure~\ref{figure_sq_allT}, we display 
together all the computed limiting curves $\mathcal{B}_L$ with $2\le L\le 7$.
For comparison, we also include the limiting curve for a strip of width 
$L=10$ with {\em cylindrical} boundary conditions \cite{transfer2}.

These computations can be taken a lot further by diagonalizing the diagonal
blocks $T_{\ell \ell}(L_\text{T})$ of the {\em numerical} transfer matrix.
Indeed it suffices to diagonalize one of its sub-blocks, e.g., the one with a
trivial bottom-row connectivity. This amounts to working with a transfer
matrix that acts on connectivities of only $L$ (rather than $2L$) points, with
exactly $\ell$ marked clusters. For the purpose of getting only the
eigenvalues, without going outside the specified sub-block, transfer matrix
elements which would amount to adjoining two {\em distinct} marked clusters are
set to zero.

Since further we need only the largest (or sometimes the first few largest)
eigenvalues of the given sub-block, we can diagonalize using a standard
iterative procedure (the so-called power method \cite{Wilkinson}) which
has the immense advantage of allowing for sparse matrix factorization
(i.e., the lattice edges are added one by one) combined with standard
hashing techniques. This allowed us to access widths $L \le 12$. In all
cases, $q_0(L)$ was found to correspond to the crossing (in modulus) of the
leading eigenvalues of the $\ell=0$ and $\ell=2$ sectors. The precise value
of $q_0(L)$ was then narrowed in by a Newton-Ralphson method. The final results
are shown in Table \ref{table_summary_sq}.\footnote{
  Indeed, for $2\leq L \leq 7$ the numerical values found using the 
  numerical transfer matrices agree with those found using the 
  symbolic matrices.
} 

{}From the curious fact that $q_0(L)=3$ exactly for $L=3,5,7,9,11$ we
conjecture that this is true for all square-lattice strips with odd widths:

\begin{conjecture}\label{conj.sq.1}
For square-lattice strips with fully periodic boundary conditions 
of widths $L=2p+1$ with $p \ge 1$, we have $q_0(L)=3$.
\end{conjecture}

In other words, we have found a particular sequence of strip graphs for which
the value of $q_0(L)$ is constant and, furthermore, equal to the expected
critical value $q_{\rm c}(\text{sq})=3$. Indeed, recall that for $q=3$, the
square-lattice chromatic polynomial is equal to three times the partition
function of the six-vertex model with all vertex weights equal to one
\cite{Lenard}; it is well-known that this latter model is critical (with
central charge $c=1$).

We next conjecture that parity effects do not affect the value of $q_0$ in the
thermodynamic limit:

\begin{conjecture}\label{conj.sq.2}
For square-lattice strips with fully periodic boundary conditions of widths
$L=2p$, the limit $q_0({\rm sq}) = \displaystyle \lim_{p\to\infty} q_0(L)$ 
exists and is equal to $q_{\rm c}({\rm sq}) = 3$.
\end{conjecture}

This conjecture is very similar to the one made in \cite{transfer2};
the only difference being the boundary conditions, cylindrical rather
than toroidal. 

As to the value of $q_0(L)$ for even $L$, the data of 
Table~\ref{table_summary_sq} is well fitted by the power-law form 
$q_0(L)=q_c(\text{sq}) + A L^{-1/\nu}$. Our estimates are 
$q_0(\text{sq})=2.999\pm 0.012$ and $\nu=0.49\pm 0.02$. These results 
strongly corroborate Conjecture~\ref{conj.sq.2}. A better estimate for 
$\nu$ can be obtained by fixing $q_0(\text{sq})=3$ in the above Ansatz:   
we obtain the critical exponent $\nu = 0.502 \pm 0.002$. 
We conjecture that the exact
value of the exponent is $\nu = 1/2$, consistent with the transition being
{\em first-order} in $q$.

The limiting curves $\mathcal{B}_L$ for $2\leq L \leq 7$ show some
regularities (see  Figure~\ref{figure_sq_allT}): 
for real $q$, we have three different phases and each of
them corresponds to a different $\ell$ sector ($\ell=0,1,2$). Furthermore,
within  each phase, the leading amplitude does not depend on $L$. All
our empirical findings can be summarized in the following conjecture:

\begin{conjecture}\label{conj.sq.3}
The phase diagram of the zero-temperature square-lattice Potts 
antiferromagnet with toroidal boundary conditions at real $q$ has three
different phases characterized by the number of bridges $\ell$ as follows:
\begin{itemize}
 \item[(a)] $q\in(-\infty,0)\cup(q_0(L),\infty)$ characterized by $\ell=0$ 
            with amplitude $b^{(0)}=1$
 \item[(b)] $q\in(0,2)$ characterized by $\ell=1$ with amplitude $b^{(1)}=q-1$
 \item[(c)] $q\in(2,q_0(L))$ characterized by $\ell=2$ 
            with amplitude $b^{(2)}_1=q(q-3)/2$
\end{itemize} 
\end{conjecture}
Note that for all widths studied, $2\leq L\leq 12$, we have that $2\leq
q_0(L)\leq 3$.

We also find a curious behavior of one of the eigenvalues belonging to
the highest sector for each strip width: 

\begin{conjecture}\label{conj.sq.4}
The diagonal sub-block $T_{L,L}$ for a square-lattice strip with fully 
toroidal boundary conditions and width $L$ contains the eigenvalue 
$\lambda=(-1)^L$ with amplitude $b^{(L)}$.
\end{conjecture}

%
%
\subsection{Triangular lattice limiting curves}
\label{sec.tri-lim-curves}

A summary of qualitative results for the limiting curves $\mathcal{B}$
is shown in Table~\ref{table_summary_tri}. 
In Figure~\ref{figure_tri_allT}, we display
together all the computed limiting curves $\mathcal{B}_L$ with $2\le L\le 7$.
For comparison, we also include the infinite-volume limiting curve 
obtained by Baxter \cite{Baxter_86_87}.
The results in Table~\ref{table_summary_tri} with $L \le 7$ were
obtained from the symbolic transfer matrices in Section~\ref{sec.tri.results},
and those with $L \le 12$ were
found by the numerical procedure outlined in Section~\ref{sec.sq.discussion}.
In all cases $q_0(L)$ is given by the crossing between the dominant
eigenvalues in the $\ell=0$ and $\ell=2$ sectors.

{}From the curious fact that $q_0(L)=4$ exactly for $L=2,4,5,7,8,10,11$
we conjecture that:

\begin{conjecture}\label{conj.tri.1}
For triangular-lattice strips with fully periodic boundary conditions,
and for all widths $L=3p \pm 1$ with $p \ge 1$, we have $q_0(L)=4$.
\end{conjecture}

A quick glance at Table~\ref{table_summary_tri} however reveals that the
values of $q_0(L=3p)$ are definitely not converging to $4$. We therefore
examine the data with $L=3p$ more closely. A first fit with
$q_0(L)=q_0(\text{tri}) + A L^{-1/\nu}$ yields $q_0(\text{tri}) \simeq 3.820$
and $\nu \simeq 0.52$ (See Table~\ref{table_fits_tri}).
This is consistent with the expected value $\nu = 1/2$ corresponding to a
first-order phase transition (in the variable $q$), as we have already found 
for the square lattice. Fixing $\nu = 1/2$ we then
arrive at the final estimate $q_0(\text{tri}) = 3.8196 \pm 0.0005$.

\bigskip

\noindent
{\bf Remark}. 
We have performed all the fits reported in this paper using the following 
method: as the data is essentially exact, we have done a standard 
least-squares fit with as many data points as the number of unknown 
parameters $N_A$ in the Ansatz, so that in each fit there are no  
effective degrees of freedom. To estimate the error bars (due to the existence
of correction-to-scaling terms not included in the Ansatz), we  
introduce the parameter $L_\text{min}$ such that each fit is performed with
$N_A$ consecutive data points with $L\geq L_\text{min}$. From the variation
of the estimates as a function of $L_\text{min}$, we deduce the corresponding
error bars.

\bigskip 

Finally, from the FSS theory of first-order phase transitions
\cite{Privman}, we expect that higher corrections-to-scaling terms 
should be integer powers of $L^{-1/\nu}=L^{-2}$. Thus, we have fitted  
our data to the improved Ansatz
\be
q_0(L) \;=\; q_0(\text{tri}) + A L^{-2} + B L^{-4} 
\label{def_ansatz_ok}
\ee
The results are displayed in Table~\ref{table_fits_tri}, and from these we
conclude that $q_0(\text{tri}) = 3.8198 \pm 0.0006$. 

We now recall that the triangular-lattice chromatic polynomial was
actually shown by Baxter \cite{Baxter_86_87} to be an integrable
model.  In particular, using coordinate Bethe Ansatz, this author
identified three different analytic expressions for the free energy in
the thermodynamic limit.

To be precise, Baxter used first the Nienhuis
mapping \cite{Nienhuis_82} between the zero-temperature $q$-state
Potts antiferromagnet on the triangular lattice and the critical
O($n$) loop model on the hexagonal lattice. That mapping includes an
exact relation between $q$ and the weight $n$ of a (bulk) loop. Baxter
then transformed the loop model into a vertex model, by the usual
trick of redistributing the loop weight $n$ as {\em local} complex
vertex weights. Unfortunately, when imposing periodic transverse
boundary conditions on the vertex model, the loops of non-trivial
homotopy (i.e., that wrap around the periodic direction) get the
incorrect weight of $2$, a fact not mentioned in Baxter's
paper. Actually, while giving a (boundary) weight of $n' = n$ to those
loops (by twisting the vertex model) would have been a nicer way to
perform the analysis of the O($n$) model, the subtleties of the
Nienhuis mapping \cite{Nienhuis_82} are such that yet a different
weight $n'(q) \neq n$ would have been needed to ensure that clusters
of non-trivial homotopy in the Potts antiferromagnet carry their
correct weight of $q$.

One might think that these subtleties have no
implications on the behavior in the thermodynamic limit, but the
reader is reminded that we are here dealing with the antiferromagnetic
regime in which boundary conditions can (and sometimes do)
matter. Nevertheless, it seems reasonable to expect that this should
have no consequence for the {\em bulk} free energies. Numerical
analysis \cite{Baxter_86_87,transfer3} of Baxter's analytically
determined free energies shows that the corresponding eigenvalues
become equimodular at \be q_\text{B} \simeq 3.8196717312 \,.
 \label{qBaxter}
\ee
The value (\ref{qBaxter}) is determined implicitly by equating two analytical
expressions given in Ref.~\cite{Baxter_86_87}. Given the excellent agreement
with our above numerical determination of $q_0(\text{tri})$ we can 
therefore conjecture that:

\begin{conjecture}\label{conj.tri.2}
For triangular-lattice strips with fully periodic boundary conditions 
of widths $L=3p$, the limit 
$q_0({\rm tri}) = \displaystyle \lim_{p\to \infty} q_0(L)$
exists and is equal to $q_\text{B}$ of Eq.~\reff{qBaxter}. Furthermore,
the sequence $q_0(3p)$ is monotonically increasing in $p$.
\end{conjecture}

The number of regions enclosed by $\mathcal{B}$ having a non-zero
intersection with the real $q$-axis is $2$ for $L=2,3,4$ 
[cf.~Figure~\ref{figure_tri_1} and
Table~\ref{table_summary_tri}]. However, for $L=5$ a further such
region appears, whose intersecting with the real $q$-axis is the
interval $(q_2(L),q_0(L))$, where $q_2(L) \simeq 3.84826329$. The value
$q_2(L)$ corresponds to the crossing of the two largest eigenvalues in
the $\ell=2$ sector and is tabulated for strip widths $L \le 12$ in
Table~\ref{table_summary_tri}. This table gives compelling evidence
that $q_2(L)$ tends to a limit, $q_2({\rm tri}) < 4$, independently of
the parity of $L \mod 3$.

As above, we first fit the data for $q_2(L)$ with a fixed parity, $L \mod 3 =
0$, to the form $q_2(L)=q_2(\text{tri}) + A L^{-1/\nu}$ (See 
Table~\ref{table_fits_tri}). This is again
favourable of $\nu \approx 1/2$. Repeating the fit with a fixed $\nu=1/2$ we
arrive at the final estimate $q_2(\text{tri}) = 3.8194 \pm 0.0016$. Using
the improved Ansatz \reff{def_ansatz_ok}, we get the value 
$q_2(\text{tri}) = 3.8199 \pm 0.0017$. We can therefore conjecture that:

\begin{conjecture}\label{conj.tri.3}
For triangular-lattice strips with fully periodic boundary conditions 
the limit $q_2({\rm tri}) = \displaystyle \lim_{L\to\infty} q_2(L)$ 
exists and is equal to $q_\text{B}$ of Eq.~\reff{qBaxter}. Furthermore,
the sequence $q_2(3p)$ is monotonically decreasing in $p$.
\end{conjecture}

The monotonic nature of the sequences $q_0(\text{tri})$ and 
$q_2(\text{tri})$, and the existence of a common limit (as stated in
Conjectures~\ref{conj.tri.2}--\ref{conj.tri.3}), means that 
this common limit is bounded by the finite-widths values: i.e.,
$q_0(3p)\leq q_\text{B} \leq q_2(3p)$ for all $p\geq 2$. The data 
for $L=12$ displayed in Table~\ref{table_summary_tri} implies that
\be
3.81314 \;\ltapprox\; q_\text{B} \;\ltapprox\; 3.82428
\label{def_qB_ineq}
\ee
In particular, this narrow interval excludes that this common limit 
would be equal to the value $q_\text{G}=B_{12}$ \reff{def_qG} proposed
in Ref.~\cite{transfer4}.

As for the square-lattice case, the limiting curves $\mathcal{B}_L$ for 
$2\leq L \leq 7$ show some regularities (see  Figure~\ref{figure_tri_allT}): 
for real $q$, we have up to
four different phases belonging to only three different $\ell$ sectors 
($\ell=0,1,2$). Again, within  each phase, the leading amplitude does not 
depend on $L$. All our empirical findings can be summarized in the following 
conjecture:

\begin{conjecture}\label{conj.tri.4}
The phase diagram of the zero-temperature triangular-lattice Potts
antiferromagnet with toroidal boundary conditions at real $q$ has four 
different phases characterized by the number of bridges $\ell$ as follows:
\begin{itemize}
 \item[(a)] $q\in(-\infty,0)\cup(q_0(L),\infty)$ characterized by $\ell=0$
            with amplitude $b^{(0)}=1$
 \item[(b)] $q\in(0,2)$ characterized by $\ell=1$ with amplitude $b^{(1)}=q-1$
 \item[(c)] $q\in(2,q_2(L))$ characterized by $\ell=2$
            with amplitude $b^{(2)}_1=q(q-3)/2$
 \item[(d)] $q\in(q_2(L),q_0(L))$ characterized by $\ell=2$
            with amplitude $b^{(2)}_1=q(q-3)/2$
\end{itemize}
\end{conjecture}

Note that the phase (d) only appears for finite width $L$ whenever
$q_0(L)>q_2(L)$ (e.g. $L=5,7$). But, according to 
Conjectures~\ref{conj.tri.1}--\ref{conj.tri.3}, this phase does survive in
the thermodynamic limit $L\to\infty$. We believe that this phase
will appear for all $L \ge 5$ such that $L \mod 3 \neq 0$.

For $5\leq L\leq 7$ we further note the appearance of two disjoint 
triangular-shaped complex-conjugate
enclosed regions, whose boundary contains three T points, one of which
belongs to the branch of $\mathcal{B}$ that contains the point $q=2$. The
dominant eigenvalue inside these enclosed regions comes from the $\ell=0$
sector. Figures~\ref{figure_tri_1}--\ref{figure_tri_2} provide some evidence
that these regions may subsist in the thermodynamic limit.

Finally, for $L=6$ we note the emergence of a pair of
complex-conjugate enclosed regions analogous to those found for the square
lattice. 
More precisely, these enclosed regions
contain points for which $\Re q > 4$, and the corresponding
dominant eigenvalue comes from the $\ell = 1$ sector. We find it
likely that such regions will appear for all $L=3p$ with $p \ge 2$,
and will merge in the $L\to\infty$ limit so as to form a branch that
intersects the real $q$-axis in $q=q_0(\text{tri})=4$.

There is a analogous conjecture for the triangular lattice to 
Conjecture~\ref{conj.sq.4} about one of the eigenvalues of the highest sector: 

\begin{conjecture}\label{conj.tri.5}
The diagonal sub-block $T_{L,L}$ for a triangular-lattice strip with fully
toroidal boundary conditions and width $L$ contains the eigenvalue
$\lambda=(-1)^L\, 2$ with amplitude $b_1^{(L)}$.
\end{conjecture}

%
%
\subsection{Triangular-lattice free energy} \label{sec.tri.free}

In this section we will discuss the free energy for the triangular-lattice
strips we have considered and its relation to the analytic eigenvalues found 
by Baxter \cite{Baxter_86_87}.

The (real part of the) free energy {\em per unit area} $f_L(q)$ for 
a triangular-lattice strip of width $L$ is given by\footnote{
  Note that in previous papers 
  \protect\cite{transfer1,transfer2,transfer3,transfer4}, we have considered 
  the free energy {\em per spin} and used the same notation $f_L(q)$.  
}
\be
f_L(q) \;=\; \frac{1}{L\, G} \, \log \left| \lambda_\star(L;q) \right| 
\label{def_f_L}
\ee
where $\lambda_\star(L;q)$ is the dominant eigenvalue of the transfer matrix
$\T(L)$, and $G$ is the geometric factor [c.f., \cite{transfer4,forests}]
\be
G(\text{tri}) \;=\; \frac{\sqrt{3}}{2} 
\label{def_G}
\ee
relating the free energy per unit area and per spin. With this normalization
the FSS of the free energy is given by
\be
f_L(q) \;=\; f_\text{bulk}(q) + \frac{c(q)\pi}{6\, L^2} + \cdots 
\label{def_fss_free}
\ee
where $f_\text{bulk}$ is the bulk free energy {\em per unit area} (which is
expected to be independent of boundary conditions), $c(q)$ is the 
(effective) central charge, and the dots stand for higher-order corrections
depending on the value of $q$ (in particular, CFT predicts a non-universal
$L^{-4}$ correction term). 

In Figure~\ref{figure_free_tri_L=12}, we have plotted the two most dominant 
free energies for each of the three first sectors $\ell=0,1,2$ 
for a triangular-lattice strip of width $L=12$ around 
$q=q_\text{B}$ \reff{def_qB}.\footnote{
 The width $L=12$ is the largest one for which this computation could be  
 carried out numerically. To keep computation times at a reasonable level
 we therefore used a rather coarse mesh, $\Delta q = 0.01$, for scanning
 through the values of $q$. This is the reason why the curves 
 in Figure~\protect\ref{figure_free_tri_L=12} does not look as smooth as they 
 should do around the eigenvalue-crossings within a given level $\ell$. This
 is particularly clear for $\ell=0$. 
}
The other sectors $\ell\geq 3$ are always sub-dominant and play no role in
this discussion. The relevant eigenvalue crossing occurs at 
$q=q_0(L=12)\approx 3.8131463701<q_\text{B}$ where the dominant eigenvalue
changes from the $\ell=2$ sector (solid black line) to the $\ell=0$ sector
(dot-dashed red line). 
There is also a crossing between the two most dominant $\ell=2$ eigenvalues at 
$q=q_2(L=12)\approx 3.82428323>q_\text{B}$. This crossing gives rise to
a new $\ell=2$ phase (between $q_0$ and $q_2$) for all strips of width 
not a multiple of three. In both cases, the $\ell=1$ sector plays no role
in the phase structure around $q=q_\text{B}$. This computation gives 
numerical evidence that our scenario is correct and no new elements are
at play.  

Another interesting question is whether the bulk free energy $f_\text{bulk}(q)$
agrees or not with Baxter's free energies {\em per unit area}
$(1/G)\log |g_i|$ ($i=1,2,3$) \cite{Baxter_86_87}. 
(See Ref.~\cite[Section 6.1]{transfer3} for a detailed description of Baxter's
solution). 
In particular, Baxter found that the free energy is given by $g_3$ in 
the region containing the interval $q\in(0,q_\text{B})$, by $g_2$ in the region 
containing the interval $q\in(q_\text{B},4)$, and by $g_1$ in the rest of 
the complex $q$-plane. 

In Figure~\ref{figure_free_tri_L=12} we observe that the agreement with 
Baxter's solution is very good: for $q\gtapprox q_\text{B}$ the free energy is 
very similar to the $g_2$ contribution (actually it seems to be almost equal to
the contribution of the $\ell=1$ sector), while for $q\ltapprox q_\text{B}$,
it is similar to the contribution of $g_3$. 

A striking difference between our phase-transition diagrams and Baxter's
solution is the transition line going through $q=2$, which is absent in the
latter. This line separates the phase characterized by $\ell=1$ from the one
characterized by
$\ell=2$ (which agrees with Baxter's solution close to $q=q_\text{B}$).  
Therefore a natural question is whether there is one eigenvalue missing in 
Baxter's solution and corresponding to the $\ell=1$ sector. To answer this
question we have plotted in Figure~\ref{figure_ratio_Baxter}(a) the ratio of 
the free energy per unit area $f_L(q)$ and Baxter's free energy 
$(1/G)\log g_3(q)$ in the interval $q\in(0,3.5)$. To avoid parity effects,
we have considered only lattices of widths $L=3p$ with $p=2,3,4,5$, and
for each width we have considered the free-energy contribution from the 
three lowest sectors $\ell=0,1,2$ (represented by solid, dashed, and 
dot-dashed curves, respectively).

Note that the eigenvalue crossings
occur at the points $q=0,2$. Another exact crossing, involving the
second and third most dominant eigenvalues, takes place at $q=1$. The fact
that the loci of these crossings are subject to no finite-size corrections
can presumably be explained using quantum group arguments.

The numerical agreement with $g_3$ is not
very good as the ratio is consistently below the expected result (i.e., $1$).
The reason is that the FSS corrections \reff{def_fss_free}
have not been taken into account. A better agreement can be obtained if we
take into account the first correction term in \reff{def_fss_free}. We 
then define the {\em corrected} free energy per unit area
\be
f_{\text{corr},L}(q) \;=\; f_L(q) - \frac{\pi c_{\rm eff}(q)}{6\, L^2} \;=\; 
                           f_\text{bulk}(q) + o\left(L^{-2}\right) 
\label{def_f_corrected}
\ee
where $c_{\rm eff}(q)$ is the effective central charge for the BK phase
discussed in the next section; see \reff{def_c_effective_BK}.
We expect that this value will be closer to the 
thermodynamic limit $f_\text{bulk}(q)$, as the correction terms are
$o(L^{-2})$. This plot is shown in Figure~\ref{figure_ratio_Baxter}(b). 
Our first observation is that the values of the ratio are much closer to the
expected value of $1$ than in Figure~\ref{figure_ratio_Baxter}(a). 
This shows that our assumption that CFT holds for this model is valid. 
Note that the eigenvalue crossings do not need to be at the 
expected points $q=0,2$, as the correction in \reff{def_f_corrected} 
depends on both $q$ and $L$.
The second observation is that the agreement of the $\ell=2$ free energy  
with the contribution of $g_3$ is very good for $q\gtapprox 1$ 
(less than $0.02\%$ for $L=9,12$). For $q\ltapprox 1$, the best agreement
corresponds to the $\ell=0$ sector. 
However, it is difficult from this data alone to draw any definitive 
conclusion about the necessity of including an extra eigenvalue to the 
set introduced by Baxter.  

%
%
\subsection{Conformal properties}

Following Refs.~\cite{Saleur_90_91,AFpaper} we expect a number of the
conformal properties of the chromatic polynomial to be given by the
BK antiferromagnetic phase. In brief, the situation
is the following.%
\footnote{We refer to Refs.~\cite{Saleur_90_91,AFpaper} for more details.}
The two-dimensional Potts model is exactly solvable (i.e., integrable)
on the curves
\begin{equation}
\begin{array}{ll}
  v = \pm \sqrt{q} & \mbox{(square lattice)} \\[2mm]
  v^3+3v^2 = q     & \mbox{(triangular lattice)} 
\end{array}
\label{critcurves}
\end{equation}

The part of the curves with $0 \le q \le 4$ and $v \ge 0$ corresponds
to the usual second-order transition between a ferromagnetic and a
paramagnetic phase. Its critical properties are given by the Coulomb
gas (free field) construction with coupling constant $g \in [0,1/2]$;
in particular the thermal operator (corresponding to a perturbation of
the critical point in the $v$ variable) is {\em relevant} in the
RG sense. Now, the analytic continuation to
the region with $g \in (1/2,1)$ describes the lower (resp.\ middle)
branch of the curves (\ref{critcurves}) for the square (resp.\
triangular) lattice. Crucially, at these critical points the thermal
operator is {\em irrelevant}. Therefore, for each value of $q \in
(0,4)$, these critical points will act as an RG attractor for a finite
interval $(v_-(q),v_+(q))$: this is the BK phase.  For
the square lattice, the end-points of these intervals are given by the
two mutually dual antiferromagnetic phase transition points \cite{Baxter_82b}
\begin{equation}
 v_\pm(q) = -2 \pm \sqrt{4-q}
 \label{AFpoints}
\end{equation}
whereas for the triangular lattice $v_\pm(q)$ are presently unknown.

The upshot is that the part of the chromatic line $v=-1$ that intersects
the BK phase will have its critical properties determined by the latter.
For the square lattice this happens for $\Re q \in (0,3)$, according to
Eq.~(\ref{AFpoints}), in perfect agreement with Conjecture~\ref{conj.sq.2}.
For the triangular lattice, we obtain a consistent scenario if we suppose that
the intersection with the BK phase is for $\Re q \in (0,q_2)$, with $q_2$
given by Conjecture~\ref{conj.tri.3}.

Within the BK phase, we can parametrize $q$ by
\be
q \;=\; 4 \cos^2 \left( \frac{\pi}{p} \right)
\label{def_p}
\ee
where  $p\in (2,\infty)$. The central charge is then
\be
 c(p) = 1-\frac{6(p-1)^2}{p}.
\label{def_c_BK}
\ee
Furthermore, the theory possesses an infinite set of bulk operators 
$\mathcal{O}_k$, with
$k=1,2,\ldots$, of conformal weight \cite{Dup_Sal_87}
\be
 h_k(p) = \frac{k^2 - (p-1)^2}{4p}.
 \label{watermelons}
\ee
Geometrically, the insertion of an $\mathcal{O}_k$ operator at either 
end of the strip corresponds to a situation in which $2k$ cluster boundaries 
propagate along the length direction of the strip. (The boundaries may of 
course also wind around the periodic transverse direction.) For $k \ge 2$ we 
may equivalently think of $k$ propagating clusters.
It should be carefully noted that the case $k=1$ is special: a single 
propagating cluster may either correspond to two or {\em zero} propagating 
cluster boundaries. The latter case will
occur when the cluster is allowed to have {\em cross-topology} \cite{FSZ_87}, i.e., 
to percolate across both
periodic lattice directions. The corresponding bulk operator is denoted
$\tilde{\mathcal{O}}_1$ and has conformal weight
\be
 \tilde{h}_1(p) = \frac{-3p^2+8p-4}{16p}.
 \label{orderparam}
\ee
Indeed, by duality this is nothing else than the order parameter 
(magnetization) operator.

In the probabilistic regime (i.e., $v>0$) the identity operator 
$\mathcal{I}$ is the most relevant (all other conformal weights are 
strictly positive). Similarly, one has $\tilde{h}_1 < h_1$, 
translating the fact that a single cluster is indeed most likely to have 
cross-topology. In the
BK phase, $v<0$ and some of these statements need no longer be true, due to the
presence of negative conformal weights. In fact, for any $p>2$ the weight  
$h_1$ is the most negative, and so one
would naively expect that the most ``likely'' configurations are those 
with two propagating
cluster boundaries. This is however {\em not true}, due to a subtle effect: the amplitudes of the corresponding transfer matrix eigenvalues vanish identically 
on the torus \cite{AFpaper}!

So the remaining candidates for the most relevant operators are 
$\tilde{\mathcal{O}}_1$ and
$\mathcal{O}_2$.  From Eqs.~(\ref{watermelons})--(\ref{orderparam}) it is 
easy to see that
the former (resp.\ the latter) will dominate for $2<p<4$ (resp.\ 
for $4<p<\infty$). This argument
is of course valid in the thermodynamic limit only, but the conclusion 
will in fact hold true in
finite size due to the underlying quantum group symmetry of the model. 
Accordingly we indeed
observe numerically, for any width $L$ and for both lattices, that branches 
of the limiting curves $\mathcal{B}$ cross the real $q$-axis at $q=0$ ($p=2$) 
and $q=2$ ($p=4$); the former (resp.\ the latter) crossing corresponds 
to the equimodularity of the dominant eigenvalues of the $\ell=0$ and 
$\ell=1$ (resp.\ the $\ell=1$ and $\ell=2$) sectors. However, the phase 
containing $q=2^+$ does not extend
to $q=4$ ($p=\infty$) for either lattice, since the chromatic line leaves 
the BK phase at $q_0=3$ (square lattice), respectively 
$q_2 \simeq 3.8196717312$ (triangular lattice).

We can further check the validity of this scenario by turning to the 
amplitudes of the dominant
eigenvalues. For the order parameter operator $\tilde{\mathcal{O}}_1$ 
the amplitude is $q-1$ (by a simple counting argument),
and this is indeed the amplitude of the dominant eigenvalue observed 
numerically for $p \in (2,4)$ for both lattices. 
(See Conjectures \ref{conj.sq.3}(b) and \ref{conj.tri.4}(b).)
For the two-cluster operator $\mathcal{O}_2$ the 
amplitude should be either
$b^{(2)}_1$ or $b^{(2)}_2$ [cf.~Eqs.~\reff{def_b21}--\reff{def_b22}]. 
And indeed the amplitude
of the eigenvalues which is dominant in the phase containing $q=2^+$ is 
observed numerically to be $b^{(2)}_1$.
(See Conjectures~\ref{conj.sq.3}(c) and \ref{conj.tri.4}(c).)
As a last check, note that from Eqs.~\reff{watermelons}--\reff{orderparam}
the effective central charge should read
\be
c_\text{eff}(q) \;=\; \begin{cases} 
   \displaystyle  \; 1 - \frac{3p}{2} & \qquad \text{for $2< p < 4$} \\[4mm] 
   \displaystyle  \; 1 - \frac{24}{p} & \qquad \text{for $4< p < p_0$} 
                   \end{cases}
\label{def_c_effective_BK}
\ee
We have checked numerically that this is true on both lattices (using strips 
of width $L \le 12$) to a very good precision; this check is essentially
the contents of Figure~\ref{figure_ratio_Baxter}(b).

A final elusive point is the value of $c$ in the interval $q \in (q_2,q_0)$ 
on the triangular lattice.
The $q_0(\text{tri})=4$ case is known to be equivalent to 
Baxter's three-colouring problem on the hexagonal
lattice and hence have $c=2$. Our numerical data seem to indicate that
the rest of the interval $(q_2,q_0)$ may also correspond to
a critical theory, the properties of which are unrelated to those of the
BK phase.

We have made an attempt to obtain an expression for the central charge 
in the interval $q \in (q_2,q_0)$ by performing some fits to our free-energy
data. Our first two-parameter Ansatz was given by Eq.~\reff{def_fss_free} 
without any correction-to-scaling term. In Figure~\ref{figure_c_tri}
we show our estimates as a function of $q$ for different values of 
$L_\text{min}=6,9,12$. Much better estimates can be obtained by including 
higher-order corrections to the above Ansatz. In previous work 
\cite{transfer3,transfer4,forests}, we have found that including a 
correction term $\sim L^{-4}$ greatly improves the estimates for the
central charge. Thus, we have used the improved three-parameter Ansatz
\be
f_L(q) \;=\; f_\text{bulk}(q) + \frac{\pi c(q)}{6\, L^2} + A \, L^{-4}
\label{def_fss_free_improved}
\ee
In Figure~\ref{figure_c_tri} we have also displayed the estimates for $c(q)$
obtained with this Ansatz and $L_\text{min}=6,9$. Their stability   
is superior to that of the two-parameter estimates. 
The solid line represents a guess for the exact
value of the (effective) central charge in this regime  
(see \reff{def_c_diff_conjectured} below)
\be
c(q) \;=\; 2 - \frac{28}{p^2 - 2p - 12}
\label{def_c_conjectured}
\ee
where $p\in(2,\infty)$ is related to $q$ by (\ref{def_p}).

Let us now explain in detail how we have obtained the expression 
\reff{def_c_conjectured}. The functional form of the Ansatz \reff{def_c_conjectured}
is motivated by already known results of the central charge in other 
regimes [c.f., \reff{def_c_BK}]. Thus, our goal is to obtain the central
charge $c(q)$ as a rational function of $p$ with integer coefficients.  
We started with the estimates coming from the three-parameter Ansatz 
\reff{def_fss_free_improved}. It is more useful to plot the results as
a function of $1/p$, so that $q=4$ ($p=\infty$) corresponds to the origin. 
In Figure~\ref{figure_c_tri_raw}(a) we display the difference $c(q=4)-c(q)$ 
versus $1/p$ for the two available data sets $L_\text{min}=6,9$. 
Both data sets fall approximately on a single curve, especially on the
interval corresponding to $q\in(q_2,q_0)$. This empirical 
result implies that the largest FSS correction to our 
estimates for $c(q)$ corresponds to a global shift.  
We then fit the data for $c(q=4)-c(q)$ to a polynomial Ansatz. Our best
result is given by
\be
c(q=4)-c(q) \;=\; 28.60506 \frac{1}{p^2} + 32.11981 \frac{1}{p^3} +
                  897.88358 \frac{1}{p^4}
\label{def_fit_poly_delta_c}
\ee
This Ansatz is motivated by several facts: by definition, the curve 
should go through the origin, and it is clear in 
Figure~\ref{figure_c_tri_raw}(a) that its slope at $1/p=0$ is zero. 
Furthermore, the addition of more terms
to \reff{def_fit_poly_delta_c} only leads to less stable estimates. 
As our goal is to obtain a rational function of $p$, we computed the 
Pad\'e approximant $[2,2]$ of the above polynomial \reff{def_fit_poly_delta_c}.
The result is given by 
\be
c(q=4)-c(q) \;=\; \frac{28.60506}{p^2 - 1.12287 p -30.12813} 
\label{def_pade_delta_c}
\ee
Finally, as we expect that the coefficients of the above rational function 
are integers, we systematically searched the set of closest integers which
best approximates the data of Figure~\ref{figure_c_tri_raw}(a). Our best 
candidates were 
\begin{subeqnarray}
c(q=4) - c(q) &=& \frac{28}{p^2 - 2p - 12} \slabel{def_c_diff_conjectured1}\\
c(q=4) - c(q) &=& \frac{29}{p^2 -  p - 20} \slabel{def_c_diff_conjectured2} 
\label{def_c_diff_conjectured} 
\end{subeqnarray}
In Figure~\ref{figure_c_tri_raw}(a) it is clear that both expressions give 
an excellent fit to the whole set of data points. 
In order to tell which integer-coefficient rational approximant was best,
we plotted the difference between our estimates and the fits. In 
Figure~\ref{figure_c_tri_raw}(b) we show this difference for the above 
approximants \reff{def_c_diff_conjectured}. We observe that the approximant
\reff{def_c_diff_conjectured1} is slightly superior to the other one
\reff{def_c_diff_conjectured2}. 

The final step is to fix the overall shift, i.e., the value of $c(q=4)$. 
As the triangular-lattice Potts antiferromagnet at zero temperature is critical
and can be mapped onto a Gaussian model with two bosons 
\cite{Baxter_70,Henley},  we expect that 
\be
c(q=4) \;=\; 2 
\ee
Thus, we arrive at the final expression \reff{def_c_conjectured}.
Indeed, this computation should be taken with a grain of salt: we have 
found a function with the expected properties (rational function of $p$ with
integer coefficients) that describes rather accurately our estimates 
for the central charge of our model on the interval $q\in(q_2,q_0)$. This
does not mean that this is unique possible solution with those properties 
nor that the right expression for $c(q)$ should satisfy those properties.%
\footnote{Accordingly, we have referred to \reff{def_c_conjectured} as
a guess rather than a conjecture.}
Another important point is that our conjectured expression for the 
central charge, although it was first devised to deal with the 
interval $q\in(q_2,q_0)$, is able to reproduce rather accurately the data
outside the latter interval. 

%
%
\subsection{The ``$\bm{q=2}$ branch''} \label{sec.q=2} 

At the end of Section~\ref{sec.tri.free}, we have already discussed the branch 
of the triangular-lattice limiting curve $\mathcal{B}_L$ going 
through the point $q=2$, and separating the $\ell=1$ and $\ell=2$ phases. 
In this subsection we will study further properties of this branch in both the
square and triangular limiting curves. 

The first question we want to address is that of the density of chromatic zeros
along this branch for finite strips of size $L\times N$. In order to estimate 
this density, we first fixed a region in the complex $q$-plane containing the
$q=2$ branch. This region is 
\begin{subeqnarray}
 \text{Sq} &\Rightarrow& 1.8 \;\leq\; \Re q \;\leq\; 2.2 \,, 
                 \qquad |\Im q| \;\leq\; 1.2 \\ 
\text{Tri} &\Rightarrow& 1.8 \;\leq\; \Re q \;\leq\; 2.6 \,, 
                 \qquad |\Im q| \;\leq\; 1.4 
\label{def_region_density}
\end{subeqnarray}
In Figure~\ref{figure_density}, we show the density of the chromatic zeros
$\rho$ in the regions \reff{def_region_density} for some finite strips of size
$L\times N$ as a function of the aspect ratio $L/N$. At fixed aspect ratio, 
we expect the density $\rho$ to converge to some (small) positive constant.
In Figure~\ref{figure_density}, we observe that for fixed $L/N$, the
density is decreasing in $L$, and it attains at $L=7$ very small values 
$\ltapprox 0.02$. If this monotonicity holds for larger values of $L$,
we expect that the density of zeros along the $q=2$ branch will converge to
zero in the thermodynamic limit (in contrast with the behavior of the
density of zeros in other parts of the limiting curve).

Another way to tackle this problem is consider the parameter $t$ introduced
in Ref.~\cite[Section~4.1.1]{transfer1}. On the limiting curve $\mathcal{B}_L$,
there are generically two equimodular leading eigenvalues $\lambda_1$ and
$\lambda_2$. Indeed, their ratio is just a phase 
\be
\frac{\lambda_1}{\lambda_2} \;=\; e^{i\, \theta} \,, \qquad \theta\in\R 
\label{def_theta}
\ee
In Ref.~\cite{transfer1}, instead of $\theta$, we found it more convenient to
work with the real parameter $t$, defined as 
\be
t \;=\; \tan\left(\frac{\theta}{2}\right) \,,\qquad t\in[0,\infty)
\label{def_t}
\ee
In practice, we use this parameter to ensure the plots of the limiting curves 
$\mathcal{B}_L$ are correct and we do not miss any small feature.  
Note that $t=\theta=0$ corresponds to two {\em identical} eigenvalues. At 
$q=2$, we have that $t=0$ for any $L\geq 3$; 
but as we move along the $q=2$ branch, $t$ increases. 
Let us now study how the parameter $t$ evolves (along this branch)
as a function of $|q-2|$. In Figure~\ref{figure_theta}, we show 
(for both the square and triangular lattices) the values
of $t$ along the $q=2$ branch for points in the complex $q$-plane 
with $\Im q =0,0.01,\ldots,0.10$. It is clear that for fixed $L$, the
relation is approximately linear. For the square lattice, the slope 
is given approximately by $1/L$, while for the triangular lattice
the slope is given approximately by $\sqrt{3}/(2L) + 0.653/L^3$. In both cases,
the discrepancies between the estimated slopes and the above results are
of order $O(10^{-4})$. Thus, we expect that in the thermodynamic limit both
leading eigenvalues $\lambda_1$ and $\lambda_2$ become equal along this branch.

Finally, we can study the behavior at $q=2$ by considering the series
expansion of the two equimodular leading eigenvalues $\lambda_1$ and
$\lambda_2$ (corresponding to the $\ell=1$ and $\ell=2$ sectors, respectively).
Indeed, such series expansions are expected to exist, as the eigenvalues 
$\lambda_i(q)$ are algebraic functions of $q$. On each phase, we can compute
the series expansion of the leading eigenvalue at $q=2+\epsilon$  
\be
\lambda_i(q+\epsilon) \;=\; \sum\limits_{k=0}^\infty \lambda_{i,k} \, 
    \epsilon^k 
\label{def_series_lambda}
\ee
For $L=2,3$, this can be done symbolically by using {\tt mathematica}. For
$4\leq L \leq 7$, we can only obtain (again with {\tt mathematica}) 
the series numerically, as the symbolic expressions for the 
coefficients are too lengthy. In practice, we are able to obtain the first 
three coefficients $\lambda_{i,k}$ ($k=0,1,2$) in 
\reff{def_series_lambda}.\footnote{
  For the triangular lattice, this computation could only be performed up to 
  width $L=6$. 
}
With these coefficients, we can obtain the corresponding series expansion 
for the free energy: 
\be
f_L(2+\epsilon) \;=\; \begin{cases}
  \frac{1}{L}\log |\lambda_1(2+\epsilon)| \;=\; 
  \sum\limits_{k=0}^\infty f_{L,k}^{(1)} \, \epsilon^k & 
  \quad \text{if $\epsilon \leq 0$} 
  \\[4mm] 
  \frac{1}{L}\log |\lambda_2(2+\epsilon)| \;=\; 
  \sum\limits_{k=0}^\infty f_{L,k}^{(2)} \, \epsilon^k & 
  \quad \text{if $\epsilon \geq 0$} 
  \end{cases}
\label{def_series_f}
\ee
For $4\leq L \leq 12$, we numerically obtained the contribution to the 
free energy $f_{L,i}(q)=(1/L)\log|\lambda_i(q)|$ of the two leading 
eigenvalues $\lambda_i$ close to $q=2$. In particular, we obtained the 
values $f_{L,i}(q=2)$ and $f_{L,i}(q=2\pm 10^{-5})$ for $i=1,2$. From these 
six values, we estimated the first three coefficients $f^{(i)}_{L,k}$ 
(with $k=0,1,2$) in \reff{def_series_f}.   
Indeed, we know that the free energy is continuous at $q=2$, thus we
expect that $\Delta f_L(2) = f_{L,0}^{(2)}-f_{L,0}^{(1)}=0$, as we find in 
the actual computation. 
We are interested in computing the increment of the first 
derivatives of the free energy $f_L(q)$ at $q=2$. In practice, we have
computed the increments of the first two derivatives:\footnote{
  For $4\le L\ltapprox 7$, we had two distinct ways of estimating the 
  coefficients $E_L$ and $C_L$, namely symbolically and numerically. 
  We checked that the difference between those estimates  
  was $\ltapprox 10^{-11}$ for $E_L$, and $\ltapprox 10^{-6}$ for $C_L$.
}
\begin{subeqnarray}
E_L &=& \Delta \left(\left. \frac{d f_L}{dq} \right|_{q=2}\right)    \;=\; 
    f_{L,1}^{(2)} - f_{L,1}^{(1)} \\
C_L &=& \Delta \left(\left. \frac{d^2 f_L}{dq^2} \right|_{q=2}\right)\;=\;
  2(f_{L,2}^{(2)} -  f_{L,2}^{(1)})
\label{def_increments_f_primes}
\end{subeqnarray}
If $E_L\neq 0$, then the transition at $q=2$ is of first order. However, 
if $E_L =0$ and $C_L\neq 0$, then the transition is of second order (and this
argument can be generalized to higher-order phase transitions). 
In Figure~\ref{figure_series} we have plotted the
coefficients $E_L$ and $C_L$ versus $1/L$ for both the square and triangular
lattices. The plots suggest that both coefficients for both lattices
tend to a limit as $L\to\infty$. 
If we fit the $E_L$--data to a power law $E_L = E_\infty + A_E L^{-\omega_E}$, 
we find that $|E_\infty| \ltapprox 10^{-4}$ is indeed very small 
(in modulus) for $L_\text{min}=10$:
\begin{subeqnarray}
\text{Sq}  &\Rightarrow& 
E_L\;\approx\;9.8\times 10^{-7} - 1.98329 \, L^{-1.99737}\\ 
\text{Tri} &\Rightarrow& 
E_L\;\approx\;-0.00010        \;- 2.12401 \, L^{-2.02506}
\label{def_fit_series_E}
\end{subeqnarray}
In particular, our square-lattice data can be fitted very well with the 
simple Ansatz: $E_L(\text{sq})=-2 L^{-2}$. These fits are depicted in the 
corresponding plots in Figure~\ref{figure_series}. We have chosen 
$L_\text{min}$ to be the largest possible value, so that there are no 
effective degrees of freedom in the fits. 

The power-law fits $C_L = C_\infty + A_C L^{-\omega_C}$ are not as stable 
as the preceding ones: we find sizeable parity effects in the estimates, so 
we have to split the whole data set into two smaller sets with 
well-defined $L$--parity. Thus, for each value of $L_\text{min}$, we have 
fitted the data with $L=L_\text{min},L_\text{min}+2,L_\text{min}+4$.
The estimates for even $L$ are given for $L_\text{min}=8$ by: 
\begin{subeqnarray}
\text{Sq}  &\Rightarrow&
C_L  \;\approx\; \phantom{-}0.00051 - 0.35387 \, L^{-1.73364} \\
\text{Tri}  &\Rightarrow&
C_L  \;\approx\;           -0.00012 - 0.99690 \, L^{-2.18046} 
\label{def_fit_series_Ceven}
\end{subeqnarray}
For odd $L$ with $L_\text{min}=7$, the corresponding estimates are: 
\begin{subeqnarray}
\text{Sq}  &\Rightarrow&
C_L  \;\approx\; \phantom{-} 0.00116 - 0.30522 \, L^{-1.62941}\\
\text{Tri}  &\Rightarrow&
C_L  \;\approx\;            -0.00092 - 1.42076 \, L^{-2.38725} 
\label{def_fit_series_Codd}
\end{subeqnarray}
Our preferred fits are those given by \reff{def_fit_series_Ceven}, as
$L_\text{min}$ is larger; these fits are depicted in the corresponding plots in
Figure~\ref{figure_series}. They also show show that $C_L$ tend to some 
small (in modulus) number in the thermodynamic limit 
$|C_\infty|\ltapprox 5\times 10^{-4}$. We conjecture that for both lattices, 
$E_L,C_L\to 0$ as $L\to \infty$. Indeed, this does not mean that there is 
no transition at all at $q=2$ in the thermodynamic limit, but that the 
transition (if any) is at least of third order.

In this section we have presented three numerical approaches to study the
thermodynamic limit along the $q=2$ branch. None of them is conclusive; but 
taken together they give strong support to the following scenario:  
even though at finite $L$ there are two different phases coexisting along the
$q=2$ branch, in the thermodynamic limit there is a {\em unique} phase. 
This scenario is consistent with the three results discussed above:
in the thermodynamic limit $L\to\infty$: 
1) The density of chromatic zeros goes to zero along this branch; 
2) The ratio of eigenvalues along this branch tend to one
$\lambda_2/\lambda_1 \to 1$; and 
3) At $q=2$ the increment in the first two derivatives of the free energy 
$f_L(q)$ (with respect to $q$) tends to zero. 
Thus, this branch seems to be somehow an artifact of the boundary conditions;
in particular, of the periodic longitudinal boundary conditions, as this
branch also appears in the cyclic case \cite{transfer4}, and it is absent
for free and cylindrical boundary conditions 
\cite{transfer1,transfer2,transfer3}.\footnote{
  A similar (but not related) phenomenon appears in a recent study of the 
  RSOS \protect\cite{RSOS} representation of the Potts model when $q=B_p$ 
  with $p=B_p$ \protect\reff{def_Bp}. In the complex $v$-plane, the large-$|v|$
  region (which corresponds to the high-temperature regime of the model and 
  is expected to be non-critical) displays several non-bounded outward branches
  separating different phases. The origin of these branches is simple:
  there are two almost-degenerate eigenvalues, each of them corresponding
  to a different topological sector. These two eigenvalues become exactly 
  equimodular along those non-bounded outward branches, whose number 
  increases with $L$. While for any finite $L$ there are several 
  phase-transition lines in the large-$v$ region, in the thermodynamic limit
  we expect that both eigenvalues become exactly identical over all this
  region. Thus, we recover the standard result that there are no 
  phase-transition lines in the large-$v$ (high-temperature) regime of this 
  model.
} 
This conclusion agrees with Baxter's
solution \cite{Baxter_86_87} in the sense that only three phases survive in
the phase diagram of the triangular-lattice chromatic polynomial. 
 
%
%
\section{Conclusions} \label{sec.conclusions}

In this paper we have extended our transfer-matrix studies  
\cite{transfer1,transfer2,transfer3,transfer4} to toroidal boundary 
conditions. We have obtained the exact expression of the chromatic
polynomial for square- and triangular-lattice strips of fixed width 
$2\leq L \leq 7$ and arbitrary length $N$. The cases $L=5,6,7$ are
new in the literature (note that going to larger widths $L\geq 8$ would
require better computer facilities and/or improved symbolic algorithms).
We have also obtained the limiting curves $\mathcal{B}_L$ and isolated
limiting points for these strips in the infinite-length limit $N\to\infty$.
The limiting curve  $\mathcal{B}_L$ for a fixed value of $L$ 
provides the exact phase diagram for this semi-infinite strip.

By studying the features of these finite-$L$ phase diagrams and using
techniques from CFT and FSS, we have
been able to determine some properties of the chromatic-polynomial phase
diagram in the thermodynamic limit $L\to\infty$. These findings are 
summarized in the following points:

\begin{itemize}

\item Toroidal boundary conditions appear to give results qualitatively and
      quantitatively closer to those of the thermodynamic limit than
      the other standard boundary conditions (free, cylindrical, and 
      cyclic) \cite{transfer1,transfer2,transfer3,transfer4}.
      In particular, the exact value of $q_c$ is obtained for all 
      square lattices of (finite) odd widths, and for all triangular 
      lattices of (finite) widths not a multiple of three.  
     
\item As for cyclic boundary conditions \cite{transfer4}, we find that 
      the limiting curves $\mathcal{B}_L$ always enclose regions of the 
      complex $q$-plane, thus determining true phases as in the expected 
      infinite-volume phase diagram. Each phase is 
      also characterized by the number of bridges $\ell$ associated to the 
      leading eigenvalue; but for toroidal boundary conditions only 
      the $\ell=0,1,2$ sectors contribute for both lattices. (For
      the triangular lattice with cyclic boundary conditions we expect
      to have contributions up to $\ell=6$ bridges). 
      The other difference with the cyclic triangular strip is that 
      there are now two different phases characterized by $\ell=2$ bridges
      (For cyclic boundary conditions, there is at most one phase for each
      value of $\ell$). 

\item It is rather remarkable that for cyclic boundary conditions
      \cite{Saleur_90_91,transfer4}, all the Beraha numbers $B_p$
      \reff{def_Bp} played a special role in the BK phase, either as
      phase transition points (even $p$) or as isolated limiting
      points (odd $p$).  In contrast, for toroidal boundary conditions
      only the finite subset of integer $q$ play such special roles:
      there are phase transitions at $q=0,2,4$ and isolated limiting
      points at $q=1,3$. This observation alone is a nice illustration
      of the crucial influence of boundary conditions on
      antiferromagnetic systems. Note however, that we still expect
      (and have to some extent numerically observed) that the full set
      of $B_p$ will be the loci of level crossings of {\em
      sub-dominant} eigenvalues.

\item Seeing that the maximum chromatic number $H(g)$ of a graph $G$
      increases with the genus $g$ of the surface that it is embedded
      in according to \reff{Heawood}, one might have expected that
      the toroidal topology studied here would have revealed further
      structure of the limiting curves for $q \in (4,7)$. This is not
      the case. The reason is that it is precisely the interplay between
      the properties of the BK phase and the representation theory of
      the quantum group $U_t(SU(2))$ for $t$ a root of unity that
      is responsible for generating the partition-function zeros. This
      mechanism is confined to the interval $q \in [0,4]$. The graphs
      having chromatic zeros for $q \in (4,7)$ must therefore be rather
      exceptional and have little to do with the recursive families of
      graphs studied here. One might even speculate that the number of
      such exceptional graphs might be small in some sense, or even
      finite. These conclusions are expected to carry over to higher
      genus as well.

\item For toroidal triangular strips we have found two phase transitions
      inside the critical region: $q=2$ and $q=q_\text{B}$ \reff{def_qB}.

      This result is rather puzzling when compared to Baxter's 
      exact solution \cite{Baxter_86_87} and the re-analysis discussed in  
      Ref.~\cite{transfer4} of Baxter's eigenvalues. 
      On one hand, the value of $q_\text{B}$ coincides with that 
      predicted by Baxter based on the result that only three eigenvalues
      are relevant in the complex $q$-plane. This conclusion stems from
      the Bethe-Ansatz solution of a closely related model \cite{Baxter_86_87}.
      The mapping between these two models is exact, except for 
      the boundary conditions. Usually boundary conditions have no effect
      in the ferromagnetic regime; but here we are deep in the 
      antiferromagnetic one.

      On the other hand, a careful and high-precision analysis of these
      three eigenvalues \cite{transfer4} revealed that, although the
      outer boundary of Baxter's infinite-volume curve $\mathcal{B}_\infty$
      was indeed correct, the
      inner structure was richer: there were infinitely many more
      branches than those included by Baxter and those indicated by the 
      finite-strip chromatic zeros. In particular, the point $q_\text{B}$
      \reff{def_qB} does not belong to the limiting curve 
      $\mathcal{B}_\infty$, and its role is now played by $q_\text{G}$ 
      \reff{def_qG}. 

\item In Section~\ref{sec.q=2} we have found some numerical evidence
      that the branch of the limiting curves $\mathcal{B}_L$ going 
      through $q=2$ might disappear in the true thermodynamic limit
      $L\to\infty$. This may provide an explanation of why this branch is
      absent from the triangular-lattice 
      limiting curve $\mathcal{B}_\infty$ proposed by Baxter
      \cite{Baxter_86_87}. However, it is still not clear how to explain the 
      existence of the $q=2$ branch for finite widths from Baxter's analysis. 
      The resolution of these discrepancies would  
      most likely involve identifying (at least) one more analytic
      expression for the free energy from the Bethe Ansatz equations
      of Ref.~\cite{Baxter_86_87}. Such a study might also shed some
      light on the possible existence of additional enclosed regions
      in $\mathcal{B}_\infty$, such as the triangular-shaped regions
      discussed towards the end of
      Section~\ref{sec.tri-lim-curves}.
      
\item In the triangular-lattice case we obtain a novel behavior: one  
      important feature of the infinite-volume phase diagram {\em does} 
      depend on the parity of the width (i.e., on $L \bmod 3$). 
      Our numerics strongly suggests that
\be
\lim_{L\to\infty} q_0(L) \;=\; \begin{cases}
            q_c(\text{tri}) \;=\; 4 & \qquad \text{if $L \bmod 3\neq0$}\\ 
            q_B \;\approx\; 3.8196717312 & \qquad \text{if $L \bmod 3= 0$}
            \end{cases}
\ee
      In the square-lattice case, we have that this limit for even widths
      is equal to $q_c(\text{sq})=q_0(L=2p+1)$ for all $p\geq 1$.

\item In the light of the numerical results given at the end of
      Section~\ref{sec.discussion}, we might finally speculate that the
      full solution of the triangular-lattice chromatic polynomial (in
      the sense of a Bethe Ansatz solution) should reveal not only two
      distinct bulk free energies for $q \in [0,4]$, but also two
      different sets of associated critical scaling levels throughout that
      interval. In other words, we suggest that there exists two
      ``superposed'' critical theories throughout the interval, one
      having the physics of the BK phase, and the other being some
      deformation of the two-boson theory \cite{Baxter_70,Henley}
      responsible for $c=2$ at $q=4$. This is a strong motivation for
      a more refined study of the Bethe equations corresponding to
      Baxter's $g_2$ solution \cite{Baxter_86_87}. Note that these
      expectations apply also to the Nienhuis loop model \cite{Nienhuis_82}.

\end{itemize} 

%
%
\appendix
\section{Two combinatorial identities} \label{sec.combin}

In this section we prove the following combinatorial identities
that were used in Section~\ref{sec.structure}.

\begin{lemma} \label{lemma_comb}
For all integer $m\geq 1$, the following identities hold
\begin{subeqnarray}
m \, \binom{2m-1}{m}^2 &=& \sum\limits_{p=0}^m \, p\, \binom{2m}{m-p}^2 
\slabel{lemma1} \\
m \, \binom{2m}{m}^2 &=& \sum\limits_{p=0}^m \, p\, \binom{2m+2}{m+1-p}\, 
\binom{2m}{m-p} 
\slabel{lemma2}
\label{lemmas12}
\end{subeqnarray}
\end{lemma}

\newpage

\proof

Let us start with the proof of \reff{lemma1}. The first goal is to make
explicit the $m$ factor in the r.h.s.\ of \reff{lemma1}
\begin{eqnarray}
\sum\limits_{p=0}^m \, p\, \binom{2m}{m-p}^2 
 &=& \sum\limits_{p=0}^m (m-p)\, \binom{2m}{p}^2 \nonumber \\[2mm]
 &=& m \, \left[ \sum\limits_{p=0}^m \binom{2m}{p}^2 - 
             2\sum\limits_{p=0}^m \binom{2m}{p} \binom{2m-1}{p-1} \right]
     \nonumber  \\[2mm]
 &=& m \,    \sum\limits_{p=0}^m \binom{2m}{p} \left[    
             \binom{2m-1}{p} - \binom{2m-1}{p-1} \right] 
\label{app.eq.1}
\end{eqnarray}
where we have first changed the summation index $p\to m-p$, then used 
the absorption identity \cite[Eq.~(5.5)]{Knuth} 
\be
\binom{a}{b} \;=\; \frac{a}{b} \, \binom{a-1}{b-1} \,, \qquad 
\text{integer $b\neq 0$} 
\label{def_absorption}
\ee  
and finally used the addition formula \cite[Eq.~(5.8)]{Knuth} 
\be
\binom{a}{b} \;=\; \binom{a-1}{b} + \binom{a-1}{b-1} \,, \qquad b\in\Z 
\label{def_addition}
\ee 
If we use \reff{def_addition} to expand $\binom{2m}{p}$ in 
\reff{app.eq.1}, we arrive at the final result:
\begin{eqnarray}
\sum\limits_{p=0}^m \, p\, \binom{2m}{m-p}^2 
 &=& m \, \sum\limits_{p=0}^m \left[ \binom{2m-1}{p-1} + \binom{2m-1}{p} \right]
              \left[ \binom{2m-1}{p} - \binom{2m-1}{p-1} \right] 
\nonumber \\[2mm]
 &=& m \, \left[ \sum\limits_{p=0}^m \binom{2m-1}{p}^2 - 
       \sum\limits_{p=0}^m \binom{2m-1}{p-1}^2 \right] \nonumber \\[2mm]
 &=& m \, \binom{2m-1}{m}^2 
\end{eqnarray} 

Let us now prove Eq.~\reff{lemma2}. The idea is similar to the above proof;
but now it is more efficient to extract a common factor $(m+1)$ rather 
than $m$. Thus, if we follow the same steps as above, we get 
\begin{eqnarray}
\sum\limits_{p=0}^m \, p\, \binom{2m+2}{m+p+1}\binom{2m}{m-p} 
 &=& \sum\limits_{p=0}^m (m-p)\, \binom{2m+2}{p+1}\binom{2m}{p} 
  \nonumber \\[2mm]
 &=& (m+1) \, \left[ \sum\limits_{p=0}^m \binom{2m+2}{p+1}\binom{2m}{p} - 
             2\sum\limits_{p=0}^m \binom{2m}{p} \binom{2m+1}{p} \right]
     \nonumber  \\[2mm]
 &=& (m+1) \, \sum\limits_{p=0}^m \binom{2m}{p} \left[    
             \binom{2m+1}{p+1} - \binom{2m+1}{p} \right] 
\label{app.eq.2}
\end{eqnarray}
We now use \reff{def_addition} to expand the two binomials inside
the brackets in \reff{app.eq.2}, and we get the desired result: 
\begin{eqnarray}
\sum\limits_{p=0}^m \, p\, \binom{2m+2}{m+p+1}\binom{2m}{m-p} 
 &=& (m+1) \, \left[ \sum\limits_{p=0}^m \binom{2m}{p} \binom{2m}{p+1}
           - \sum\limits_{p=0}^m \binom{2m}{p} \binom{2m}{p-1} \right] 
   \nonumber \\[2mm]
  &=& (m+1) \, \binom{2m}{m}\binom{2m}{m+1} \nonumber \\[2mm]
  &=& m \, \binom{2m}{m}^2  
\end{eqnarray} 
where the last equality follows from the fact that $m$ is a positive integer.
\qed

%
%
\section*{Acknowledgments}

\begin{figure}
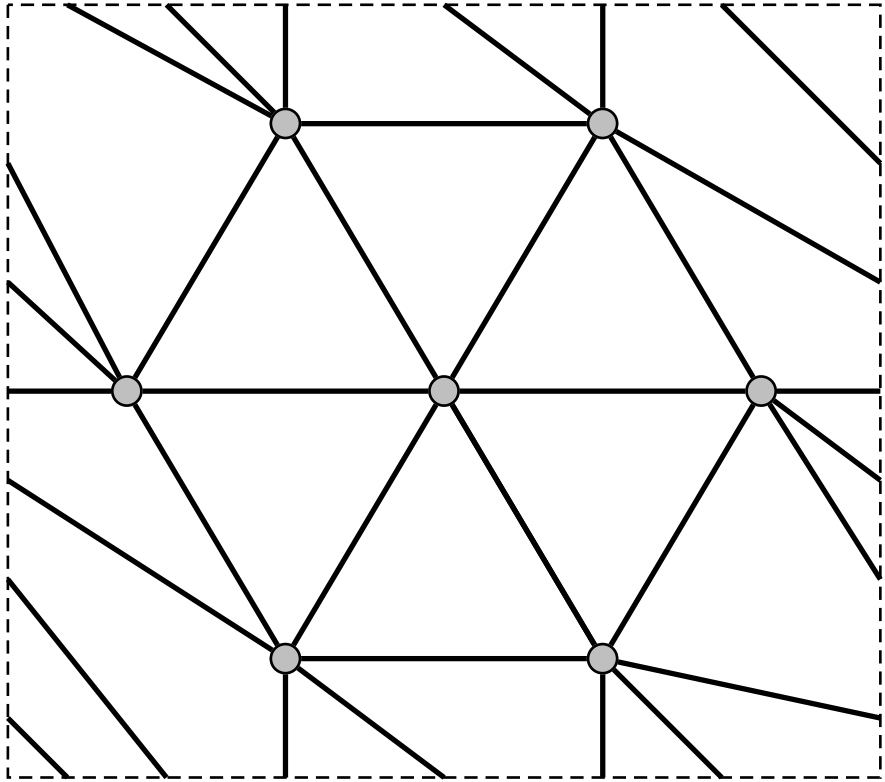

\centering
\psset{xunit=0.05pt}
\psset{yunit=0.05pt}
\pspicture(0,-10)(6708,5973)
\psline[linewidth=2pt](3354,2979)(4554,954)
\psline[linewidth=2pt](3354,2979)(4554,954)
\psline[linewidth=2pt](2154,5004)(4554,5004)
\psline[linewidth=2pt](4554,5004)(5754,2979)
\psline[linewidth=2pt](5754,2979)(4554,954)
\psline[linewidth=2pt](4554,954)(2154,954)
\psline[linewidth=2pt](2154,954)(954,2979)
\psline[linewidth=2pt](954,2979)(2154,5004)
\psline[linewidth=2pt](2154,5004)(3354,2979)
\psline[linewidth=2pt](3354,2979)(954,2979)
\psline[linewidth=2pt](4554,5004)(3354,2979)
\psline[linewidth=2pt](3354,2979)(5754,2979)
\psline[linewidth=2pt](3354,2979)(2154,954)
\psline[linewidth=2pt](2154,954)(2154,54)
\psline[linewidth=2pt](2154,5004)(2154,5904)
\psline[linewidth=2pt](4554,5904)(4554,5004)
\psline[linewidth=2pt](4554,954)(4554,54)
\psline[linewidth=2pt](4554,954)(5454,54)
\psline[linewidth=2pt](5454,5904)(6654,4704)
\psline[linewidth=2pt](54,4704)(954,2979)
\psline[linewidth=2pt](5754,2979)(6654,2979)
\psline[linewidth=2pt](54,2979)(954,2979)
\psline[linewidth=2pt](54,3804)(954,2979)
\psline[linewidth=2pt](6654,3804)(4554,5004)
\psline[linewidth=2pt](2154,954)(54,2304)
\psline[linewidth=2pt](6654,2304)(5754,2979)
\psline[linewidth=2pt](5754,2979)(6654,1554)
\psline[linewidth=2pt](54,1554)(1254,54)
\psline[linewidth=2pt](1254,5904)(2154,5004)
\psline[linewidth=2pt](2154,5004)(504,5904)
\psline[linewidth=2pt](504,54)(54,504)
\psline[linewidth=2pt](6654,504)(4554,954)
\psline[linewidth=2pt](4554,5004)(3354,5904)
\psline[linewidth=2pt](3354,54)(2154,954)
\psline[linewidth=1pt,linestyle=dashed](54,5904)(54,54)(6654,54)%
(6654,5904)(54,5904)
\rput{0}(2154,954) {\pscircle*[linecolor=lightgray](0,0){6pt}}
\rput{0}(4554,954) {\pscircle*[linecolor=lightgray](0,0){6pt}}
\rput{0}(3354,2979){\pscircle*[linecolor=lightgray](0,0){6pt}}
\rput{0}(954,2979) {\pscircle*[linecolor=lightgray](0,0){6pt}}
\rput{0}(5754,2979){\pscircle*[linecolor=lightgray](0,0){6pt}}
\rput{0}(2154,5004){\pscircle*[linecolor=lightgray](0,0){6pt}}
\rput{0}(4554,5004){\pscircle*[linecolor=lightgray](0,0){6pt}}
\rput(2154,954) {\pscircle[linewidth=1pt](0,0){6pt}}
\rput(4554,954) {\pscircle[linewidth=1pt](0,0){6pt}}
\rput(3354,2979){\pscircle[linewidth=1pt](0,0){6pt}}
\rput(954,2979) {\pscircle[linewidth=1pt](0,0){6pt}}
\rput(5754,2979){\pscircle[linewidth=1pt](0,0){6pt}}
\rput(2154,5004){\pscircle[linewidth=1pt](0,0){6pt}}
\rput(4554,5004){\pscircle[linewidth=1pt](0,0){6pt}}
\endpspicture
   \caption{\label{figure_k7}
    The complete graph $K_7$ embedded on a torus. The solid lines represent 
    the edges of the graph. To build the torus, we should glue together
    opposite dashed lines.
    We thank David Wilson for helping us preparing this plot. 
   }
\end{figure}

\begin{figure}
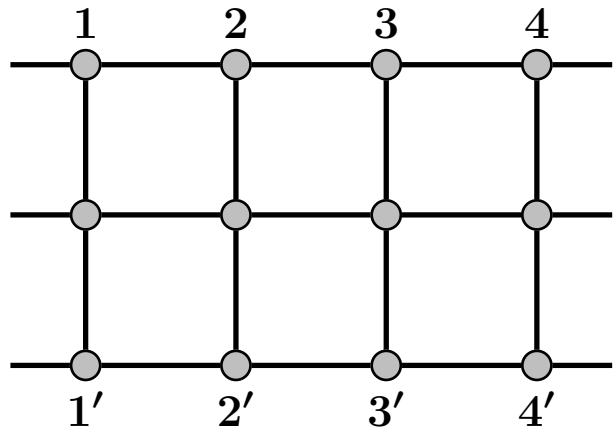

   \centering
   \psset{labelsep=10pt}
   \psset{xunit=2cm}
   \psset{yunit=2cm}
   \pspicture(-0.5,-0.5)(3.5,2,5) 
   \multirput(0,0)(1,0){4}{%
     \psline[linecolor=black,linewidth=2.pt](0,0)(0,2)
   }
   \multirput(0,0)(0,1){3}{%
     \psline[linecolor=black,linewidth=2.pt](-0.5,0)(3.5,0)
     \multirput(0,0)(1,0){4}{\pscircle*[linecolor=lightgray](0,0){6pt}}
     \multirput(0,0)(1,0){4}{\pscircle[linewidth=1pt](0,0){6pt}}
   }
   \uput[270](0,0){\Large $\bm{1'}$}
   \uput[270](1,0){\Large $\bm{2'}$}
   \uput[270](2,0){\Large $\bm{3'}$}
   \uput[270](3,0){\Large $\bm{4'}$}

   \uput[90](0,2){\Large $\bm{1}$}
   \uput[90](1,2){\Large $\bm{2}$}
   \uput[90](2,2){\Large $\bm{3}$}
   \uput[90](3,2){\Large $\bm{4}$}
   \endpspicture

   \caption{\label{figure_sq_cyclic_bc}
     Square lattice with toroidal boundary conditions of size $4\times 2$. 
     This lattice is obtained
     from a square lattice with cyclic boundary conditions of size $4\times 3$
     by identifying the top and bottom rows 
     $i \leftrightarrow i'$ ($i=1,\ldots,4=m$). 
   }
\end{figure}

\begin{figure}
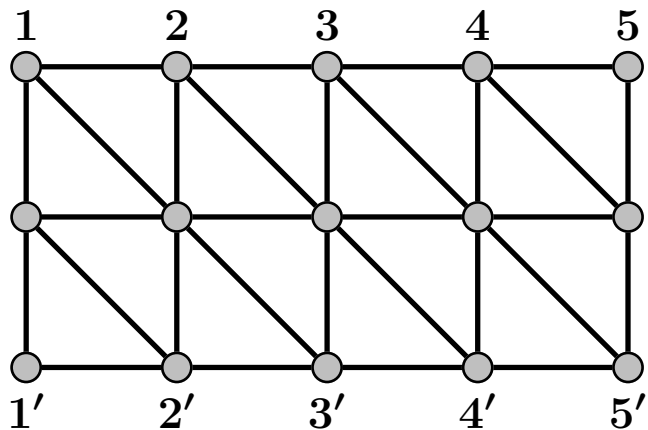

   \centering
   \psset{labelsep=10pt}
   \psset{xunit=2cm}
   \psset{yunit=2cm}
   \pspicture(-0.5,-0.5)(4.5,2,5)
   \multirput(0,0)(1,0){5}{%
     \psline[linecolor=black,linewidth=2.pt](0,0)(0,2)
   }
   \psline[linecolor=black,linewidth=2.pt](0,1)(1,0)
   \psline[linecolor=black,linewidth=2.pt](0,2)(2,0)
   \psline[linecolor=black,linewidth=2.pt](1,2)(3,0)
   \psline[linecolor=black,linewidth=2.pt](2,2)(4,0)
   \psline[linecolor=black,linewidth=2.pt](3,2)(4,1)
   \multirput(0,0)(0,1){3}{%
     \psline[linecolor=black,linewidth=2.pt](0,0)(4,0)
     \multirput(0,0)(1,0){5}{\pscircle*[linecolor=lightgray](0,0){6pt}}
     \multirput(0,0)(1,0){5}{\pscircle[linewidth=1pt](0,0){6pt}}
   }
   \uput[270](0,0){\Large $\bm{1'}$}
   \uput[270](1,0){\Large $\bm{2'}$}
   \uput[270](2,0){\Large $\bm{3'}$}
   \uput[270](3,0){\Large $\bm{4'}$}
   \uput[270](4,0){\Large $\bm{5'}$}

   \uput[90](0,2){\Large $\bm{1}$}
   \uput[90](1,2){\Large $\bm{2}$}
   \uput[90](2,2){\Large $\bm{3}$}
   \uput[90](3,2){\Large $\bm{4}$}
   \uput[90](4,2){\Large $\bm{5}$}
   \endpspicture

   \caption{\label{figure_tri_cyclic_bc}
     Triangular lattice with toroidal boundary conditions of size $4\times 2$.
     This lattice is obtained
     from a triangular lattice with {\em free} boundary conditions of 
     size $5\times 3$ by first identifying the first and last columns
     (i.e., $1\leftrightarrow 5$ and $1'\leftrightarrow 5'$), so we get
     a triangular lattice of size $4\times 3$ with cylindrical boundary 
     conditions. Then, as for the square lattice,  
     we identify the top and bottom rows
     $i \leftrightarrow i'$ ($i=1,\ldots,4=m$).
   }
\end{figure}

We wish to thank J.-F.~Richard, H.~Saleur and A.~D.~Sokal for interesting 
discussions and for collaborations on related projects. We also thank 
D. Wilson for helping us in preparing Figure~\ref{figure_k7}. 
This research has been partially supported by U.S. National Science 
Foundation grants PHY-0116590 and PHY-0424082, and by MEC (Spain)
grants MTM2005-08618-C02-01 and FIS2004-03767.

\clearpage
%
%

\clearpage
%
%
%
%
\begin{table}
\centering 
\begin{tabular}{|r|r|r|r|r|r|r|r|}
\hline\hline
$m$&$C_{m,m}$&$C_{m+1,m}$&$D_{m,m}$& 
                           TriTorus$'$ & SqTorus$'$& 
                           TriTorus & SqTorus\\
\hline\hline
 1 &        2 &        5 &     2 &     2 &      2 &     2 &   2 \\
 2 &       15 &       52 &     7 &     5 &      4 &     4 &   4 \\
 3 &      200 &      780 &    31 &    13 &      8 &    11 &   8 \\
 4 &     3185 &    12838 &   361 &    99 &     68 &    38 &  33 \\
 5 &    52920 &   214956 &  3241 &   653 &    347 &   148 &  90 \\
 6 &   884268 &  3593700 & 30916 &  5194 &   2789 &   476 & 325 \\
 7 & 14723280 & 59751978 &286924 & 40996 &  20766 &       &     \\
\hline
\end{tabular}
\caption{\label{table_dim_T_torus}
Dimensionality of the transfer matrix for toroidal boundary
conditions. For each width $m$, we give the number of
non-crossing connectivities $C_{m,m}$ \protect\reff{def_CmmBis},
the number of non-crossing connectivities $C_{m+1,m}$ 
\protect\reff{def_Cmm-1}, the number of non-crossing non-nearest-neighbor
connectivities $D_{m,m}$, the number of generated connectivity classes modulo
reflections (resp.\ reflections and translations) of non-crossing 
non-nearest-neighbor connectivities TriTorus$'$ (resp.\ SqTorus$'$), 
and the number of distinct eigenvalues 
for a triangular-lattice (resp.\ square-lattice) strip with toroidal 
boundary conditions TriTorus (resp.\ SqTorus). 
}
\end{table} 

%
%
\begin{table}
\centering
\begin{tabular}{|r|rrrrrrrr|r|}
\hline\hline
$m$  &$N_0$& $N_1$& $N_2$& $N_3$& $N_4$& $N_5$& $N_6$& $N_7$ & SqTorus$'$ \\
\hline\hline
  2  &   1 &    2 &    1 &      &      &      &      &    &     4 \\
  3  &   1 &    2 &    4 &    1 &      &      &      &    &     8 \\
  4  &   5 &   22 &   33 &    7 &   1  &      &      &    &    68 \\
  5  &   6 &   67 &  190 &   72 &  11  &    1 &      &    &   347 \\
  6  &  39 &  494 & 1432 &  644 & 163  &   16 &    1 &    &  2789 \\
  7  & 111 & 2794 &10224 & 5615 &2711  &  288 &   22 &  1 & 20766 \\
\hline
\end{tabular}
\caption{\label{table_blocks_sq}
  Block structure of the transfer matrix of a toroidal square-lattice
  strip of width $m$ as a function of number of bridges $\ell$.
  For each strip width $m$, we quote the dimension  
  $N_\ell$ of the diagonal block $T_{\ell,\ell}(m_\text{T})$ 
  for a given number of bridges $\ell$, and
  the total dimension of the transfer matrix 
  $\text{SqTorus}'(m)=\dim \T(m_\text{T})$. 
}
\end{table}

%
%
\begin{table}
\centering
\begin{tabular}{|r|rrrrrrrr|r|}
\hline\hline
$m$  &$N_0$& $N_1$& $N_2$& $N_3$& $N_4$& $N_5$& $N_6$& $N_7$& TriTorus$'$ \\
\hline\hline
  2  &   1 &    2 &    2 &      &      &      &      &   &     5 \\
  3  &   1 &    3 &    6 &    3 &      &      &      &   &    13 \\
  4  &   5 &   26 &   52 &   12 &   4  &      &      &   &    99 \\
  5  &   8 &  125 &  360 &  135 &  20  &   5  &      &   &   653 \\
  6  &  49 &  843 & 2712 & 1254 & 300  &  30  &   6  &   &  5194 \\
  7  & 186 & 5488 &20216 &11109 &3388  & 560  &  42  & 7 & 40996 \\
\hline
\end{tabular}
\caption{\label{table_blocks_tri}
  Block structure of the transfer matrix of a toroidal triangular-lattice
  strip of width $m$ as a function of number of bridges $\ell$.
  For each strip width $m$, we quote the dimension  
  $N_\ell$ of the diagonal block $T_{\ell,\ell}(m_\text{T})$ 
  for a given number of bridges $\ell$, and
  the total dimension of the transfer matrix 
  $\text{SqTorus}'(m)=\dim \T(m_\text{T})$. 
}
\end{table}

%
%
\begin{table}
\centering
\begin{tabular}{|r|rrrrrrrr|r|}
\hline\hline
$m$  &$L_0$& $L_1$& $L_2$& $L_3$& $L_4$& $L_5$& $L_6$& $L_7$& SqTorus\\
\hline\hline
  2  &   1 &    2 &    1 &      &      &      &      &     &   4 \\
  3  &   1 &    2 &    4 &    1 &      &      &      &     &   8 \\
  4  &   3 &    8 &   15 &    7 &   1  &      &      &     &  33 \\
  5  &   4 &   15 &   36 &   24 &  11  &    1 &      &     &  90 \\
  6  &  11 &   48 &  120 &   84 &  49  &   16 &    1 &     & 325 \\
  7  &  21 &  112 & $304^\dagger$
                         & $253^\dagger$ 
                                & 165  &   72 &   22 &   1 &     \\
\hline
\end{tabular}
\caption{\label{table_eigen_sq}
  Number of distinct eigenvalues for a toroidal square--lattice strip 
  of width $m$ as a function of number of bridges $\ell$.
  For each strip width $m$, we quote the number of
  different eigenvalues $L_\ell$ for a given number of bridges $\ell$, and
  the total number of distinct eigenvalues SqTorus$(m)$. Notice that 
  the sum $\sum_{\ell=0}^m L_\ell(m)$ may be greater than SqTorus$(m)$,
  as a few eigenvalues can appear on different $\ell$ blocks. 
  The results marked with a $\mbox{}^\dagger$ are conjectures based on the
  numerical evaluation of the corresponding eigenvalues.
}
\end{table}

%
%
\begin{table}
\centering
\begin{tabular}{|r|rrrrrrrr|r|}
\hline\hline
$m$  &$L_0$& $L_1$& $L_2$& $L_3$& $L_4$& $L_5$& $L_6$& $L_7$& TriTorus\\
\hline\hline
  2  &   1 &    2 &    2 &      &      &      &      &     &    4 \\
  3  &   1 &    3 &    6 &    3 &      &      &      &     &   11 \\
  4  &   3 &    8 &   18 &   11 &   4  &      &      &     &   38 \\
  5  &   6 &   24 &   57 &   42 &  19  &   5  &      &     &  148 \\
  6  &  14 &   62 &  165 &  136 &  76  &  28  &    6 &     &  476 \\
  7  &  36 &  196 & $508^\dagger$
                         & $454^\dagger$ 
                                & 289  & 122  &   40 &  7  &      \\
\hline
\end{tabular}
\caption{\label{table_eigen_tri}
  Number of distinct eigenvalues for a toroidal triangular--lattice strip 
  of width $m$ as a function of number of bridges $\ell$.
  For each strip width $m$, we quote the number of
  different eigenvalues $L_\ell$ for a given number of bridges $\ell$, and
  the total number of distinct eigenvalues TriTorus$(m)$. Notice that 
  the sum $\sum_{\ell=0}^m L_\ell(m)$ may be greater than TriTorus$(m)$,
  as a few eigenvalues can appear on different $\ell$ blocks. 
  The results marked with a $\mbox{}^\dagger$ are conjectures based on the
  numerical evaluation of the corresponding eigenvalues.
}
\end{table}

%
%
\def\kk{\phantom{$1$}}
\begin{table}
\centering
\begin{tabular}{|r||c|c|c|c|c|l||c|}
\hline\hline
Lattice    & \# C & \# E & \# T & \# D & \# ER &\multicolumn{1}{|c||}{$q_0$}
           & \# RI \\
\hline\hline
 $2_\text{T}$&  1 & 0 &\kk2& 1 & 3 & 2            &  1 \\
 $3_\text{T}$&  1 & 0 &\kk2& 0 & 2 & 3            &  1 \\
 $4_\text{T}$&  1 & 0 &\kk6& 0 & 4 & 2.7827657401 &  1 \\
 $5_\text{T}$&  1 & 2 &\kk4& 0 & 2 & 3            &  1 \\
 $6_\text{T}$&  1 & 2 &\kk8& 0 & 4 & 2.9078023424 &  1 \\
 $7_\text{T}$&  1 & 2 & 12 & 0 & 6 & 3            &  1 \\
 $8_\text{T}$&    &   &    &   &   & 2.9489357617 &    \\
 $9_\text{T}$&    &   &    &   &   & 3            &    \\
$10_\text{T}$&    &   &    &   &   & 2.9674224455 &    \\
$11_\text{T}$&    &   &    &   &   & 3            &    \\
$12_\text{T}$&    &   &    &   &   & 2.9773394917 &    \\
\hline\hline
\end{tabular}
\caption{\label{table_summary_sq}
   Summary of qualitative results for the square-lattice
   eigenvalue-crossing curves $\mathcal{B}$
   and for the isolated limiting points of zeros.
   For each toroidal square-lattice strip considered in this paper,
   we give the number of connected components of $\mathcal{B}$ (\# C),
   the number of endpoints (\# E),
   the number of T points (\# T),
   the number of double points (\# D),
   and the number of enclosed regions (\# ER).
   We also give the value $q_0(L)$ which is the largest real value where
   $\mathcal{B}$ intersects the real axis, and the
   number of real isolated limiting points of zeros (\# RI).
}
\end{table}

%
%
\def\kk{\phantom{$1$}}
\begin{table}
\centering
\begin{tabular}{|r||c|c|c|c|c|l|l||c|}
\hline\hline
Lattice    & \# C & \# E & \# T & \# D & \# ER &\multicolumn{1}{|c|}{$q_0$}
                                               &\multicolumn{1}{|c||}{$q_2$}
           & \# RI \\
\hline\hline
$2_\text{T}$&  1 & 0 &\kk0& 1 & 2 & 4            &            &  2 \\
$3_\text{T}$&  1 & 0 &\kk2& 0 & 2 & 3.7240755514 & 4          &  2 \\
$4_\text{T}$&  1 & 0 &\kk2& 0 & 2 & 4            & 4          &  2 \\
$5_\text{T}$&  1 & 0 &\kk8& 0 & 5 & 4            & 3.84826329 &  2 \\
$6_\text{T}$&  1 & 0 & 10 & 0 & 6 & 3.7943886378 & 3.84085775 &  2 \\
$7_\text{T}$&  1 & 0 & 12 & 0 & 7 & 4            & 3.86055083 &  2 \\
$8_\text{T}$&    &   &    &   &   & 4            & 3.83354219 &    \\ 
$9_\text{T}$&    &   &    &   &   & 3.8081293156 & 3.82806660 &    \\ 
$10_\text{T}$&   &   &    &   &   & 4            & 3.83762301 &    \\ 
$11_\text{T}$&   &   &    &   &   & 4            & 3.82801206 &    \\ 
$12_\text{T}$&   &   &    &   &   & 3.8131463701 & 3.82428323 &    \\ 
\hline\hline
\end{tabular}
\caption{\label{table_summary_tri}
   Summary of qualitative results for the triangular-lattice
   eigenvalue-crossing curves $\mathcal{B}$
   and for the isolated limiting points of zeros.
   For each toroidal square-lattice strip considered in this paper,
   we give the number of connected components of $\mathcal{B}$ (\# C),
   the number of endpoints (\# E),
   the number of T points (\# T),
   the number of double points (\# D),
   and the number of enclosed regions (\# ER).
   We also give the value $q_0(L)$ which is the largest real value where
   $\mathcal{B}$ intersects the real axis, 
   the value $q_2(L)$ where there is a crossing between two dominant 
   eigenvalues that belong to the $\ell=2$ sector, and the 
   number of real isolated limiting points of zeros (\# RI).
}
\end{table}

%
%
\renewcommand{\arraystretch}{1.2}
\begin{table}
\centering
\begin{tabular}{|r||r|r|r||r|r|r|}
\cline{2-7}
\multicolumn{1}{c||}{\mbox{}} & 
\multicolumn{3}{c||}{$q_0(L)$} & \multicolumn{3}{c|}{$q_2(L)$} \\ 
\cline{2-7}
\multicolumn{2}{c}{\mbox{}} & 
\multicolumn{5}{l}%
            {\qquad Ansatz: $q_k(L)\;=\; q_0(\text{tri}) + A L^{-1/\nu}$}\\
\hline
$L_\text{min}$ & \multicolumn{1}{|c|}{$q_0(\text{tri})$} & 
                 \multicolumn{1}{|c|}{$A$} & 
                 \multicolumn{1}{|c||}{$1/\nu$} &  
                 \multicolumn{1}{ c|}{$q_0(\text{tri})$} & 
                 \multicolumn{1}{|c|}{$A$} & 
                 \multicolumn{1}{|c|}{$1/\nu$} \\ 
\hline
$3$ & $3.819878$ & $-0.78118$ & $1.91015$& $3.823659$ & $7.05559$ & $3.35802$\\
$6$ & $3.820128$ & $-0.75050$ & $1.88235$& $3.820619$ & $1.67816$ & $2.46566$\\
\hline
\multicolumn{2}{c}{\mbox{}} & 
\multicolumn{5}{l}%
            {\qquad Ansatz: $q_k(L)\;=\; q_0(\text{tri}) + A L^{-2}$}\\
\hline
$L_\text{min}$ & \multicolumn{1}{|c|}{$q_0(\text{tri})$} &
                 \multicolumn{1}{|c||}{$A$} &  & 
                 \multicolumn{1}{ c|}{$q_0(\text{tri})$} &
                 \multicolumn{1}{|c|}{$A$} &  \\  
\hline
 $3$ & $3.817826$ & $-0.84376$ &  & $3.787810$ & $1.90971$ & \\
 $6$ & $3.819122$ & $-0.89040$ &  & $3.817834$ & $0.82887$ & \\
 $9$ & $3.819597$ & $-0.92887$ &  & $3.819419$ & $0.70046$ & \\
\hline
\multicolumn{2}{c}{\mbox{}} & 
\multicolumn{5}{l}%
  {\qquad Ansatz: $q_k(L)\;=\; q_0(\text{tri}) + A L^{-2} + BL^{-4}$}\\
\hline
$L_\text{min}$ & \multicolumn{1}{|c|}{$q_0(\text{tri})$} &
                 \multicolumn{1}{|c|}{$A$} &
                 \multicolumn{1}{|c||}{$B$} & 
                 \multicolumn{1}{ c|}{$q_0(\text{tri})$} &
                 \multicolumn{1}{|c|}{$A$} &
                 \multicolumn{1}{|c|}{$B$} \\ 
\hline
$3$& $3.819284$ & $-0.90934$ & $0.47222$& $3.821587$ & $ 0.38978$ & $10.94351$\\
$6$& $3.819755$ & $-0.96450$ & $1.84684$& $3.819947$ & $ 0.58157$ & $ 6.16332$\\
\hline\hline
\end{tabular}
\caption{\label{table_fits_tri}
Summary of fits of the $q_0(3p)$ and $q_2(3p)$ data ($1\leq p \leq 4$)
for the zero-temperature triangular-lattice Potts antiferromagnet.
For each data set and Ansatz, we show the different estimates as a function 
of the parameter $L_\text{min}$. 
See text for an explanation of how the fits were performed.
}
\end{table}
\renewcommand{\arraystretch}{1}

\clearpage
%
%
%
%
\clearpage

%
%

%
%

%
%
\begin{figure}
   \centering
   \psset{labelsep=10pt}
   \psset{xunit=2cm}
   \psset{yunit=2cm}
   \begin{tabular}{c}
   \pspicture(-0.5,-0.5)(3.5,2.5) 
   \psline[linecolor=black,linewidth=1.pt](-0.5,0)(3.5,0)
   \psline[linecolor=black,linewidth=1.pt](-0.5,2)(3.5,2)
   \psline[linecolor=black,linewidth=2.pt](0,0)(0,0.25)(2,0.25)(2,0) 
   \psline[linecolor=black,linewidth=2.pt](1,2)(1,1.75)(3,1.75)(3,2) 
   \psline[linecolor=black,linewidth=2.pt,linestyle=dashed](3,0)(3,2)
   \multirput(0,0)(0,2){2}{%
     \multirput(0,0)(1,0){4}{\pscircle*[linecolor=lightgray](0,0){6pt}}
     \multirput(0,0)(1,0){4}{\pscircle[linewidth=1pt](0,0){6pt}}
   }
   \uput[270](0,0){\Large $\bm{1'}$}
   \uput[270](1,0){\Large $\bm{2'}$}
   \uput[270](2,0){\Large $\bm{3'}$}
   \uput[270](3,0){\Large $\bm{4'}$}

   \uput[90](0,2){\Large $\bm{1}$}
   \uput[90](1,2){\Large $\bm{2}$}
   \uput[90](2,2){\Large $\bm{3}$}
   \uput[90](3,2){\Large $\bm{4}$}
   \psline[linecolor=black,linewidth=4pt]{->}(3.5,0.6)(3.5,1.4)
   \endpspicture
   \\[2mm] 
   $(a)$ 
   \\[2mm] 
   \pspicture(-2.5,-2.5)(2.5,2.5) 
   \pscircle[linecolor=black,linewidth=1.pt](0,0){2}
   \pscircle[linecolor=black,linewidth=1.pt](0,0){4}
   \psarc[linecolor=black,linewidth=2.pt](0,0){2.5}{90}{270}
   \psarc[linecolor=black,linewidth=2.pt](0,0){3.5}{0}{180}
   \psline[linecolor=black,linewidth=2.pt](1.73,0)(2,0)
   \psline[linecolor=black,linewidth=2.pt](-1.73,0)(-2,0)
   \psline[linecolor=black,linewidth=2.pt](0,1)(0,1.265)
   \psline[linecolor=black,linewidth=2.pt](0,-1)(0,-1.265)
   \psline[linecolor=black,linewidth=2.pt,linestyle=dashed](1,0)(2,0)
   \rput( 1, 0){\pscircle*[linecolor=lightgray](0,0){6pt}} 
   \rput( 0, 1){\pscircle*[linecolor=lightgray](0,0){6pt}} 
   \rput(-1, 0){\pscircle*[linecolor=lightgray](0,0){6pt}} 
   \rput( 0,-1){\pscircle*[linecolor=lightgray](0,0){6pt}} 
   \rput( 2, 0){\pscircle*[linecolor=lightgray](0,0){6pt}} 
   \rput( 0, 2){\pscircle*[linecolor=lightgray](0,0){6pt}} 
   \rput(-2, 0){\pscircle*[linecolor=lightgray](0,0){6pt}} 
   \rput( 0,-2){\pscircle*[linecolor=lightgray](0,0){6pt}} 
   \rput( 1, 0){\pscircle[linewidth=1pt](0,0){6pt}} 
   \rput( 0, 1){\pscircle[linewidth=1pt](0,0){6pt}} 
   \rput(-1, 0){\pscircle[linewidth=1pt](0,0){6pt}} 
   \rput( 0,-1){\pscircle[linewidth=1pt](0,0){6pt}} 
   \rput( 2, 0){\pscircle[linewidth=1pt](0,0){6pt}} 
   \rput( 0, 2){\pscircle[linewidth=1pt](0,0){6pt}} 
   \rput(-2, 0){\pscircle[linewidth=1pt](0,0){6pt}} 
   \rput( 0,-2){\pscircle[linewidth=1pt](0,0){6pt}} 
   \uput[270](0,1){\Large $\bm{1'}$}
   \uput[180](1,0){\Large $\bm{4'}$}
   \uput[0](-1,0){\Large $\bm{2'}$}
   \uput[90](0,-1){\Large $\bm{3'}$}
   \uput[90](0,2){\Large $\bm{1}$}
   \uput[0](2,0){\Large $\bm{4}$}
   \uput[180](-2,0){\Large $\bm{2}$}
   \uput[270](0,-2){\Large $\bm{3}$}
   \psline[linecolor=black,linewidth=4pt]{->}(1.6,1.6)(2.2,2.2)
   \endpspicture
   \\[2mm]
   $(b)$
   \\[2mm]
   \end{tabular}

   \caption{\label{figure_sq_cyclic_bc_bis}
     Connectivity state for a toroidal strip of width $4$. (a) 
     We show the connectivity state ${\cal P}=\delta_{1',3'}\delta_{2,4,4'}$,  
     which can be seen as a bottom-row connectivity state 
     ${\cal P}_\text{bot} =\delta_{1',3'}$ (lower solid thick line), 
     a top-row connectivity state ${\cal P}_\text{top} =\delta_{2,4}$ 
     (upper solid thick line), and a bridge connecting both rows 
     $\delta_{4,4'}$ 
     (dashed thick line). The thin solid lines show the top (resp.\ bottom)
     row containing the unprimed (resp.\ primed) sites.  
     The thick arrow shows the transfer direction (upwards).
     (b) The same connectivity state depicted in such a way that the 
     transverse periodic boundary conditions are clearer. 
     The transfer direction is outwards (thick arrow). 
   }
\end{figure}
\clearpage
%
%
\begin{figure}
\centering
\begin{tabular}{cc}
   \includegraphics[width=210pt]{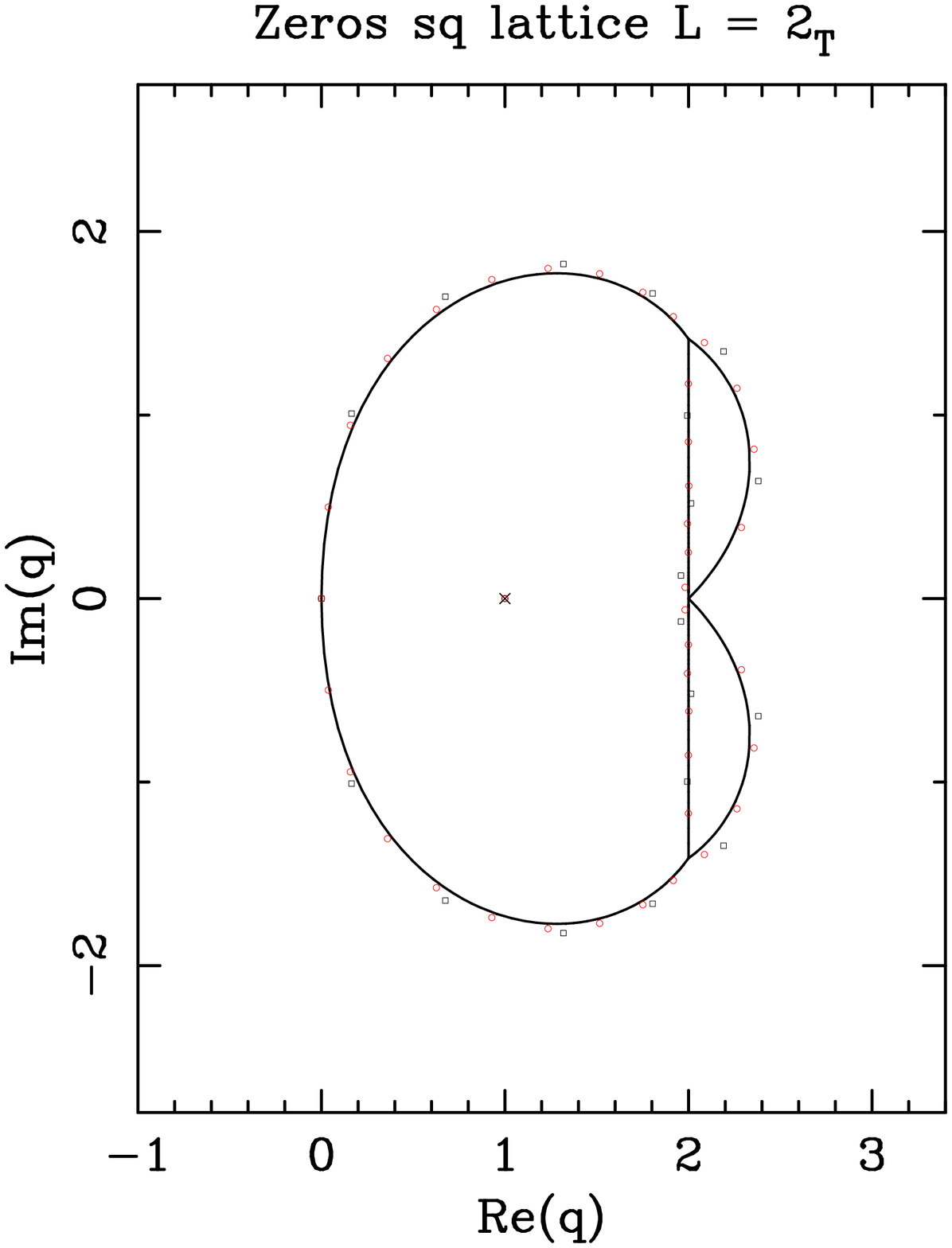} &
   \includegraphics[width=210pt]{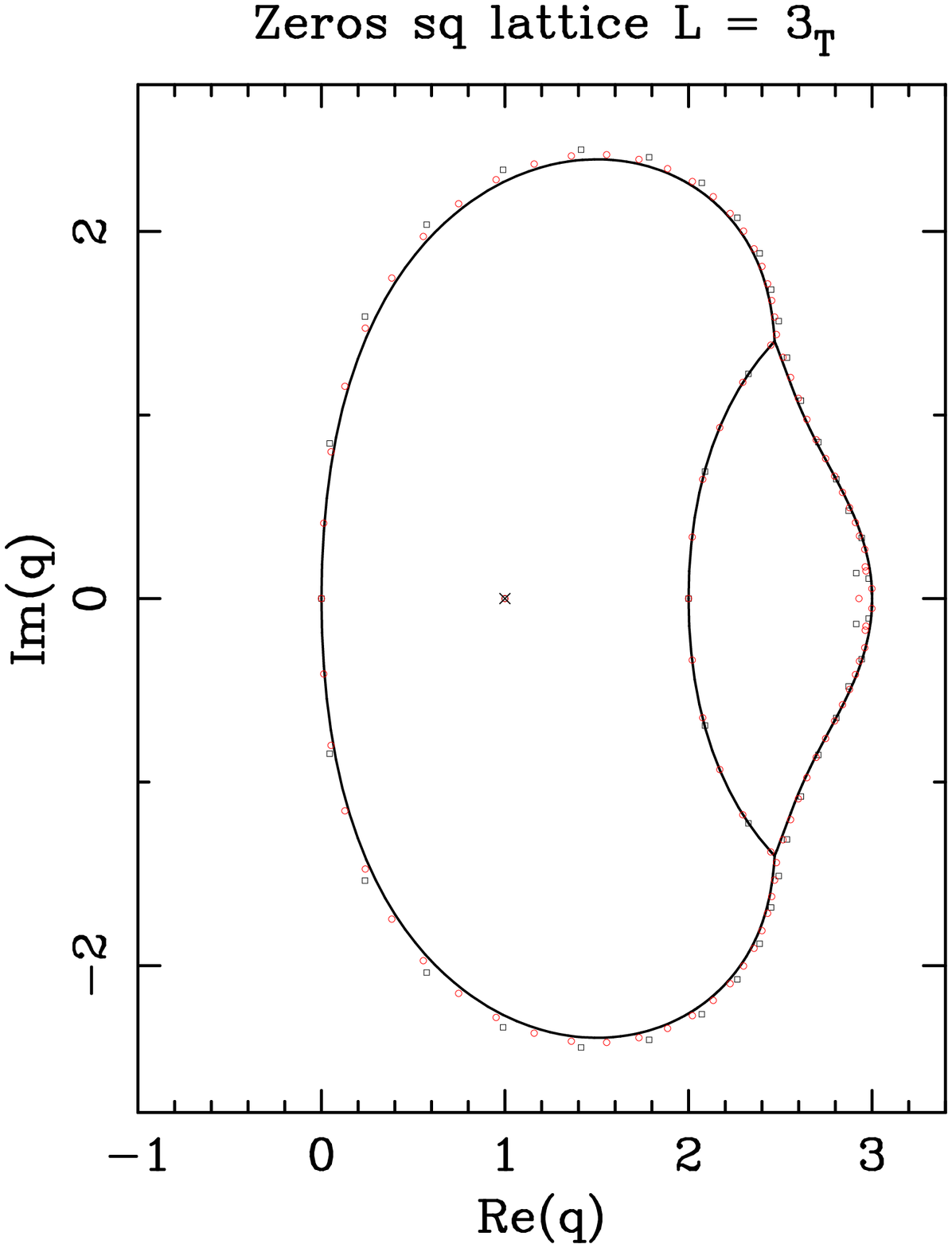}
   \\[1mm]
   \phantom{(((a)}(a)    & \phantom{(((a)}(b) \\[5mm]
   \includegraphics[width=210pt]{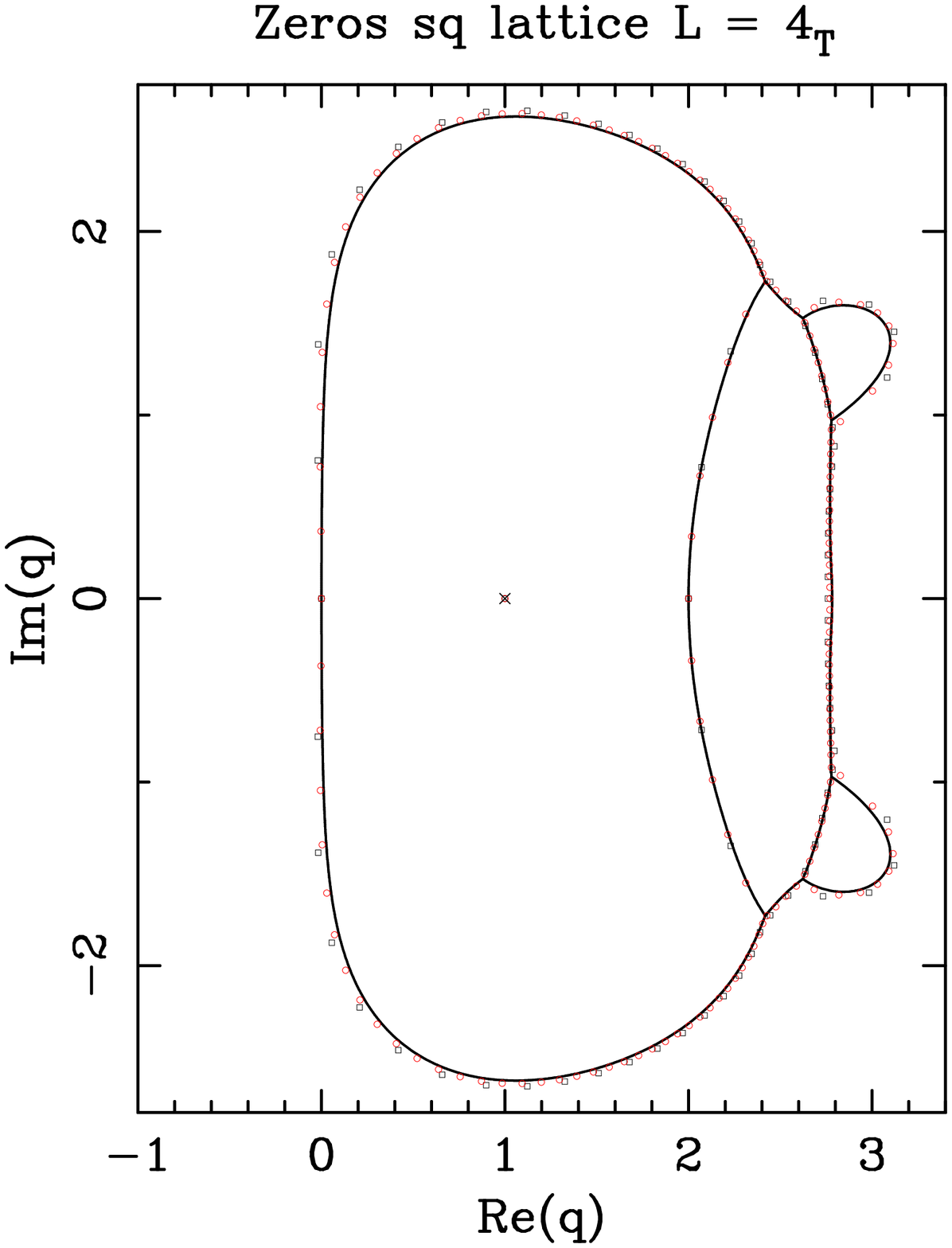} &  
   \includegraphics[width=210pt]{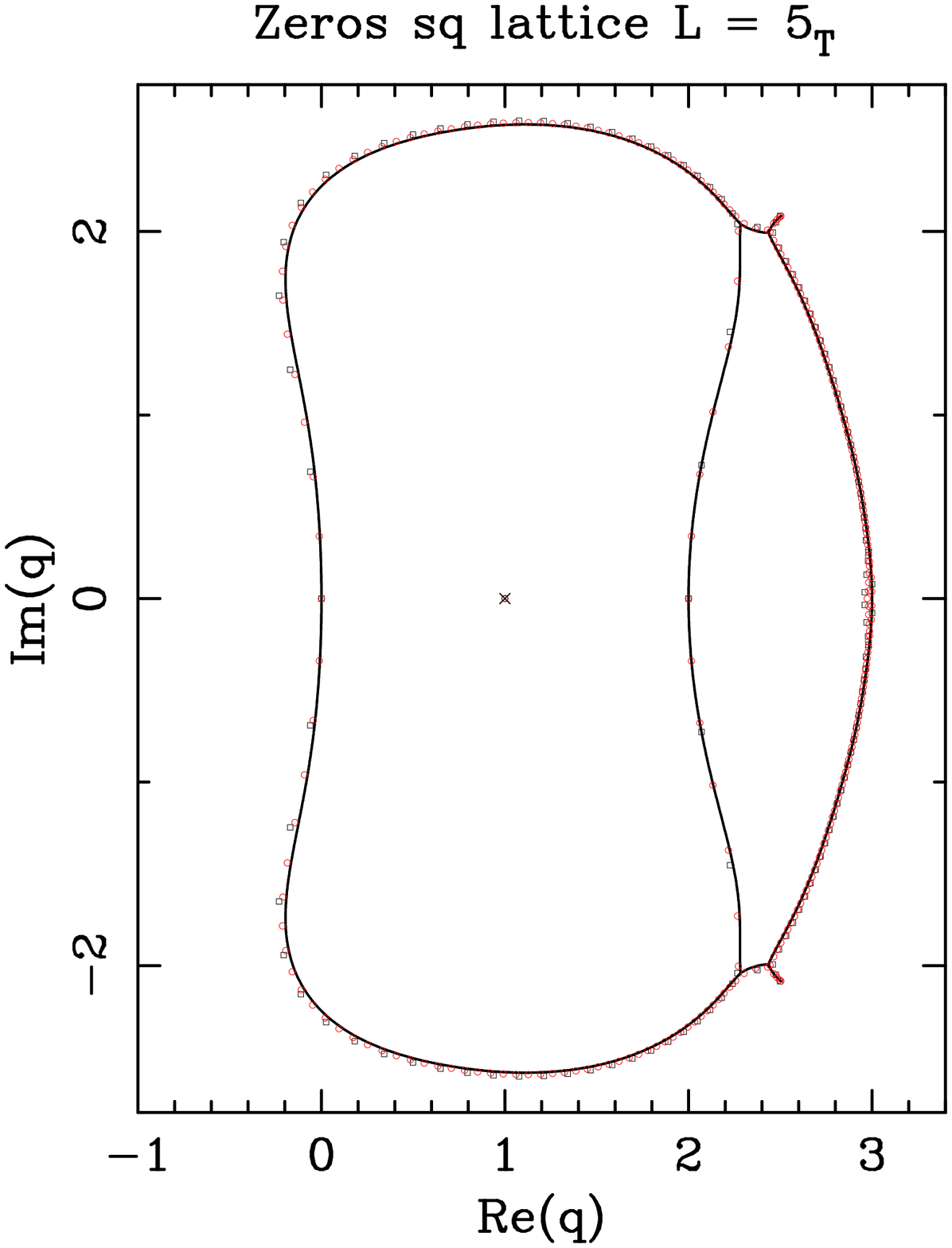} 
   \\[1mm]  
   \phantom{(((a)}(c)    & \phantom{(((a)}(d) \\
\end{tabular}
\caption{\label{figure_sq_1}
Limiting curves for square-lattice strips of width (a) $L=2$, (b) $L=3$,
(c) $L=4$, and (d) $L=5$  with toroidal boundary conditions. We also show the
zeros for the strips $L_\text{F} \times (5L)_\text{P}$ (black $\Box$) and
$L_\text{F} \times (10L)_\text{P}$ (red $\circ$) for the same values of $L$.
}
\end{figure}

\clearpage
%
%
\begin{figure}
\centering
\begin{tabular}{cc}
   \includegraphics[width=210pt]{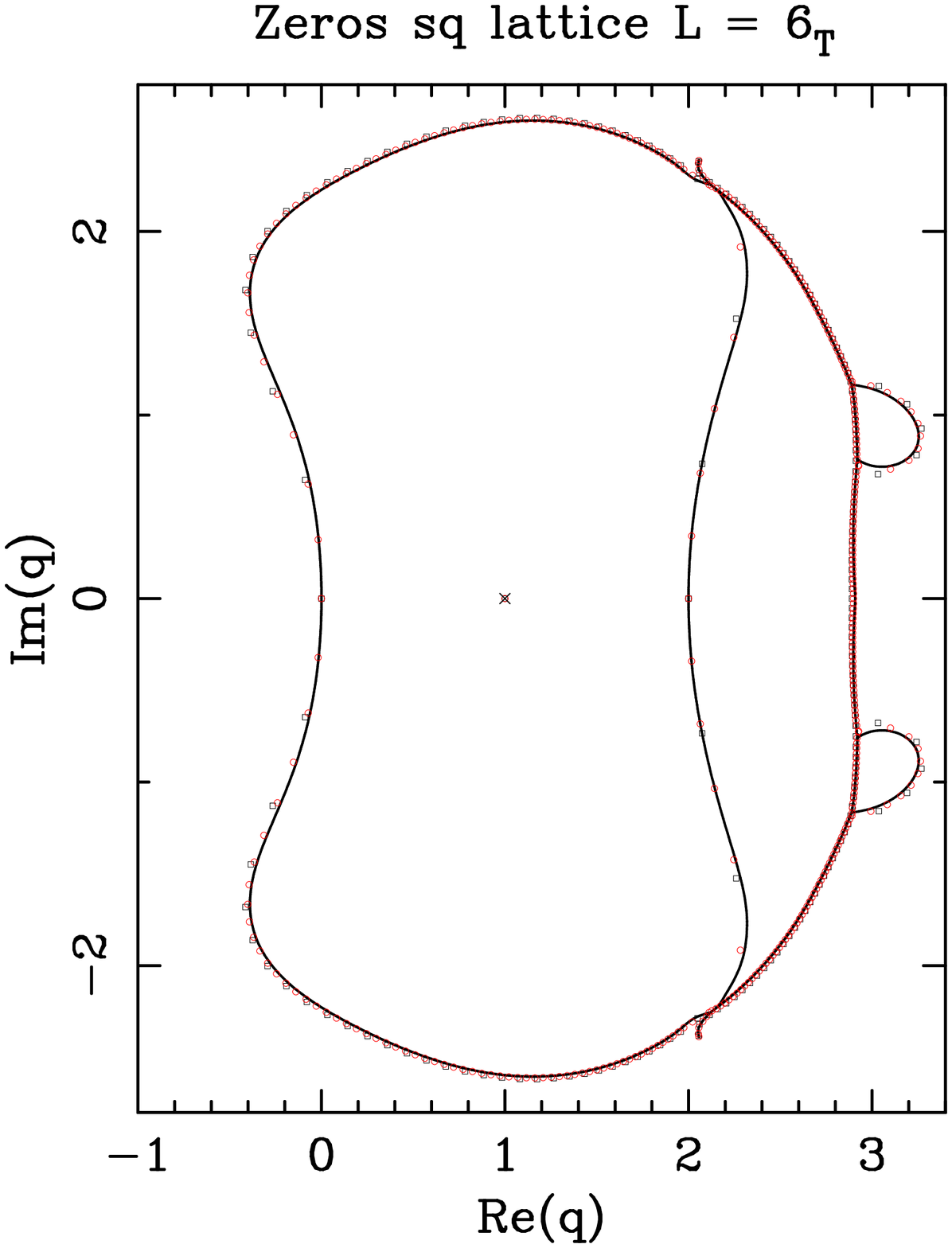} 
&
   \includegraphics[width=210pt]{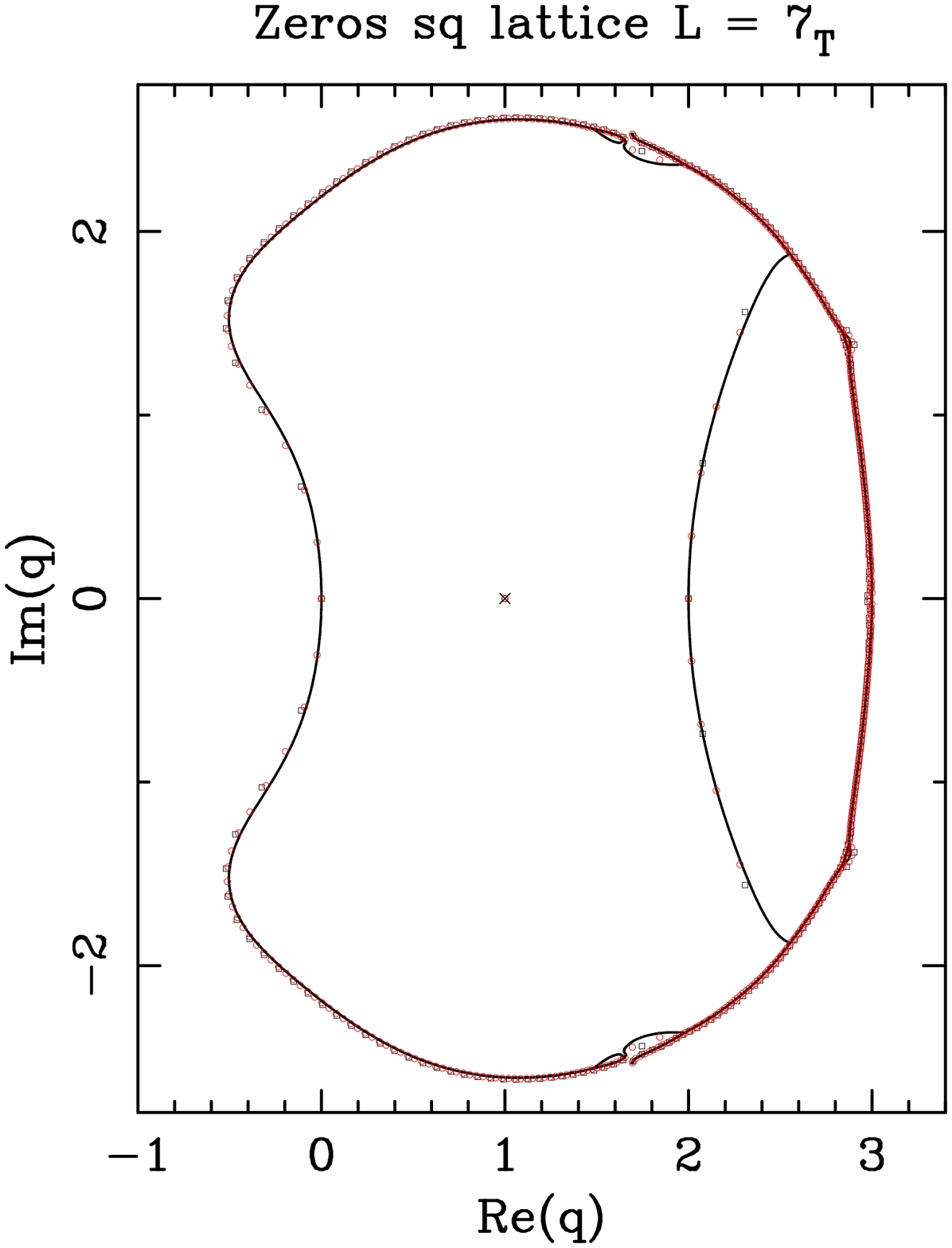}
   \\[1mm]
   \phantom{(((a)}(a)    & \phantom{(((a)}(b)\\
\end{tabular}
\caption{\label{figure_sq_2}
Limiting curves for square-lattice strips of width (a) $L=6$, and (b) $L=7$ 
with toroidal boundary conditions. 
We also show the
zeros for the strips $L_\text{F} \times (5L)_\text{P}$ (black $\Box$) and
$L_\text{F} \times (10L)_\text{P}$ (red $\circ$) for the same values of $L$.
}
\end{figure}

\clearpage
%
%
\begin{figure}
\centering
\begin{tabular}{cc}
   \includegraphics[width=210pt]{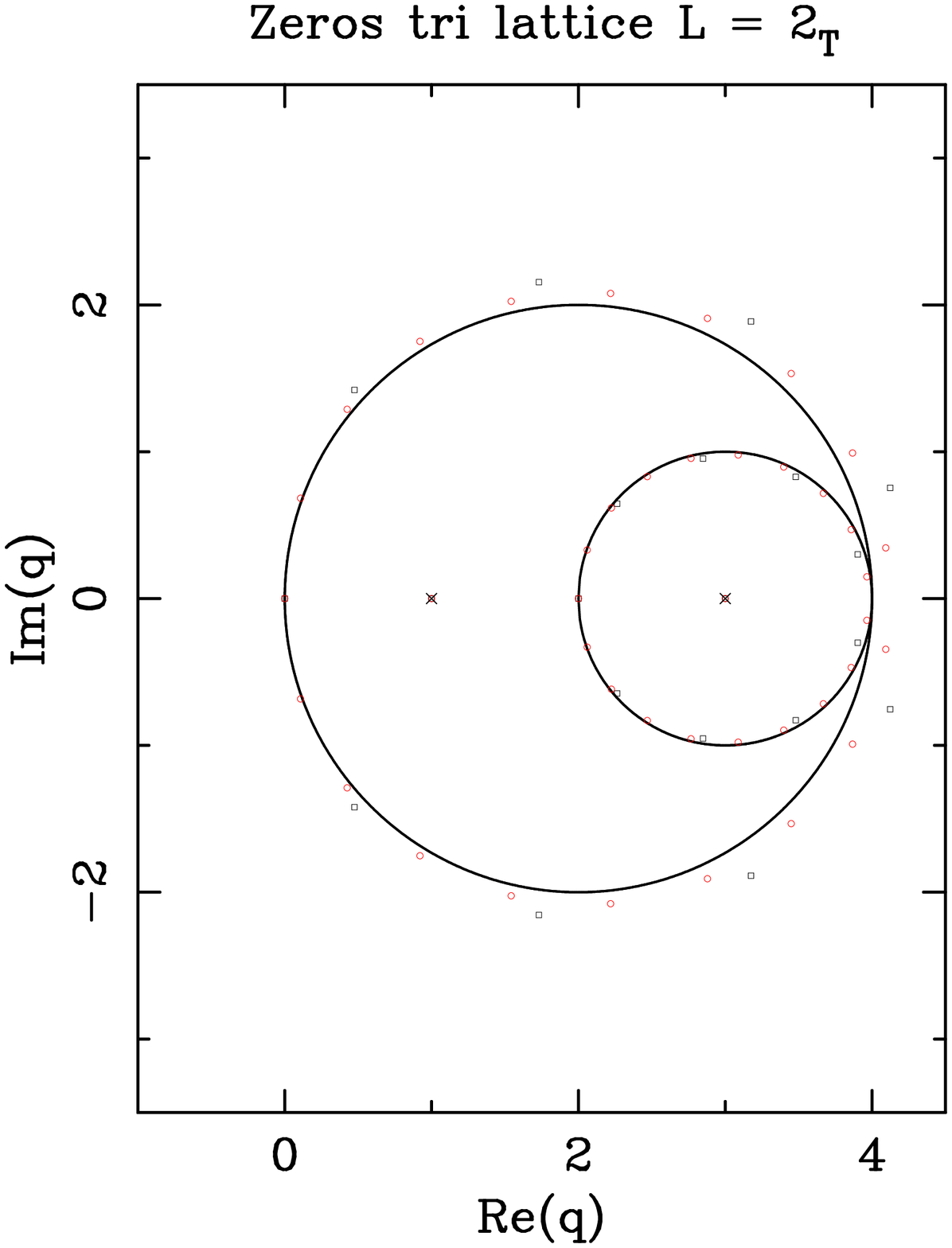} &
   \includegraphics[width=210pt]{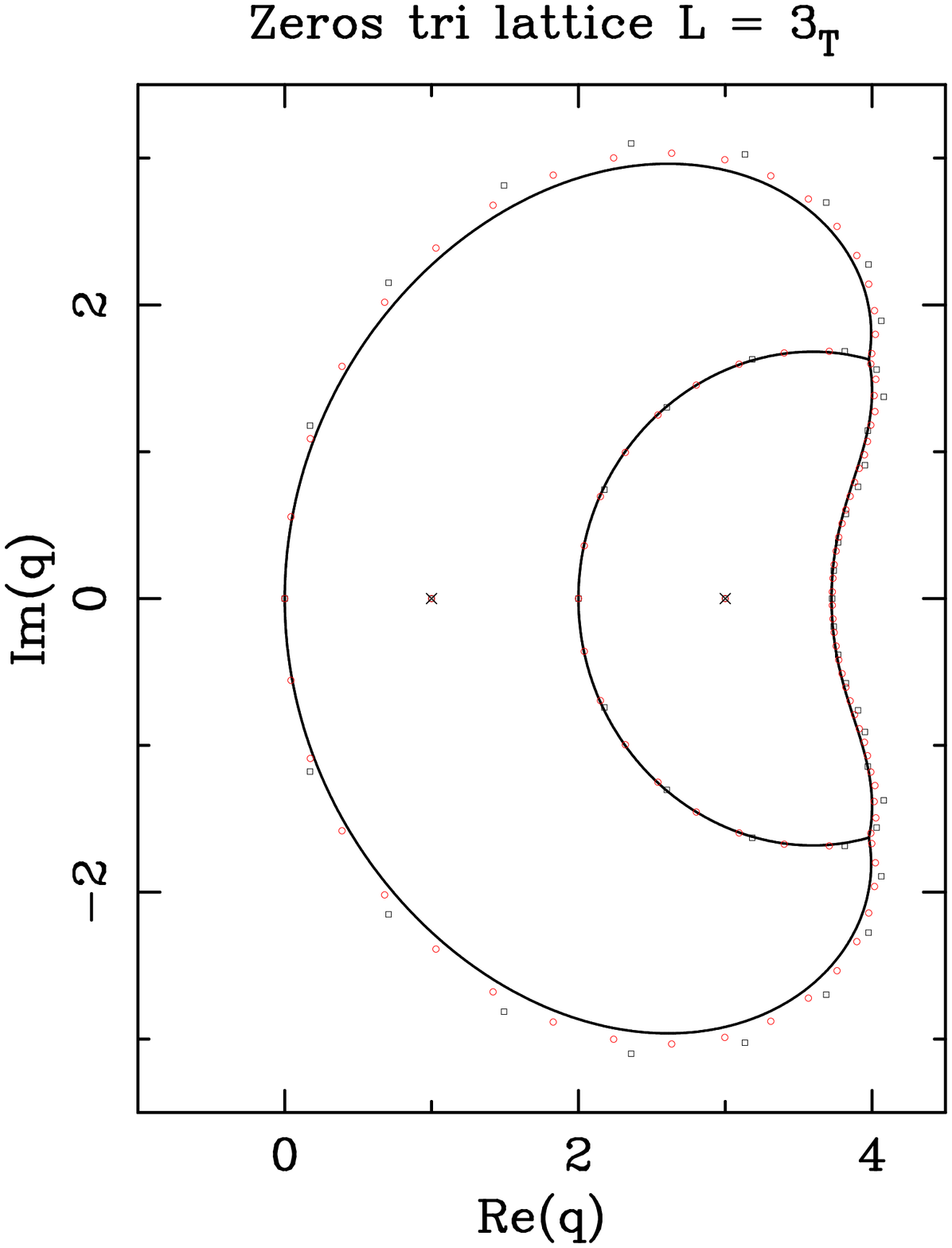}
   \\[1mm]
   \phantom{(((a)}(a)    & \phantom{(((a)}(b) \\[5mm]
   \includegraphics[width=210pt]{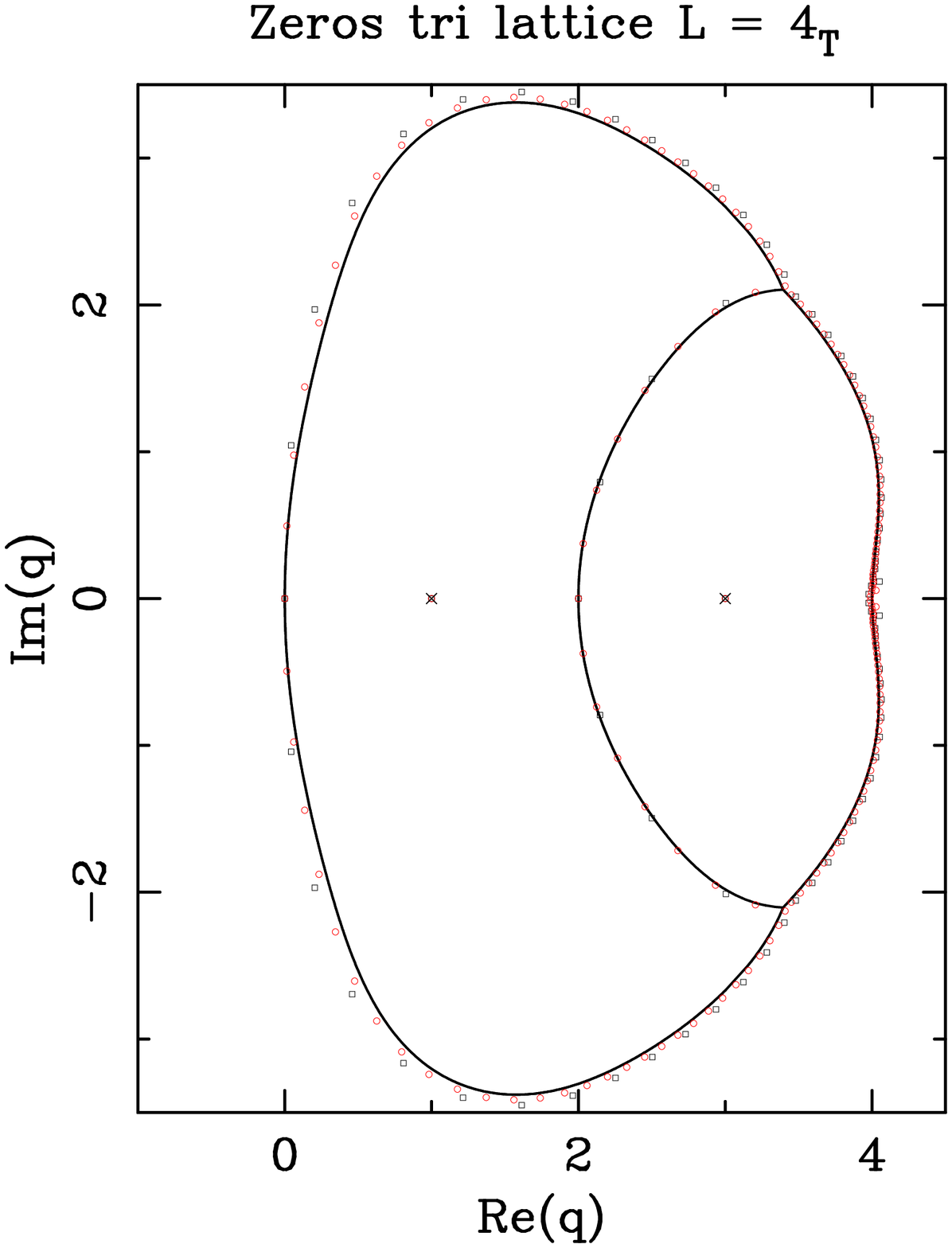} &  
   \includegraphics[width=210pt]{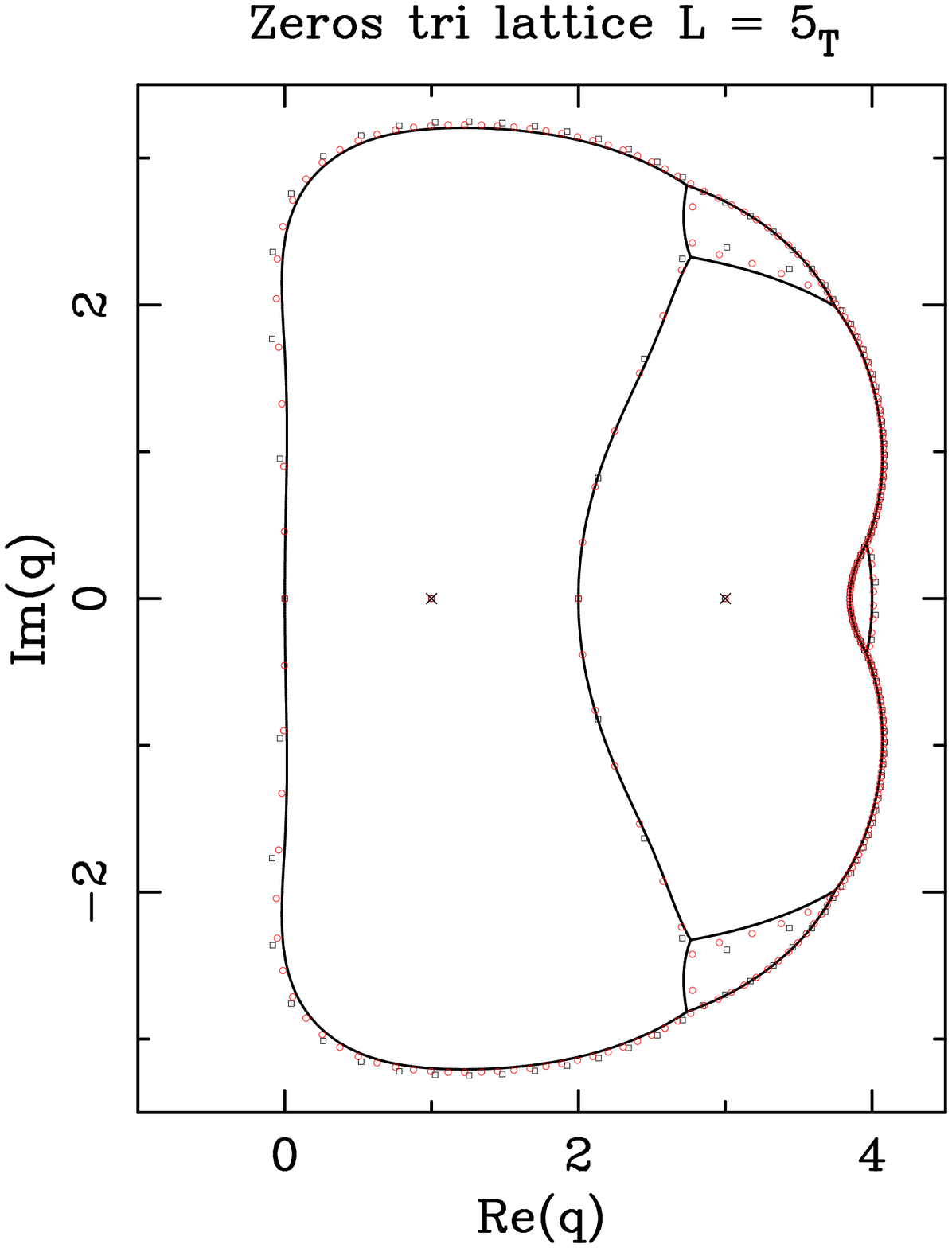} 
   \\[1mm]  
   \phantom{(((a)}(c)    & \phantom{(((a)}(d) \\
\end{tabular}
\caption{\label{figure_tri_1}
Limiting curves for triangular-lattice strips of width (a) $L=2$, (b) $L=3$,
(c) $L=4$, and (d) $L=5$  with toroidal boundary conditions. We also show the
zeros for the strips $L_\text{F} \times (5L)_\text{P}$ (black $\Box$) and
$L_\text{F} \times (10L)_\text{P}$ (red $\circ$) for the same values of $L$.
}
\end{figure}

%
%
\begin{figure}
\centering
\begin{tabular}{cc}
   \includegraphics[width=210pt]{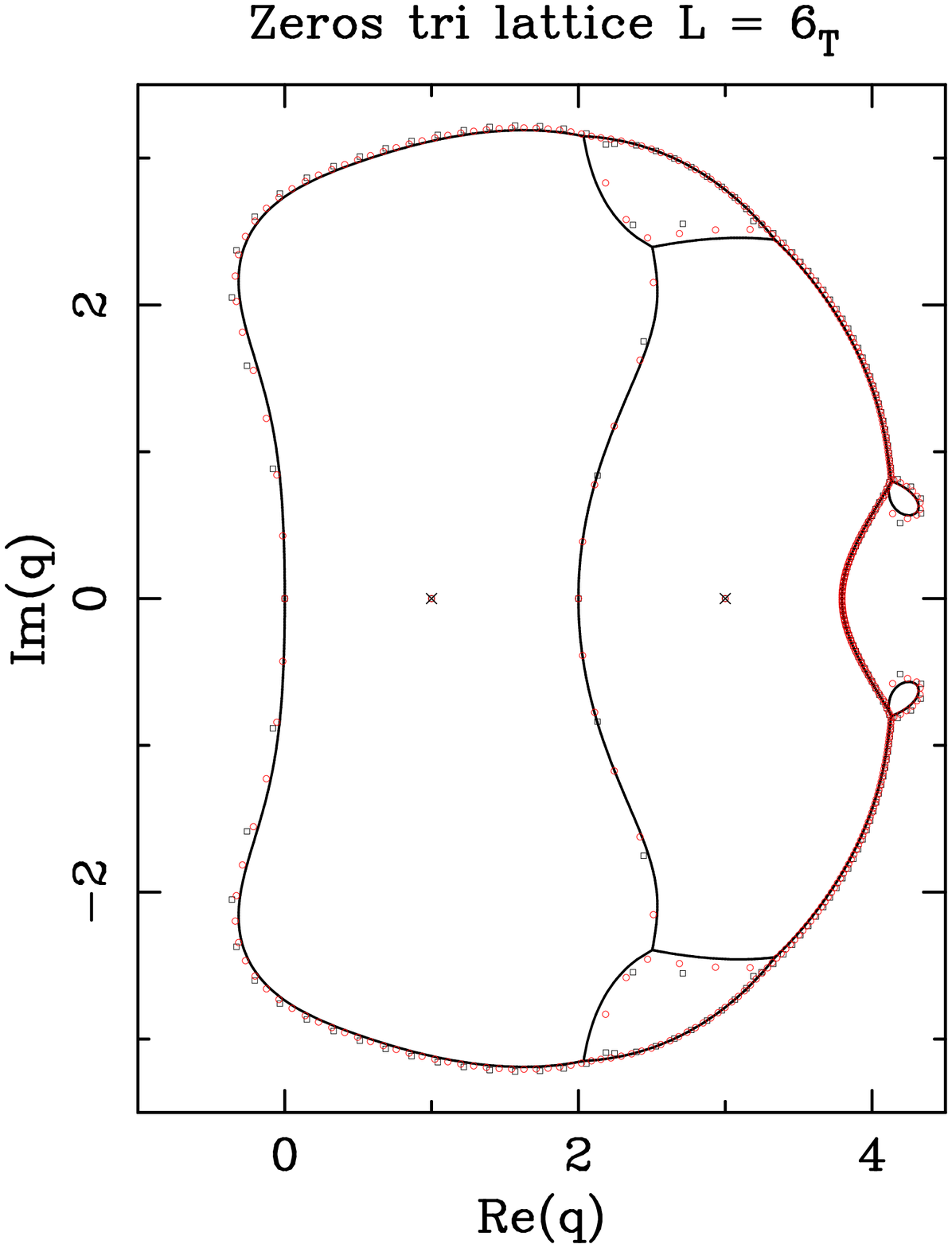} &
   \includegraphics[width=210pt]{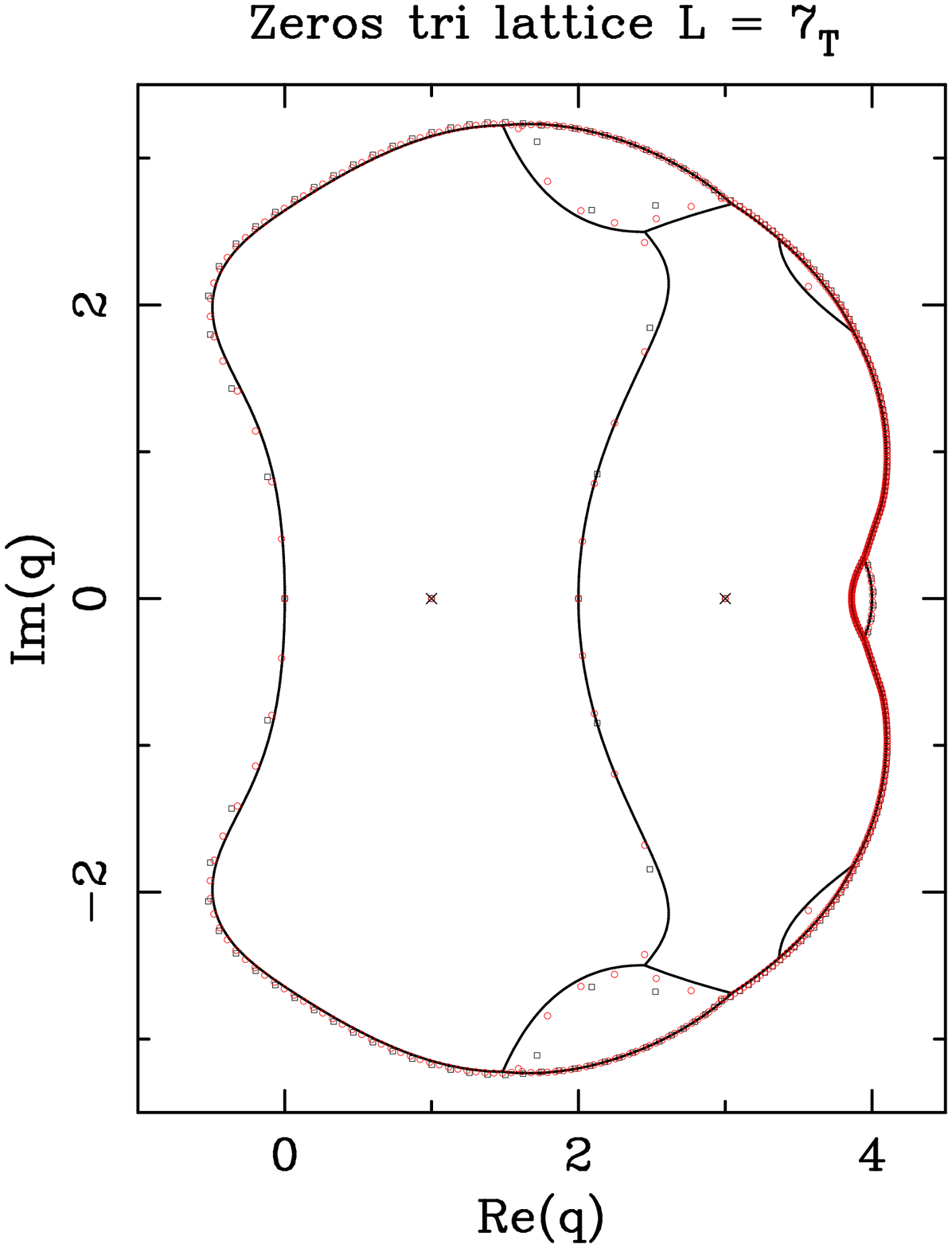}
   \\[1mm]
   \phantom{(((a)}(a)    & \phantom{(((a)}(b) \\[5mm]
\end{tabular}
\caption{\label{figure_tri_2}
Limiting curves for triangular-lattice strips of width (a) $L=6$, and 
(b) $L=7$ with toroidal boundary conditions. We also show the
zeros for the strips $L_\text{F} \times (5L)_\text{P}$ (black $\Box$) and
$L_\text{F} \times (10L)_\text{P}$ (red $\circ$) for the same values of $L$.
}

\end{figure}
\clearpage
%
%
\begin{figure}
  \centering
  \includegraphics[width=400pt]{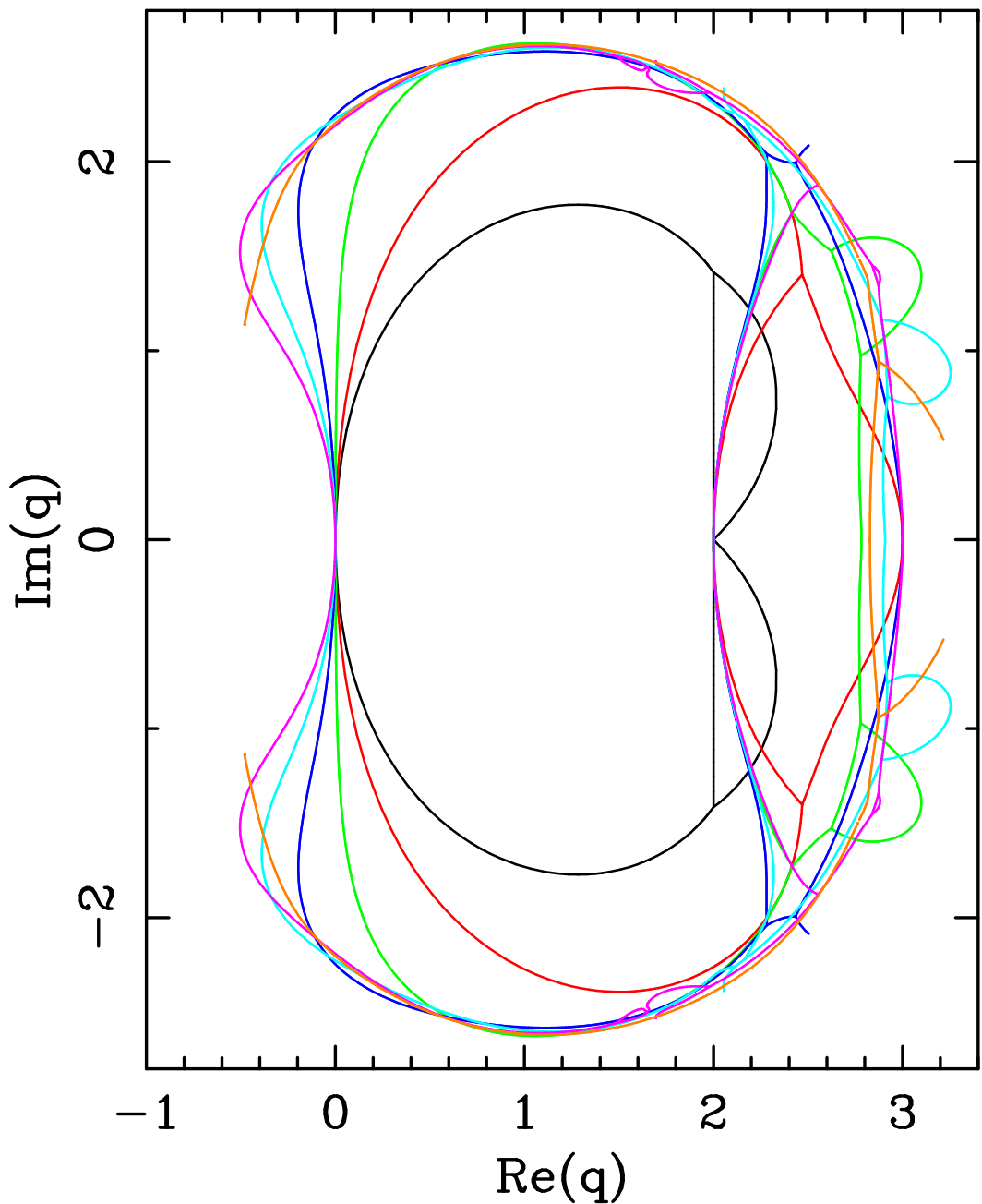}
  \caption{\label{figure_sq_allT}
   Limiting curves for square-lattice strips of width $L$ and toroidal 
   boundary conditions. We show the results for $L=2$ (black), $L=3$ (red), 
   $L=4$ (green), $L=5$ (blue), $L=6$ (light blue), and $L=7$ (pink). 
   We also show the limiting curve for a square-lattice strip of width $L=10$
   with cylindrical boundary conditions (orange) \protect\cite{transfer2}.
  }
\end{figure}

\clearpage
%
%
\begin{figure}
  \centering
  \includegraphics[width=400pt]{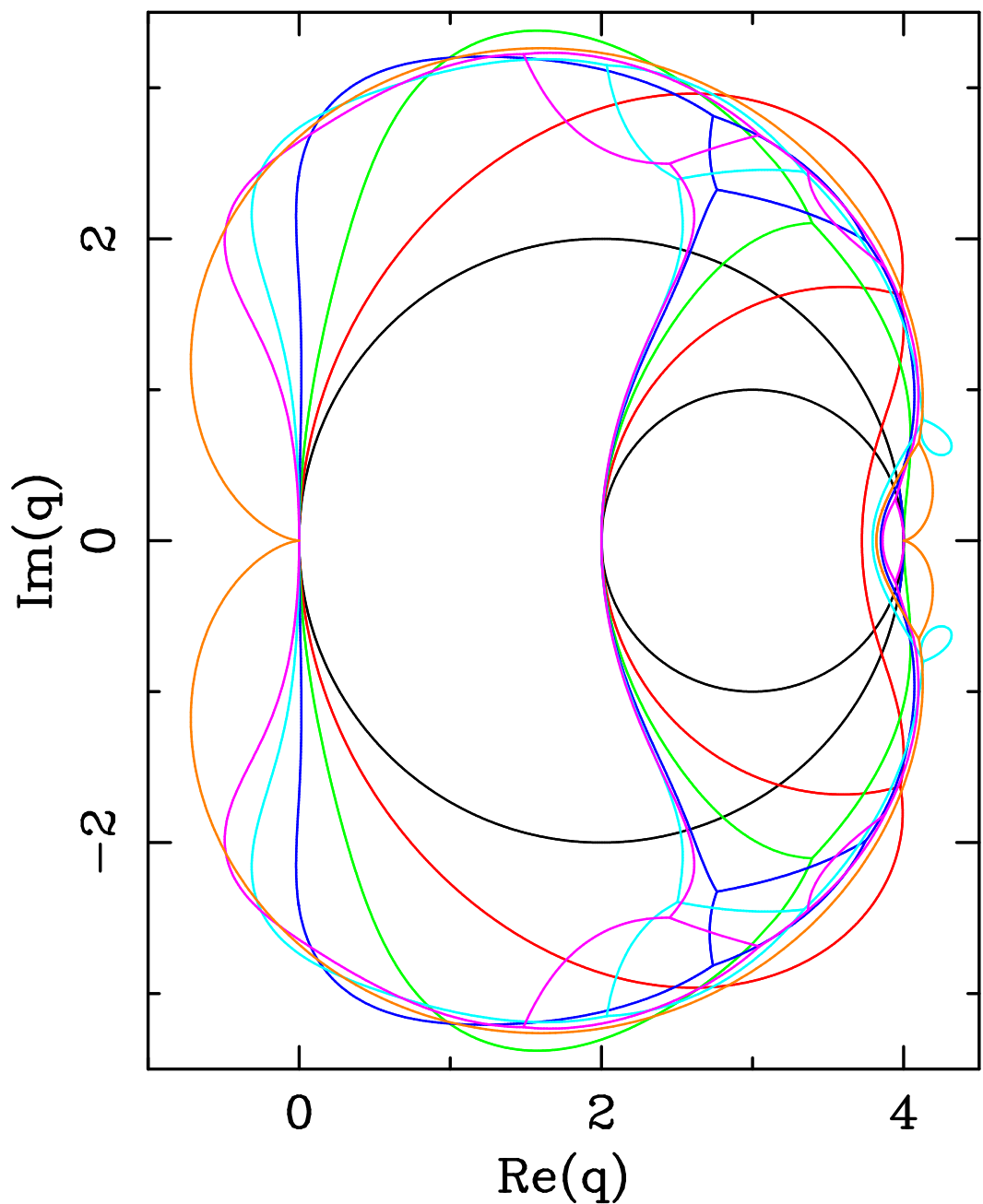}
  \caption{\label{figure_tri_allT}
   Limiting curves for triangular-lattice strips of width $L$ and toroidal 
   boundary conditions. We show the results for $L=2$ (black), $L=3$ (red), 
   $L=4$ (green), $L=5$ (blue), $L=6$ (light blue), and $L=7$ (pink).
   We also show the infinite-width limiting curve $\mathcal{B}_\infty$ 
   obtained by Baxter \protect\cite{Baxter_86_87} (orange).
  }
\end{figure}

\clearpage
%
%
\begin{figure}
  \centering
  \includegraphics[width=400pt]{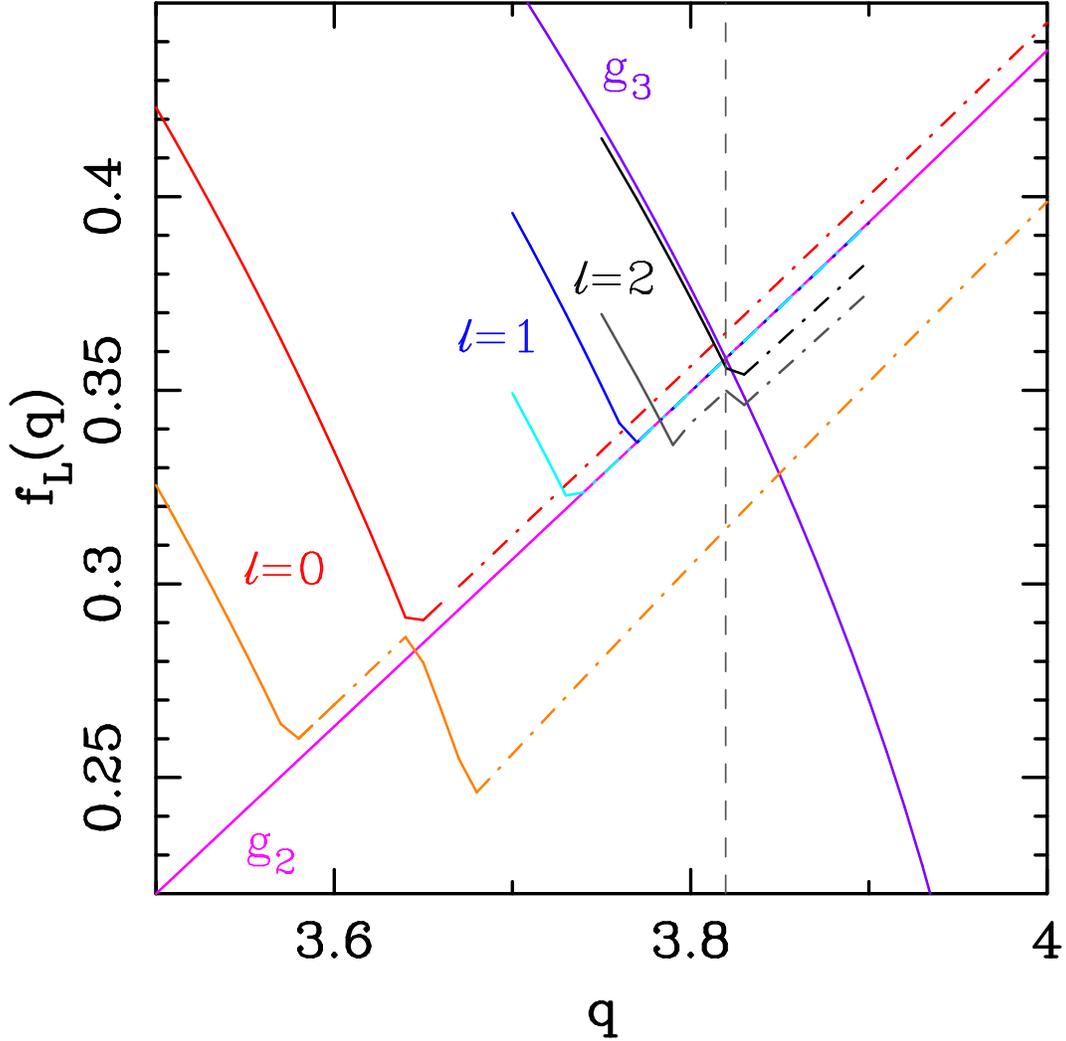}
  \caption{\label{figure_free_tri_L=12}
  Free energy {\em per unit area} $f_L(q)$  
  for the triangular-lattice chromatic polynomial on an 
  infinitely long strip of width $L=12$ and toroidal boundary conditions.
  We show the free energy $f_L(q)$ associated to a selected set of 
  eigenvalues of the transfer matrix $\T(12_\text{T})$ as a function of 
  $q\in(3.5,4)$.  
  For each sector $\ell=0,1,2$, we show the dominant and the first sub-dominant 
  eigenvalues respectively with colors red/orange for $\ell=0$, 
  blue/cyan for $\ell=1$, and black/dark gray for $\ell=3$. Those eigenvalues
  that are decreasing (resp.\  increasing) in $q$ are depicted with a 
  solid (resp.\ dot-dashed) curve.  
  For comparison, we also show Baxter's free energies $(1/G)\log g_2$ and 
  $(1/G)\log g_3$ \protect\cite{Baxter_86_87} normalized {\em per unit area} 
  and depicted as solid pink and violet, respectively. 
  The vertical dashed dark gray line shows Baxter's value $q_\text{B}$
  \protect\reff{def_qB}. 
}
\end{figure}

\clearpage
%
%
\begin{figure}
  \centering
  \begin{tabular}{cc}
  \includegraphics[width=230pt]{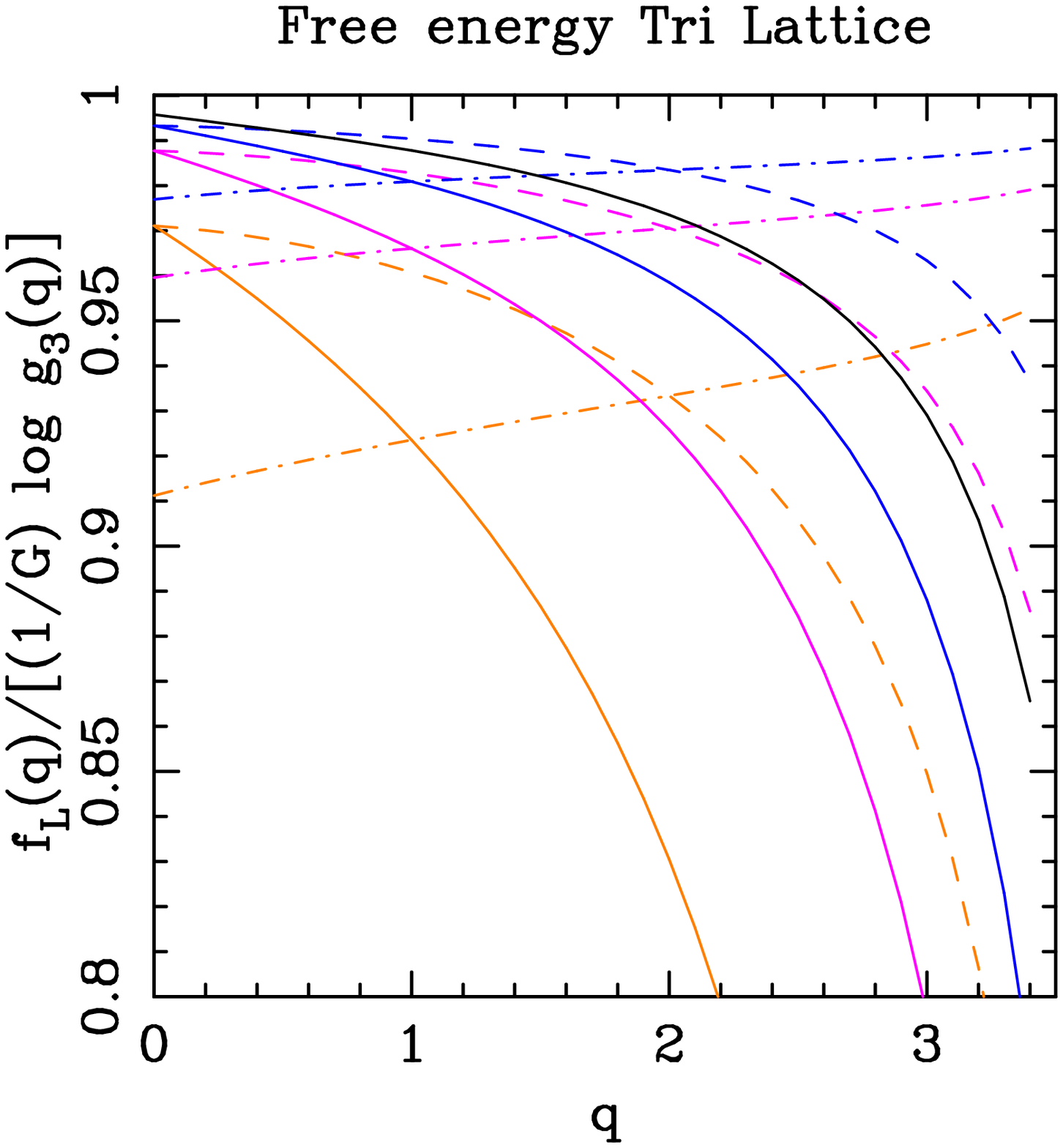} &
  \includegraphics[width=230pt]{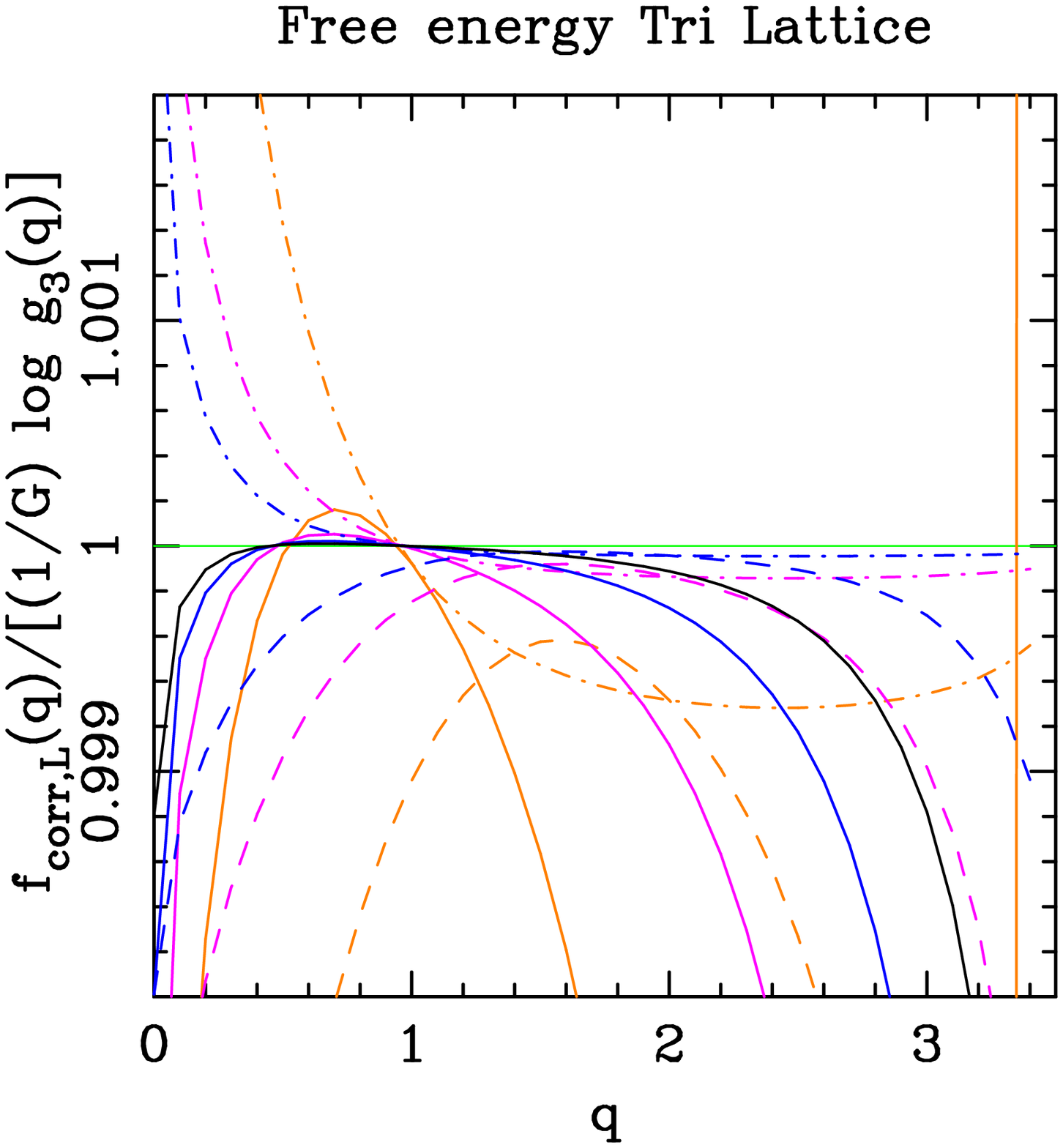} \\[1mm]
   \phantom{(((a)}(a)    & \phantom{(((a)}(b)    \\[5mm]
  \end{tabular}
  \caption{\label{figure_ratio_Baxter}
  Ratios of the free energy {\em per unit area} for an infinitely 
  long triangular-lattice strip of width $L$ and toroidal boundary conditions
  and Baxter's free energy {\em per unit area} $(1/G)\log g_3(q)$
  \protect\cite{Baxter_86_87}.   
  (a) Ratio of the free energy {\em per unit area} 
  $f_{L}(q)$ \protect\reff{def_f_L}. 
  (b) Ratio of the {\em corrected} free energy 
  {\em per unit area} $f_{\text{corr},L}(q)$ \protect\reff{def_f_corrected}. 
  For each strip width $L=6$ (orange), $L=9$ (pink), $L=12$ (blue), and $L=15$
  (black), we show the ratio of the free energies associated to the first three
  levels $\ell=0$ (solid line), $\ell=1$ (dashed line), and 
  $\ell=2$ (dot-dashed lines).  
}
\end{figure}

\clearpage
%
%
\begin{figure}
  \centering
  \includegraphics[width=400pt]{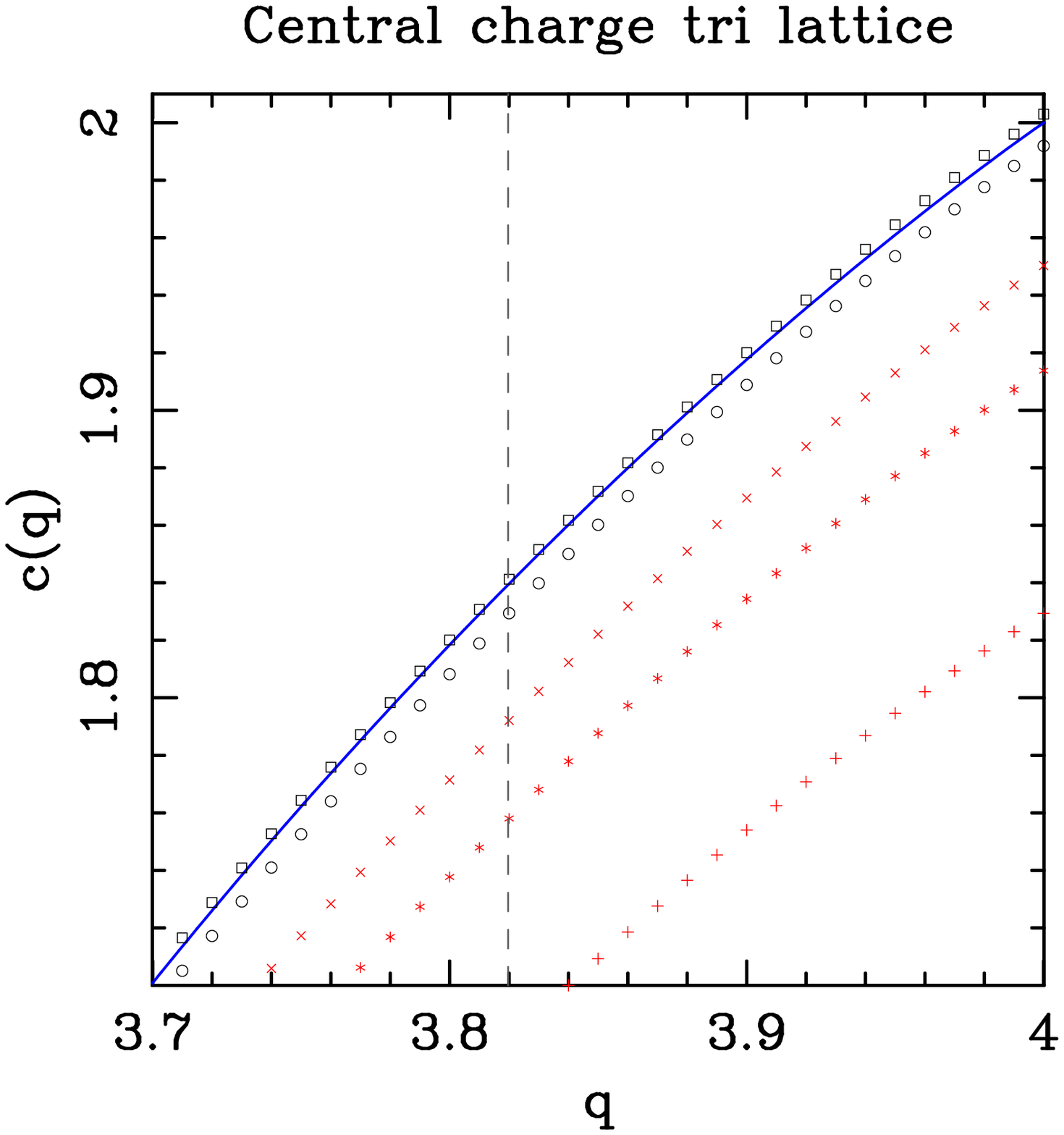}
  \caption{\label{figure_c_tri}
  Central charge for the triangular-lattice chromatic polynomial on 
  infinitely long strips with toroidal boundary conditions.
  We show several two-parameter fits (depicted in red): $L=6,9$ ($+$),
  $L=9,12$ ($\star$), and $L=12,15$ ($\times$). We also show 
  three-parameter fits using the Ansatz \protect\reff{def_fss_free_improved}
  (depicted in black): $L=6,9,12$ ($\circ$),
  and $L=9,12,15$ ($\square$). The solid blue line represents our 
  guess for the central charge 
  \protect\reff{def_c_conjectured}. 
  The vertical dashed dark gray line shows Baxter's value $q_\text{B}$
  \protect\reff{def_qB}. 
}
\end{figure}
   
\clearpage
%
%
\begin{figure}
  \centering
  \begin{tabular}{cc}
  \includegraphics[width=230pt]{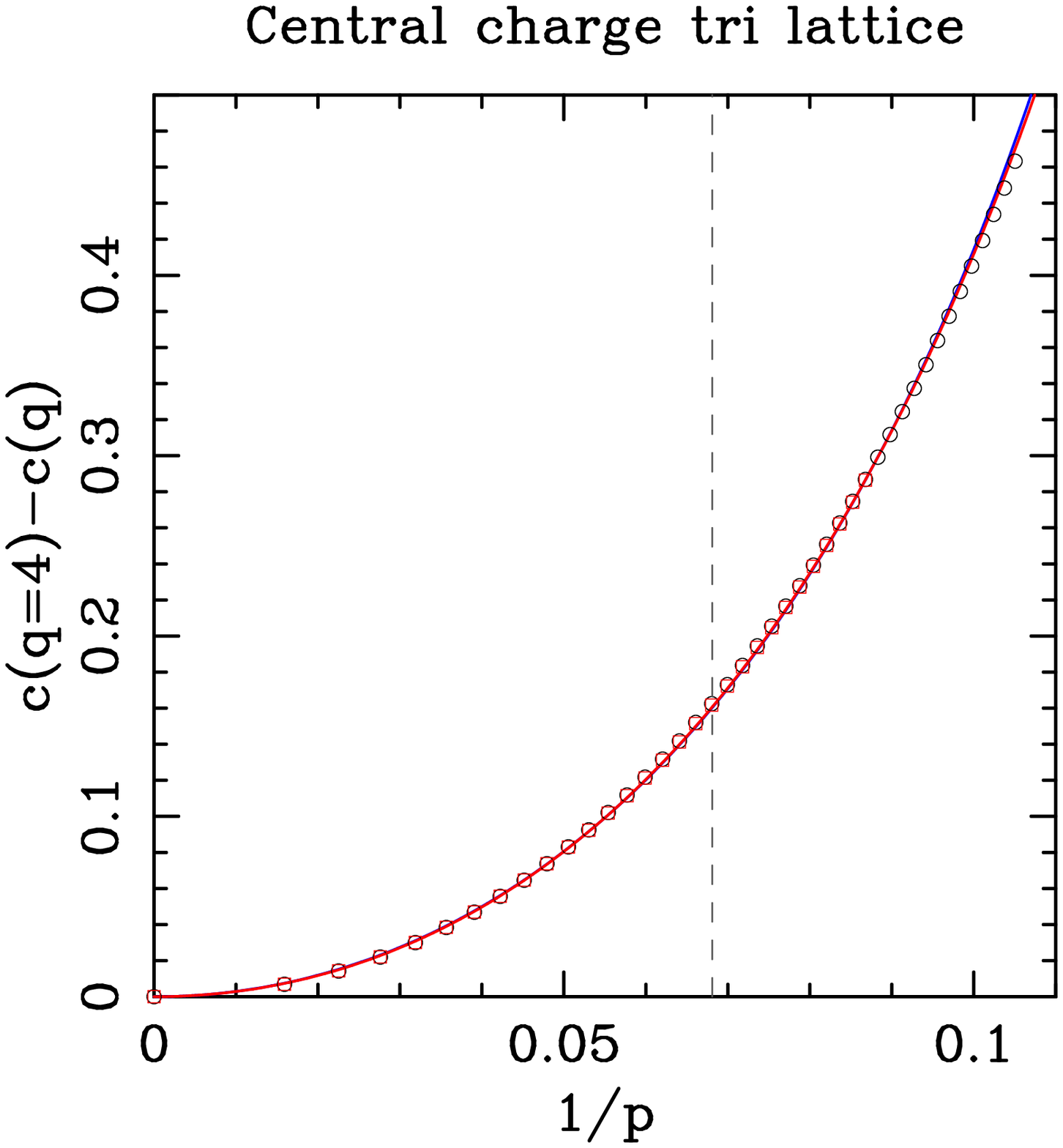} &
  \includegraphics[width=230pt]{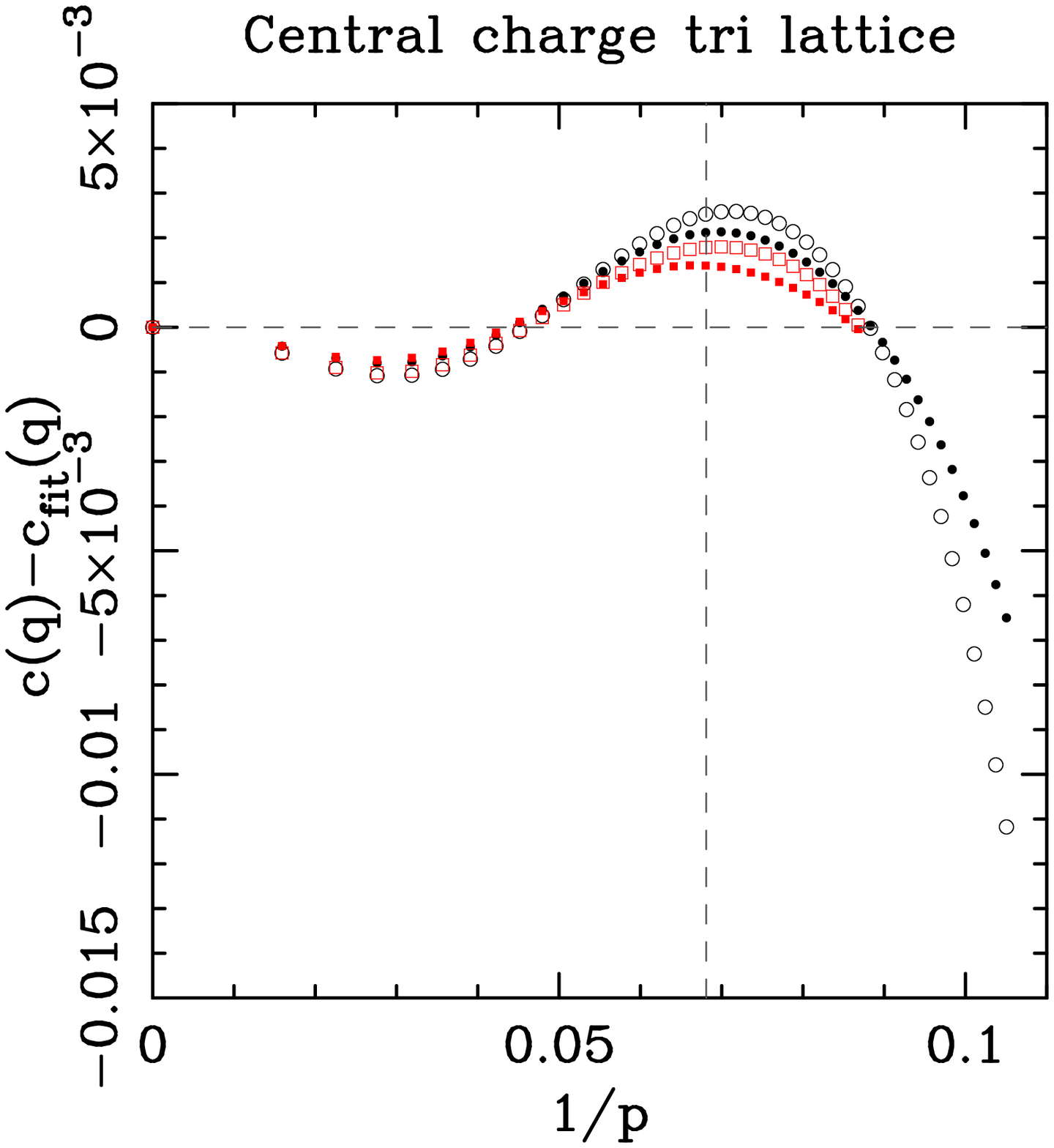} \\[1mm]
   \phantom{(((a)}(a)    & \phantom{(((a)}(b)      \\[5mm]
  \end{tabular}
  \caption{\label{figure_c_tri_raw}
  Central charge for the triangular-lattice chromatic polynomial on 
  infinitely long strips with toroidal boundary conditions. 
  (a) Difference $c(q=4)-c(q)$ as a function of $1/p$ 
  [c.f., \protect\reff{def_p}]. We show two estimates coming from the 
  three-parameter fits \protect\reff{def_fss_free_improved} with 
  $L=6,9,12$ (black $\circ$) and $L=9,12,15$ (red $\square$). 
  The (almost coincident) solid lines represent our two best 
  candidates for the central charge \protect\reff{def_c_diff_conjectured1} 
  (blue), and \protect\reff{def_c_diff_conjectured2} (red). 
  (b) Error between our numerical estimates for the conformal charge and
  our conjectured expressions \protect\reff{def_c_conjectured}. 
  We show the difference $c(q)-c_\text{fit}(q)$ between our 
  three-parameter fits and our conjectured expressions $c_\text{fit}(q)$.
  Points in black (resp.\ red) are obtained from the fits with $L=6,9,12$ 
  (resp.\ $L=9,12,15$). 
  Solid [resp.\  empty] symbols are obtained with $c_\text{fit}$ given by 
  \protect\reff{def_c_diff_conjectured1} 
  [resp.\  \protect\reff{def_c_diff_conjectured2}]. 
  In both plots, the vertical dashed dark gray line shows Baxter's value
  $1/p_\text{B}\approx 0.0681036688$ obtained from
  \protect\reff{def_qB}/\protect\reff{def_p}.
}
\end{figure}

\clearpage
%
%
\begin{figure}
  \centering
  \begin{tabular}{cc}
  \includegraphics[width=230pt]{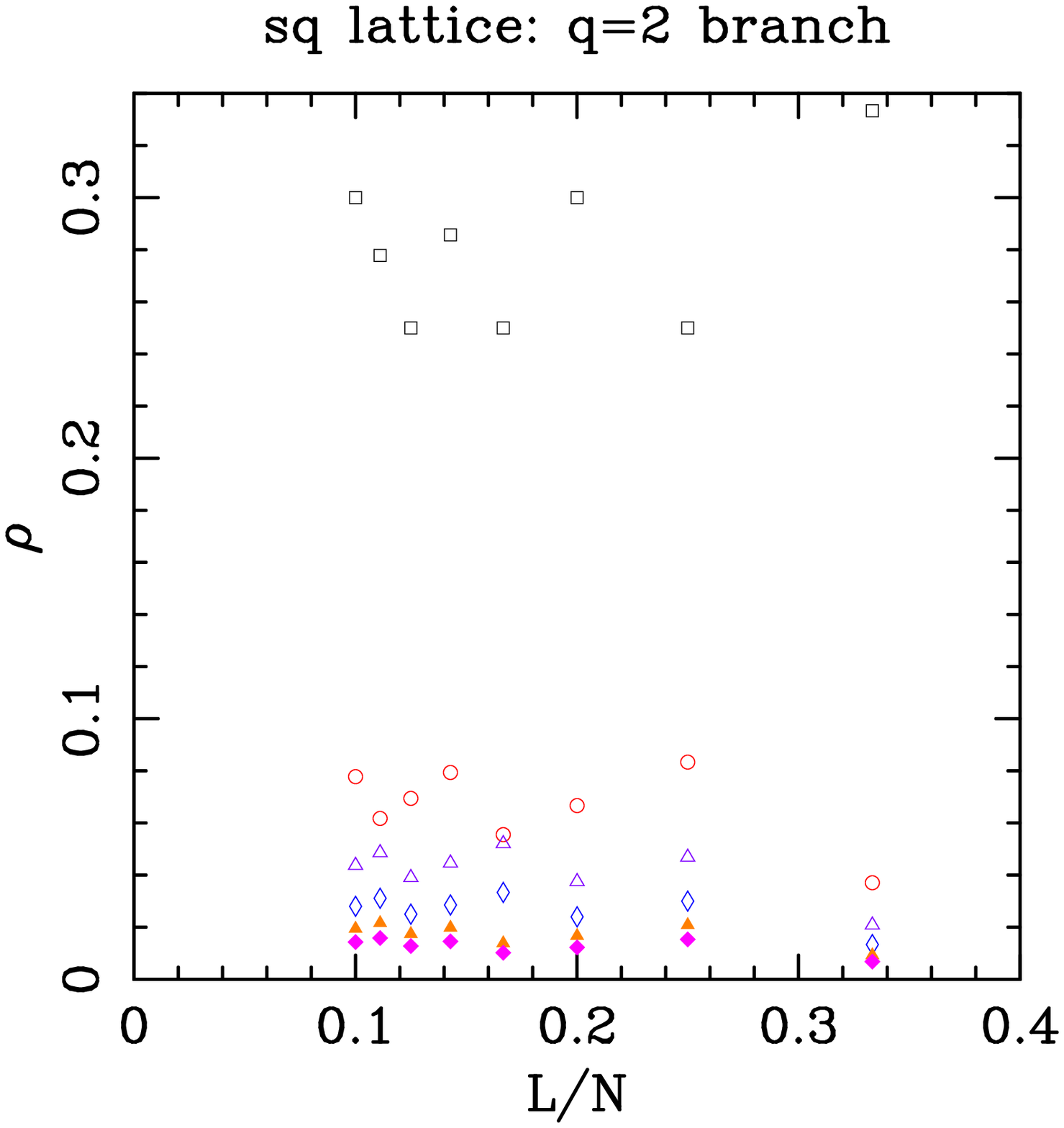} & 
  \includegraphics[width=230pt]{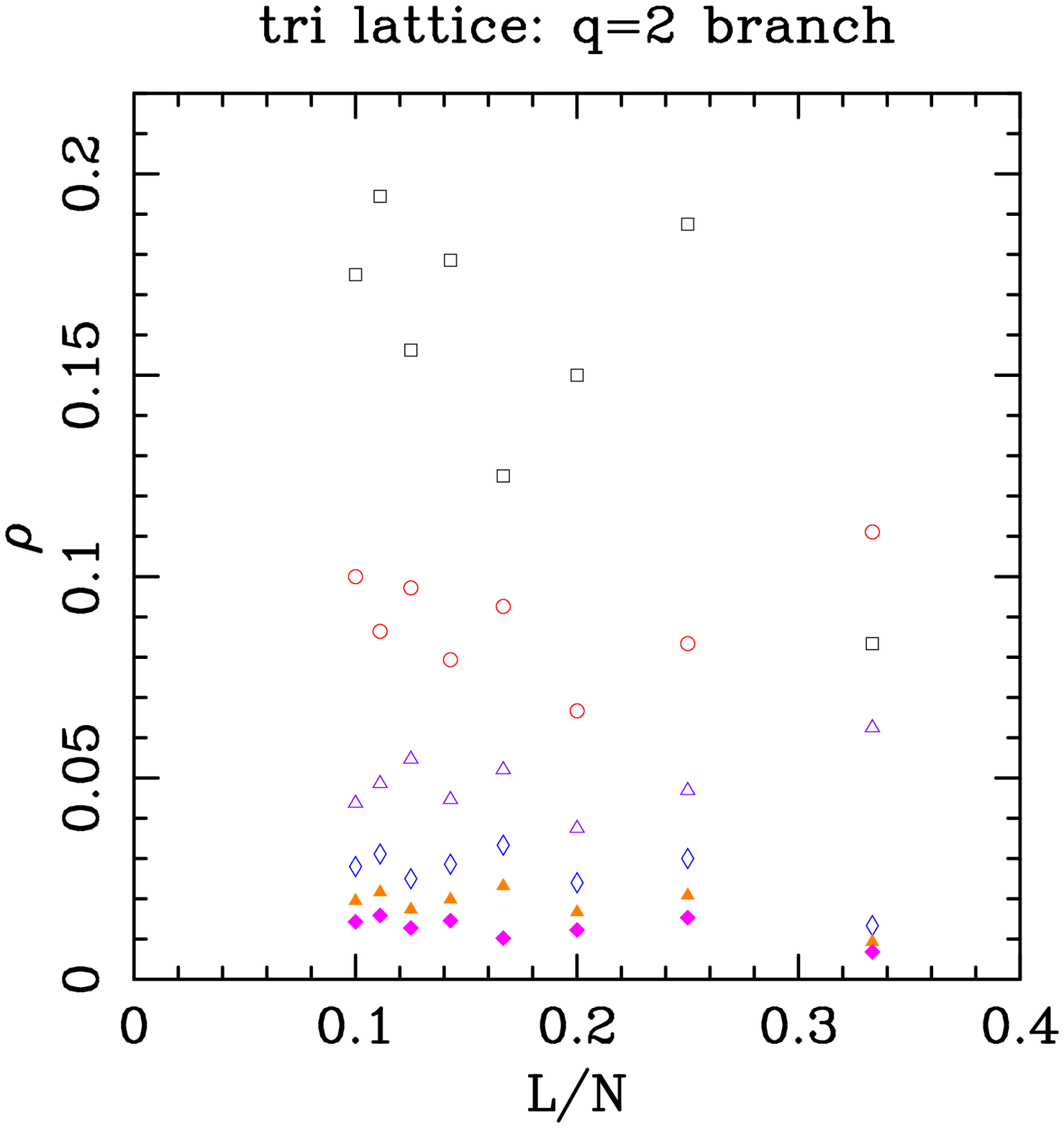} \\[1mm]
   \phantom{(((a)}(a)    & \phantom{(((a)}(b) \\[5mm]
  \end{tabular}
  \caption{\label{figure_density}
  Density of chromatic zeros for (a) square- and 
  (b) triangular-lattice strips with toroidal 
  boundary conditions along the $q=2$ branch of the limiting curve 
  $\mathcal{B}_L$. In order to compute the density of zeros, 
  We have considered the regions given by 
  \protect\reff{def_region_density}. 
  For a given lattice strip of width $L$ and length $N$, we plot the 
  ratio of zeros falling in the respective region as a function of the 
  inverse aspect ratio (i.e., $L/N$).  The widths are $L=2$ (black $\square$),
  $L=3$ (red $\circ$), $L=4$ (violet $\triangle$), $L=5$ (blue $\diamond$),
  $L=6$ (orange $\blacktriangle$), and $L=7$ (pink $\blacklozenge$). 
}
\end{figure}

\clearpage
%
%
\begin{figure}
  \centering
  \begin{tabular}{cc}
  \includegraphics[width=230pt]{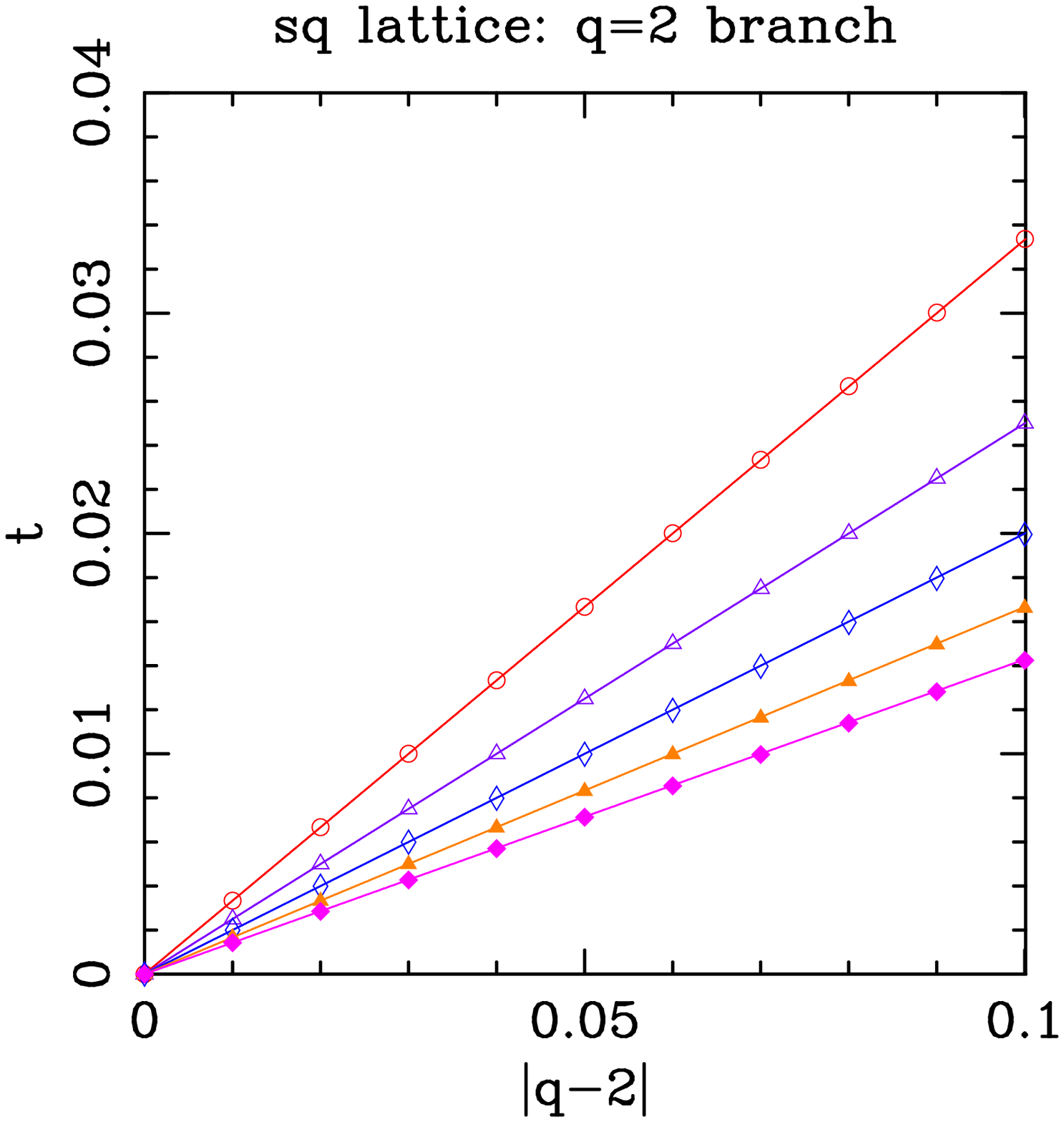} & 
  \includegraphics[width=230pt]{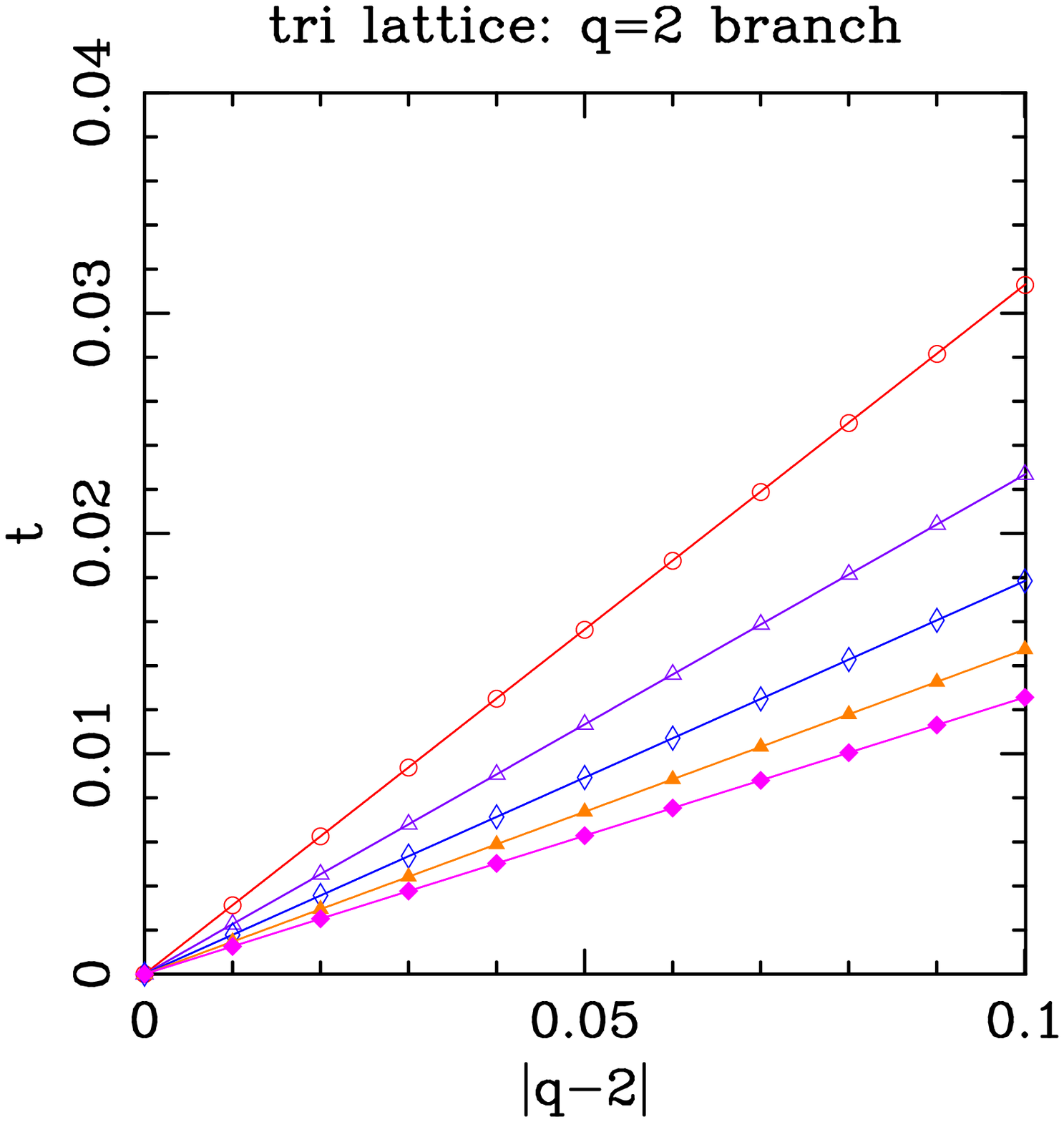} \\[1mm]
   \phantom{(((a)}(a)    & \phantom{(((a)}(b) \\[5mm]
  \end{tabular}
  \caption{\label{figure_theta}
  Values of the parameter $t$ \protect\reff{def_t} 
  along the $q=2$ branch of the limiting curve $\mathcal{B}_L$ for the 
  (a) square- and (b) triangular-lattice strip.
  For a lattice strip of width $L$, we show the value of $t$ as a function
  of $|q-2|$. The widths are 
  $L=3$ (red $\circ$), $L=4$ (violet $\triangle$), $L=5$ (blue $\diamond$),
  $L=6$ (orange $\blacktriangle$), and $L=7$ (pink $\blacklozenge$). 
  The slopes of the depicted lines are given approximately by $1/L$ (square) 
  and $\sqrt{3}/(2L) + 0.653/L^3$ (triangular).
}
\end{figure}

\clearpage
%
%
\begin{figure}
  \centering
  \begin{tabular}{cc}
  \includegraphics[width=230pt]{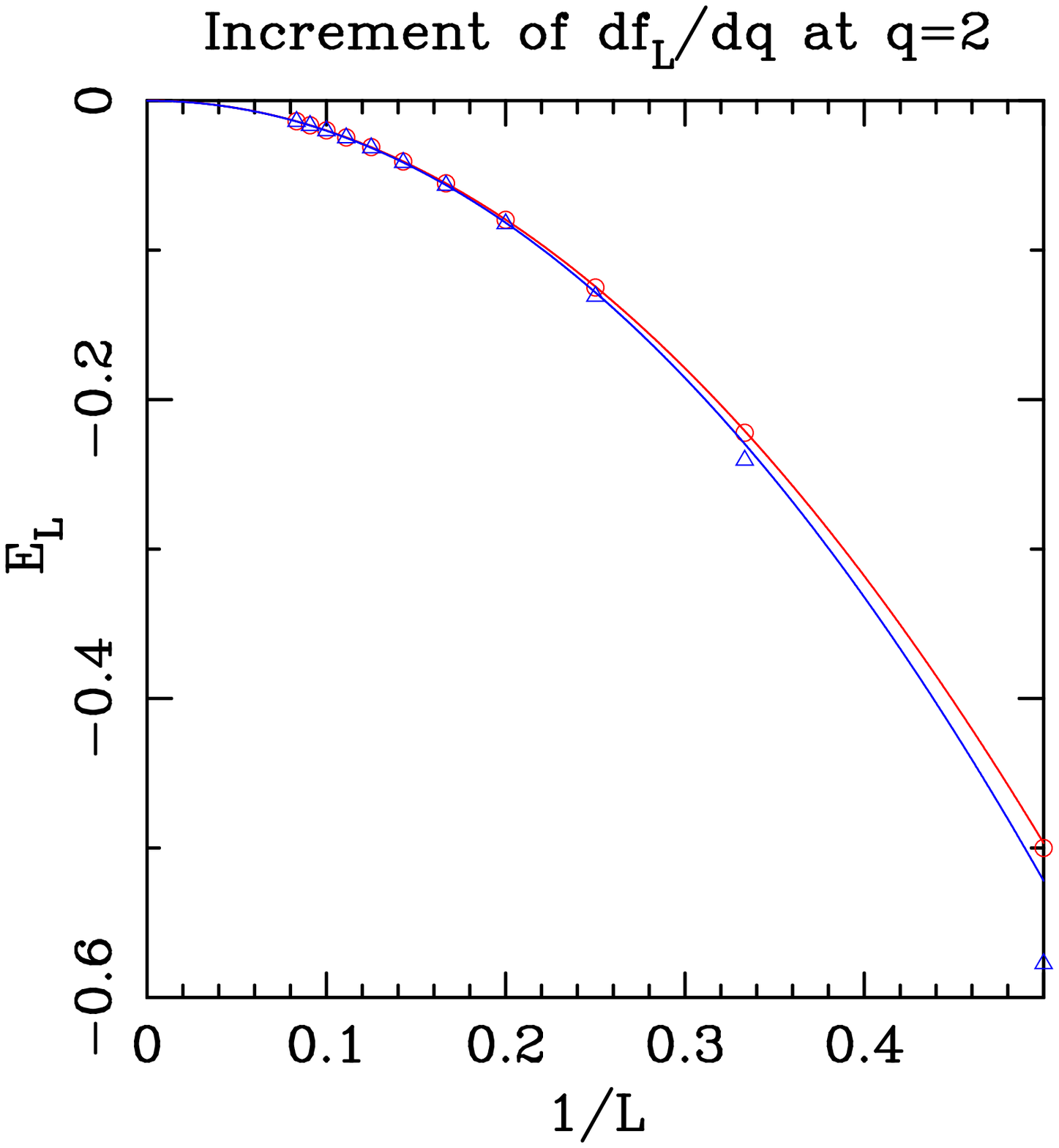} & 
  \includegraphics[width=230pt]{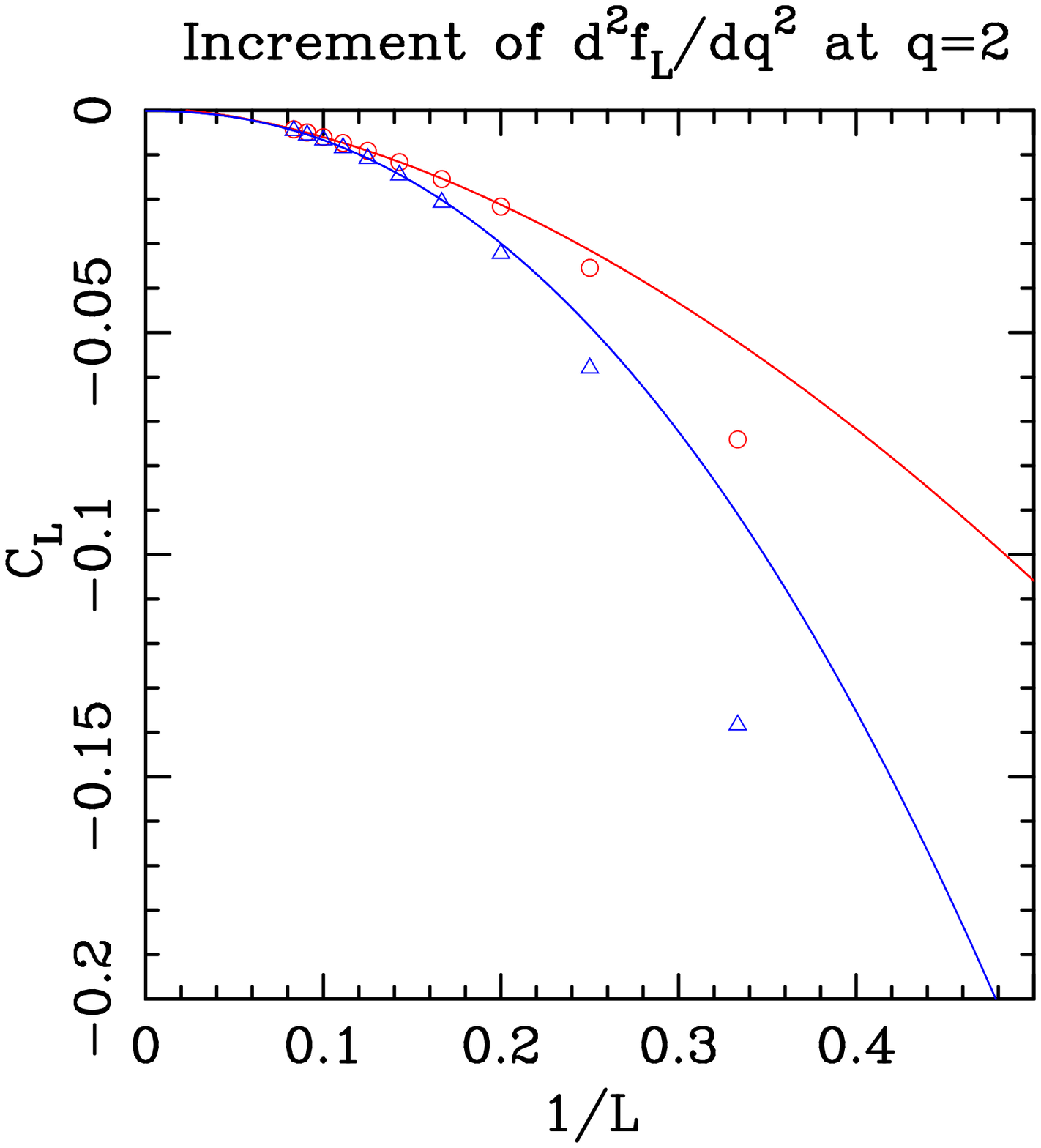} \\[1mm]
   \phantom{(((a)}(a)    & \phantom{(((a)}(b) \\[5mm]
  \end{tabular}
  \caption{\label{figure_series}
  Coefficients of the series expansions for the increment of the first (a)
  and second (b) derivatives of the free energy $f_L(q)$ with respect to
  $q$ evaluated at $q=2$. In each plot we show the corresponding coefficient
  for the square (red $\circ$) and triangular (blue $\triangle$) lattices as a 
  function of $1/L$. The solid lines depict the best power-law fits to 
  the corresponding data set 
  \protect\reff{def_fit_series_E}/\protect\reff{def_fit_series_Ceven}.  
}
\end{figure}

\end{document}